\documentclass{ar-1col-S2O}

\usepackage[comma]{natbib}
\usepackage{url}

\usepackage{graphicx}
\usepackage{pxfonts}
\usepackage{hyperref}
\usepackage{enumitem}
\usepackage[textsize=small]{todonotes}

\newcommand{\Msun}{\ensuremath{M_\odot}}

\newcommand{\Rsun}{\ensuremath{R_\odot}}

\newcommand{\vsini}{\ensuremath{v\,\sin i}}
\newcommand{\lsim}{\mathrel{\hbox{\rlap{\lower.55ex \hbox {$\sim$}}
 \kern-.3em \raise.4ex \hbox{$<$}}}}
\newcommand{\gsim}{\mathrel{\hbox{\rlap{\lower.55ex \hbox {$\sim$}}
 \kern-.3em \raise.4ex \hbox{$>$}}}}

\renewcommand\figurename{Supplemental Figure}
\jname{Annu.~Rev.~Astron.~Astrophys.}
\jvol{63}
\jyear{2025}
\doi{10.1146/annurev-astro-071221-054402}

\begin{document}

\markboth{Mathieu \& Pols}{Evolutionary pathways}

\title{Supplemental Appendix: \\ Blue Stragglers and Friends: Initial Evolutionary Pathways in Close Low-Mass Binaries - Literature Review}

\author{Robert D. Mathieu$^1$ and Onno R. Pols$^2$
\affil{$^1$Department of Astronomy, University of Wisconsin - Madison, Madison WI, USA; email: mathieu@astro.wisc.edu}
\affil{$^2$Department of Astrophysics/IMAPP, Radboud University, Nijmegen, The Netherlands; email: o.pols@astro.ru.nl}}

\maketitle

\tableofcontents

\section*{OBSERVED PROPERTIES OF FIRST-STAGE BINARY EVOLUTION PRODUCTS - LITERATURE REVIEW}\label{sec:properties}

This Supplemental Appendix provides an extensive literature review of observations of the products of first-stage binary evolution. It is intended to support and expand on Section 3 of \textit{Blue Stragglers and Friends: Initial Evolutionary Pathways in Close Low-Mass Binaries} (ARAA 63), where an integrated perspective on the common and contrasting astrophysical properties
of these products is provided. We suggest first reading Sections 1-3 in the Main Text to provide context for the detailed information here. Also, figures used in the Main Text to highlight key observational results are referenced here.

The scope of this literature review is the evolution of close binary stars having components with $M < 2~\Msun$. An observational taxonomy for the products of the first stage of evolution of these binaries is presented in Section 2 of the Main Text and comprises dwarfs ("blue stragglers"), giants ("yellow stragglers"), subdwarf B stars, and giant-like stars ("sub-subgiants" and "red stragglers"). This Supplemental Appendix is organized according to this taxonomy within three distinct environments – open star clusters, globular star clusters and the Galactic field. 

The closing date of this review is January 17, 2025, with some citations subsequently updated.

\section{Open Clusters}\label{sec:properties_OC}

\subsection{Catalogs and Populations}\label{sec:properties_OC_catalogs}

Current studies of blue straggler (BSS) populations in open clusters owe a major debt to the pioneering catalog work of \cite{Ahumada+Lapasset1995,Ahumada+Lapasset2007}. With the \emph{Gaia} data releases, this work has been redone---including yellow straggler (YSS) populations---for 408 open clusters of all ages by \cite{Rain+2021}, with particular advances in the quality of cluster proper-motion membership determinations. They identify 897 BSSs in these clusters. The number of open clusters with at least one BSS is 111 (27\%). \citeauthor{Rain+2021} find a rapid increase in the BSS frequency (relative to main-sequence turnoff (MSTO) stars) for clusters older than log t (yr) \textgreater{} 8.7, and indeed conclude that BSSs are absent in very young open clusters (Main Text Figure 2). Greatly expanding knowledge of the YSS population in open clusters, \cite{Rain+2021} also identified 77 YSSs in 43 clusters (11\%), with most hosting only one YSS.

\cite{Cordoni+2023} compare the BSS populations in 78 open clusters with binary populations determined from photometry. They too find that BSSs are not present in clusters with ages of less than $\sim$300 Myr. They also suggest a positive correlation between the fraction of BSSs and the fraction of binaries among 8 open clusters with higher central densities.


A similar analysis of BSS and YSS populations in open clusters has been done by \cite{Jadhav+Subramaniam2021} for 670 open clusters with ages of log t (yr) \textgreater{} 8.5, also using the \emph{Gaia} DR2 data release. They find 868 BSSs in 228 clusters, along with an additional 508 probable BSSs, or in total 1368 BSS candidates in 304 clusters. \cite{Jadhav+Subramaniam2021} find an increase in BSS number with cluster age and mass, and perhaps cluster density. They do not find meaningful correlations of BSS frequency with any other cluster parameters (including age), other than relaxation time.

The WIYN Open Cluster Study (WOCS; \citep{Mathieu2000}) has completed comprehensive studies of 7 rich open clusters older than 1 Gyr---NGC 7789 \citep[1.6 Gyr;][]{Nine+2020}, NGC 2506 \citep[2.0 Gyr; ][]{Linck+2024a}, NGC 6819 \citep[2.5 Gyr;][]{Milliman+2014}, NGC 2682/M67 \citep[4 Gyr;][]{Geller+2015, Geller+2021a}, NGC 188 \citep[7 Gyr;][]{Geller+2008,  Geller+2009,Mathieu+Geller2015}, and NGC 6791 \citep[7 Gyr;][]{tofflemire+2014, Milliman+2016}. These studies provide time-series high-precision radial-velocity measurements over long time baselines (up to several decades) for the BSSs, YSSs and sub-subgiants (SSGs) in these clusters. These radial-velocity catalogs yield both a third dimension for kinematic membership probabilities and nearly complete detection of binaries with P \textless{} 10\textsuperscript{4} days, with many having orbit solutions.

A comprehensive compilation of SSGs and red stragglers (RSSs) in clusters, and their associated data, was done by \cite{Geller+2017}, including such stars in 6 open clusters. Interestingly, the specific frequency of SSGs seems to increase with lower-mass clusters, with SSGs far more frequent in open clusters than in globular clusters. 

As an overarching closing note, because binary evolution products lie---by definition for many---off of the single- and (non-interacting) binary-star isochrones for their associated clusters, field contamination is a constant concern. Typically cluster membership is secured by 2- or 3-dimensional kinematic association and parallax for the closer clusters, with very substantial improvement from Gaia measurements. Recently hierarchical density-based clustering analyses have also been applied. Even so, these techniques yield only membership \emph{probabilities}, and this particularly must be remembered when intensively studying any one object.


\subsection{Blue Stragglers}\label{sec:properties_OC_BSS}

\subsubsection{Color-Magnitude Diagrams}\label{sec:properties_OC_BSS_CMD} 
There are many open cluster color-magnitude diagrams (CMDs) in the literature showing the presence of BSSs and other binary evolution products. We do not attempt a comprehensive review of them here. Rather, we draw from \cite{Leiner+Geller2021} who provide an in-depth analysis of the CMDs of 16 old (1--10 Gyr), nearby (d \textless{} 3500 pc) open clusters with masses greater than 200 \Msun, building from Gaia proper-motion membership probabilities. The age range of these open clusters corresponds to MSTO masses of 1--2 \Msun. Main Text Figure 2 shows the cluster CMDs as a function of age. 
(The individual CMDs for each of the 16 clusters can be found in the appendix of \citeauthor{Leiner+Geller2021}.)


\citeauthor{Leiner+Geller2021} note the presence of a gap among the CMD distribution of the BSSs in the middle-aged (2-4 Gyr) open clusters, and a possible gap among the oldest open clusters (\textgreater{} 4 Gyr). No gap is evident among the younger clusters. 70\% of the BSSs in the older clusters fall on the fainter side of the gap, with their frequency peaking at a mass of 0.3~\Msun\ above the MSTO mass. (Here BSS mass is determined from MIST evolutionary tracks \citep{Choi+2016}.) More broadly, \citeauthor{Leiner+Geller2021} find the BSS mass distribution to be skewed toward lower masses. They find only 5 BSSs with masses more than 1~\Msun\ above the associated cluster MSTO mass, and none with masses more than 1.5 \Msun~above the MSTO. Thus they note that none of the BSSs have masses greater than can be produced by mass transfer (which would require merger or collisional formation, or multiple events). Finally, they find that the frequency of BSSs relative to red giant branch (RGB) stars increases with cluster age until $\sim$4 Gyr, after which the ratio flattens. They suggest that older clusters produce more BSSs, and/or that their BSS populations are longer lived.

We emphasize that many classical BSSs in these open clusters do not fall on the zero-age main sequence (ZAMS) of their respective coeval populations (see also Supplemental Figure~\ref{fig:NGC188UVCMD}). In the context of single-star evolution theory, this is a reflection of increasing (and varying) helium abundances of their cores. If so, their distribution in CMDs can reveal the evolutionary state of their progenitor secondary stars in the mass-transfer formation scenario or of their progenitor primary stars in the merger or collision formation scenarios.

Classical BSSs lie in photometrically distinguishing locations of CMDs. However, lower-mass BSSs may be embedded photometrically within main sequences, and thus require additional diagnostics for identification. Rapid rotation is predicted by all of mass transfer, merger, and collision formation mechanisms. Using K2 light
curves and $\vsini$ measurements, \cite{Leiner+2019} identified 11 blue lurker (BL) candidates in M67 as rapidly rotating main-sequence (MS) stars (2 days $< P_\mathrm{rot} <$ 8 days), without nearby companions that might provide tidal spin-up. With subsequent Gaia cluster memberships removing one star and using Gaia photometry, 6 of these stars are securely embedded within the M67 MS as BLs (Main Text Figure 1). These stars have much in common with the BSSs, including a high fraction of long-period spectroscopic binary companions; a number of long-period, modest-eccentricity orbits; and in one case a UV excess consistent with the presence of a hot white dwarf (WD) companion \citep{Nine+2023, Leiner+2025}.

\subsubsection{Binary Properties}\label{sec:properties_OC_BSS_binary}

\paragraph{Binary Frequency} BSSs have long been suggested to be the products of binary evolution via processes either internal or external to the progenitor binary \citep{McCrea1964,Leonard1989}. However, systematic surveys of BSS binarity in open clusters have only begun in recent decades. Radial-velocity time-series studies of BSSs in open clusters were launched by D. Latham in the 1980's, with observations in 6 clusters (1.6--5 Gyr) summarized by \cite{Milone+Latham1994} and with
findings for BSSs in M67 presented by \cite{Milone+Latham1992}, \cite{Milone+1992} and \cite{Latham2007}. These observations continue today.

Recent WOCS long-baseline time-series spectroscopic observations and follow-up studies of the 20 BSSs in the old open cluster NGC 188 (7 Gyr) have born fruit in a wealth of astrophysical insights, which we focus on here as an exemplar open cluster. The BSS binary frequency (P \textless{} 10\textsuperscript{4} days) of 80\% $\pm$ 20\% is more than three times larger than the NGC 188 solar-type MS binary frequency of 23\% $\pm$ 2\% in the same period range \citep{Mathieu+Geller2009,Mathieu+Geller2015,Geller+Mathieu2012}. Incompleteness studies leave open the possibility that some of the four non-velocity-variable BSSs are long-period binaries, perhaps even within this period regime. In M67 the BSS binary frequency is very similar at 79\% $\pm$ 24\% (as is the solar-type MS binary frequency; \cite{Geller+2015}).

\paragraph{Periods and Eccentricities}

\begin{figure}
\centerline{\includegraphics[width=1.0\textwidth]{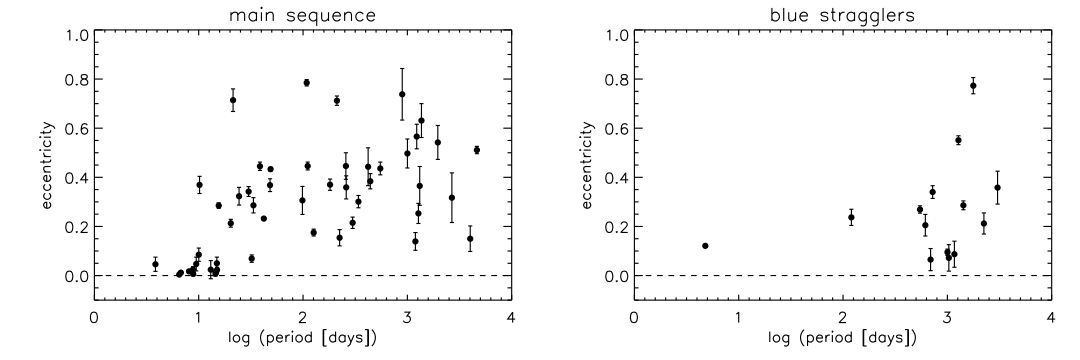}}
\caption{Orbital eccentricity against log period for 1.1 \Msun ~MS binary stars (left panel) and for BSS binary stars (right panel) in the open cluster NGC 188. Figure adapted from \cite{Mathieu+Geller2015} (CC BY 4.0).}
\label{fig:MathieuGeller_elogP}
\end{figure}

The distributions of two astrophysically important orbital parameters---period P and eccentricity e---are shown in e-P diagrams for both the BSS and the solar-type MS binary stars of NGC 188 (Supplemental Figure~\ref{fig:MathieuGeller_elogP}). The period distribution of the BSS binaries is evidently different from that of the MS binaries. Most of the BSS binaries have orbital periods within a half-decade of 1000 days, whereas the MS binaries in NGC 188 (and in other open clusters and the Galactic field) continuously populate all periods from only a few days to 10,000 days. The upper limit on the BSS orbital periods is an observational limit.

The eccentricity distribution of the long-period (P
\textgreater{} 100 days) BSSs also differs from the MS binaries; the mean eccentricity for the long-period BSS binaries is 0.27 ± 0.05, while the mean eccentricity for the MS binaries is higher at 0.42 $\pm$ 0.04. The formal likelihood of the two distributions being the same is
\textless{} 1\% \citep{Mathieu+Geller2015}.

In detail, the long-period eccentricity distribution has several facets likely to shed light on the physical origin of these BSSs. Three long-period BSS binaries have eccentricity measurements consistent with circular orbits. None of the solar-type MS binaries with periods longer than the tidal circularization period of 14.5 days show such low orbital eccentricities, and they are similarly rare among solar-type binaries in other open clusters \citep{Meibom+Mathieu2005} and the field \citep{Raghavan+2010}. The BSS circular orbits suggest tidal or other dissipative processes, as often assumed for mass transfer from an earlier evolved companion. However, most of the long-period BSS binaries have significant orbital eccentricities, with some quite high, which poses an interesting challenge to formation scenarios for BSSs (Main Text Section 4.2.3). The e-P diagram for the BSSs in M67 tells much the same story \citep{Latham2007, Latham+Milone1996}.

The one short-period BSS binary in Supplemental Figure~\ref{fig:MathieuGeller_elogP} is the double-lined binary WOCS 5078,\footnote{\cite{Mathieu+Geller2009} also report WOCS 7782 as a double-lined short-period BSS binary with two BSSs companions. However more precise Gaia proper-motions show it to be a non-member.} with an orbital eccentricity of 0.12 at an orbital period of 4.8 days, well below the tidal circularization period. Curiously, the BSS binary WOCS 1007 in M67 has rather similar orbital properties, with an orbital period of 4.2 days and an eccentricity of 0.21 \citep{Milone+Latham1992}. From light curve analyses \cite{vernekarPhotometricVariabilityBlue2023} find this latter binary to have a hot WD companion with a mass of 0.22 $\pm$ 0.05 $\Msun$.  That the only BSSs in these two rich open clusters with periods less than 100 days both have eccentric orbits at such short periods is notable.

In the context of short-period BSS binaries, the findings of \cite{Rain+2021a} in the somewhat younger cluster Trumpler 5 are interesting. Using radial-velocity variation amplitudes, \citeauthor{Rain+2021} found that 3 of 4 BSS were close binaries with amplitudes of $\sim$40 km s\textsuperscript{-1} or more. All three are rapid rotators, with \vsini\ greater than 100 km s\textsuperscript{-1}. \citeauthor{Rain+2021} suggest these to be ``probable contact'' binaries. Given more than 40 BSS candidates in this cluster, comprehensive radial-velocity and photometric surveys clearly are in order. In the old (7-9 Gyr) open cluster Collinder 261, \cite{Rain+2020} also find 5 of 9 BSSs with multiple epochs to be close binaries. Three of these are cited in the literature as eclipsing, two with sub-day periods and one semi-detached. Upper limits on their projected rotation velocities suggest that they do not show the rapid rotations seen in Trumpler 5.

Finally, the shortest period binaries in NGC 188 are four long-known W UMa stars \citep{Baliunas+Guinan1985} and three more discovered by \cite{Kaluzny+Shara1987}, all with periods of less than 0.6 days. Yet more were identified by \cite{Zhang+2004}. This rich population of contact binaries in an old population such as NGC 188 has been often noted. (M67 for example has only three identified contact binaries; but see also \cite{DeMarchi+2007} regarding NGC 6791.) These are likely progenitors of BSSs formed in the merger channel; a modern population analysis is merited.

\paragraph{Companions}

The statistically derived companion masses of the NGC 188 single-lined BSSs were striking for the narrow peak of their mass distribution between 0.4 \Msun~and 0.6 \Msun~  \citep{Geller+Mathieu2011}. These are masses of He or CO WDs, indicative of the cores of evolved stars left behind after RGB or asymptotic giant branch (AGB) mass transfer, respectively.

\begin{figure}
\centerline{\includegraphics[width=1.3\textwidth]{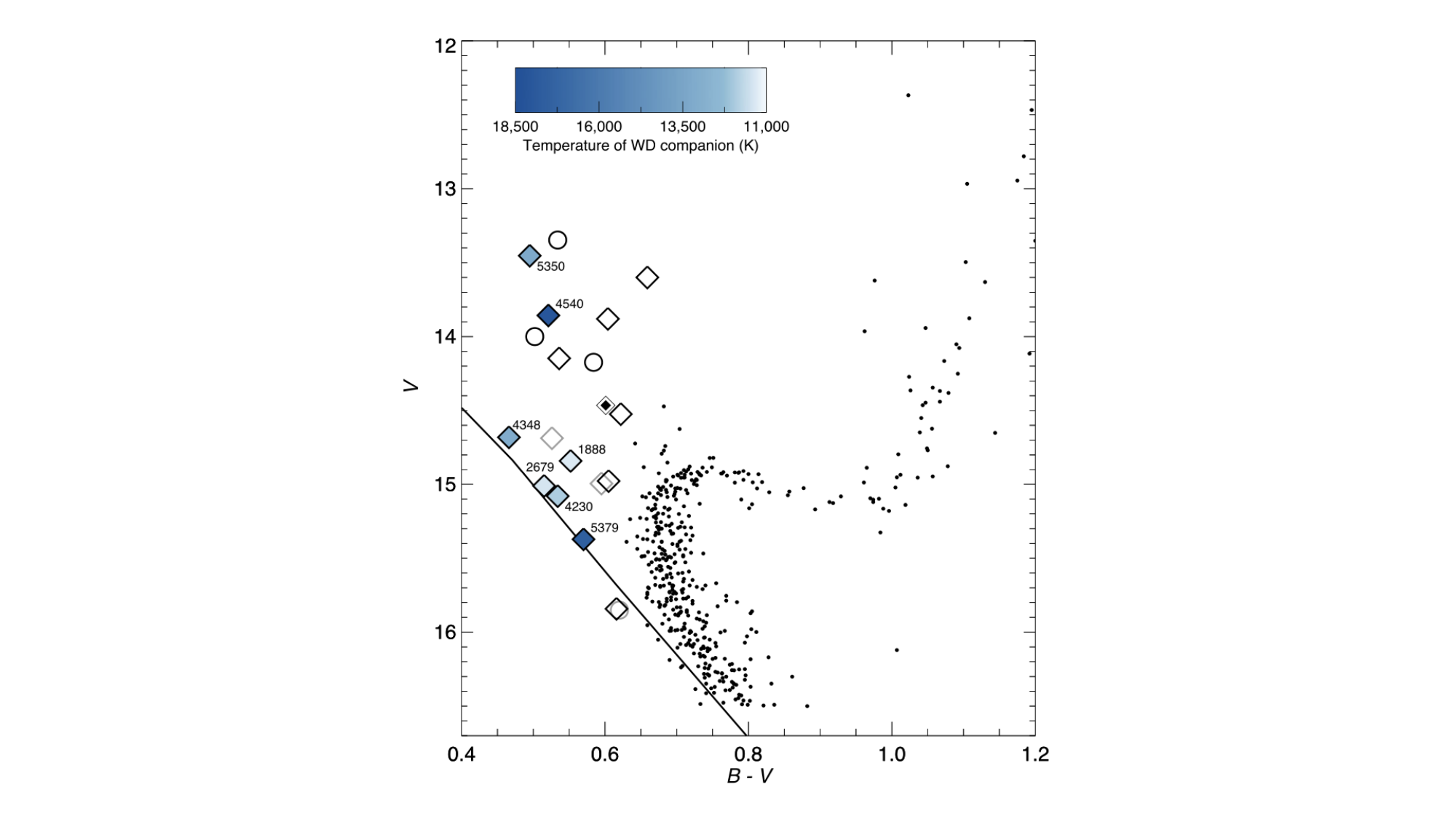}}
\caption{Optical CMD of NGC 188 cluster members with the BSS population highlighted according to binarity and the measured temperatures of WD companions. The solid black line is the ZAMS for NGC 188. Binary BSSs are shown as diamonds and non-velocity variable BSSs are shown as circles. The BSSs with FUV WD detections are shown with a color from dark blue to light blue representing the temperature of the WD companion, as indicated with the color bar. The outlined solid black diamond is a double-lined BSS binary. Figure adapted from \cite{Gosnell+2015}, with permission from the AAS.}
\label{fig:NGC188UVCMD}
\end{figure}

Ensuing \textit{Hubble Space Telescope} (HST)/ACS far-ultraviolet (FUV) photometry revealed hot WD companions to four long-period BSSs in NGC 188, with three more detected statistically (\cite{Gosnell+2014,Gosnell+2015}; Supplemental Figure \ref{fig:NGC188UVCMD}; see also \cite{Rani+2021}). With effective temperatures higher than 13,000 K, the four BSSs with directly detected WDs have transformation ages of less than 250 Myr, extremely recent in comparison to the 7 Gyr age of NGC 188. Importantly, two of the detections, including the BSS with the hottest and youngest WD, lie far off the ZAMS (Supplemental Figure \ref{fig:NGC188UVCMD}), implying that these BSSs formed recently with already evolved cores. Given that only the hottest WDs were detectable and other cooler WD companions must also be present, \cite{Gosnell+2015} argue that of order two-thirds of the NGC 188 BSSs have WD companions and formed from mass transfer.

\cite{Nine+2023} observed 8 BL candidates identified by \cite{Leiner+2019} in M67 with HST/ACS, detecting FUV excesses in WOCS 3001 and WOCS 14020. Both of these binaries have periods of less than 400 days, suggesting RGB mass transfer. The WD temperatures are cooler than found for the NGC 188 BSSs, ranging from 10,500 K to 12,000 K depending on physical assumptions for a He WD. Consequently the transformation ages (Main Text Section 3) are somewhat larger, ranging from 300 Myr to 900 Myr.

FUV high-resolution spectra were obtained with HST/COS for two of the UV-excess BSS in NGC 188 \citep{Gosnell+2019}, from which masses and cooling ages of the WD companions were determined: WOCS 4540, 0.53 \Msun~$\pm$ 0.03 \Msun, 105 Myr $\pm$ 6 Myr and WOCS 5379, 0.42 \Msun~± 0.02 \Msun, 250 Myr $\pm$ 20 Myr. \citeauthor{Gosnell+2019} conclude that WOCS 4540 contains a CO WD resulting from AGB mass transfer and that WOCS 5379 contains a He WD from RGB mass transfer, the latter of which they find to challenge current instability expectations for RGB mass transfer.

Given these precise WD mass measurements and the comprehensive observational definition of the BSSs and their binary orbits, mass-transfer evolutionary histories of these BSSs are constrained in detail. Both have been modeled with \textit{Modules for Experiments in Stellar Astrophysics} \citep[MESA;][]{paxton+2011}  as AGB and RGB mass transfer, respectively, reproducing in much detail both systems \citep{ Sun+2021, Sun+Mathieu2023} (see sidebars in Main Text).

\cite{Leiner+2025} followed the photometric FUV detection of the BL WOCS 14020 with an HST/COS spectrum. Unexpectedly, they measured a WD companion mass of 0.72 $\Msun$ with a cooling age of 400 Myr. A WD of such mass requires a $\approx 3 \Msun$ progenitor, more than twice the M67 MSTO mass. Leiner et al. argue that the evolution of the system began with a merger of the inner binary in a triple, when the MSTO mass was larger, followed by an AGB common envelope event with the original tertiary becoming the current BL.

The UVIT Open Cluster Study (UOCS) has reported a large number of UV detections in 10 open clusters with the \textit{Ultraviolet Imaging Telescope}. They detect UV excesses among BSSs, YSSs, red clump (RC) stars and MS stars near MSTOs (BLs). UOCS classifications and analyses of inferred hot companions are founded on UV-NIR two-component spectral-energy-distribution (SED) fitting. Effective temperatures and luminosities of the companions are derived from UV excesses. Masses and ages are determined photometrically from (primarily) WD evolutionary tracks. The determined masses span extremely low mass (ELM) WDs to He WDs to massive CO WDs. Indeed \cite{Panthi+2024} identify the entire range of these WD types among the BSS companions in the old cluster NGC 7142. They conclude
that the companion frequency in NGC 7142 implies that mass transfer and mergers are responsible for 60\% of the BSSs. 

\begin{figure}[h!]
\centerline{\includegraphics[width=0.6\textwidth]{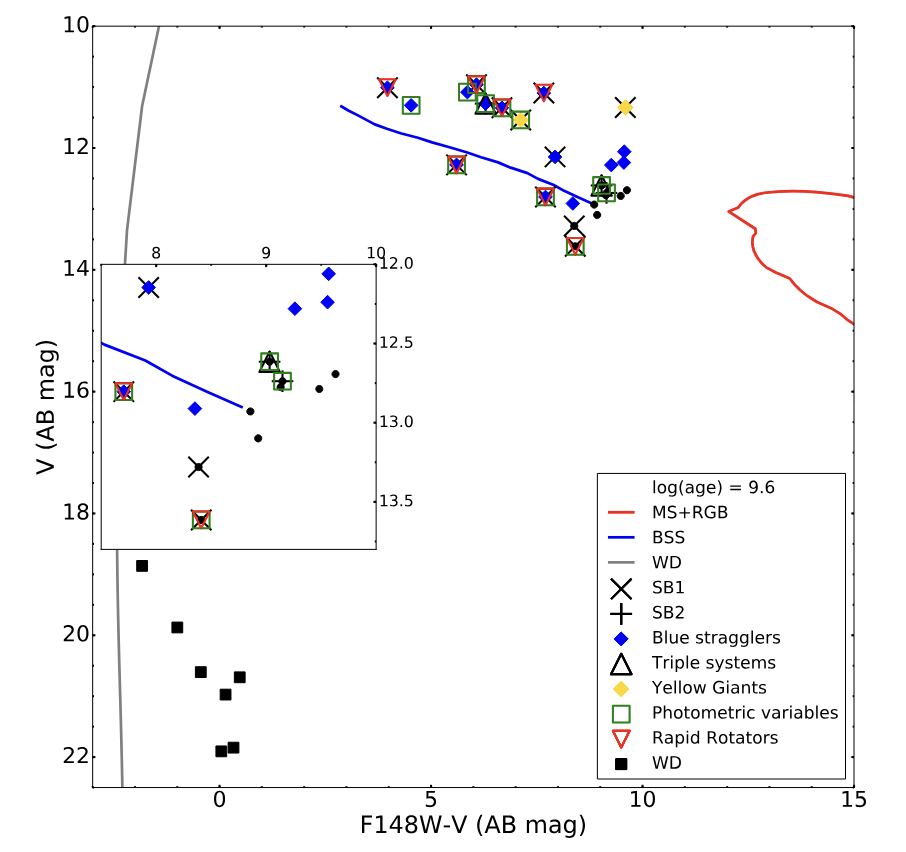}}
\caption{UV–optical CMD of M67. The inset shows the expanded view of the fainter end of the BSS sequence. UVIT detections are shown with classifications. The isochrone (log age 9.6) is based on the BaSTI model. The MS and subgiant branch is shown in red; expected locations of BSSs and WDs are shown in blue and gray, respectively. UOCS identifies a subset of 11 detections as having secure FUV excesses - 6 BSSs, 2 YSSs and 3 MS stars. Figure adapted from \cite{Jadhav+2019}, with permission of the AAS.} 
\label{fig:M67UVcmd}
\end{figure}

As an exemplar cluster, UOCS detected 41 members of M67 in the FUV \citep{Jadhav+2019}. Supplemental Figure~\ref{fig:M67UVcmd} shows both optical and optical-FUV CMDs for M67. UOCS identified UV excesses for 6 BSSs, 2 YSSs and 3 MS stars \citep{Sindhu+2019, Jadhav+2019, Subramaniam+2020, Pandey+2021} Again due to the UV detection bias to hotter temperatures, all of the WD companions are very young with transformation ages of $\approx$ 200 Myr or less. In contrast to the NGC 188 spectroscopic mass measurements, these UV photometric studies of M67 have found only low-mass ($< 0.3~\Msun$) He or ELM WD companions. Combining the UOCS M67 studies, \cite{Pandey+2021} find that 5 of the 10 analyzed M67vBSSs have WD companions, from which they place a lower limit on mass-transfer formation of M67 BSSs of 36$\%$.

Across their open cluster sample UOCS finds numerous ELM companions associated with BSSs. Such companions to BSSs are notable given frequent expectations of common envelope or other stripping mechanisms for ELM origins, processes which are not thought to greatly increase the mass of the progenitor secondary (Main Text Section 4.1). These processes are also thought to produce short-period binary products. One UOCS ELM is in a 4190-day period binary.
Others show no radial-velocity variation over long timescales at precisions lower than 1 km s\textsuperscript{-1}. \cite{Subramaniam+2020} suggest that the stars WOCS 6006 and WOCS 11005 in M67 both may be BLs with ELM companions, but neither show radial-velocity variation over $\approx$ 20 years \citep{Geller+2015}.

E. Linck (private communication) analyzed the M67 BSSs discussed by \cite{Pandey+2021}, using the orbital elements of \cite{Latham+Milone1996}. Adopting BSS masses of 1.6 \Msun, the dynamical minimum masses of WOCS 5005 and WOCS 3013 do not permit the small WD masses determined by Pandey et al. Pandey et al.~also note that the masses of their three ELMs are undermassive compared to the theoretical mass relation for stable mass transfer of \cite{Rappaport+1995} (Main Text Section 4.2.3.1). 

\cite{Linck+2024a} perform a detailed comparison of photometrically determined hot-companion masses to BSSs in NGC 2506 \citep{Panthi+2022} with extensive time-series radial-velocity measurements over at least 14 years, including binary orbit solutions when available. Two FUV-excess BSSs interpreted as ELM WD companions show no radial-velocity variation at precisions of less than 1 km s\textsuperscript{-1}. These measurements do not permit ELM companions in short-period orbits unless both cases have low inclination angles. Another suggested low-mass (0.2 \Msun) WD companion to a BSS is marginally permitted by the dynamical mass function of its orbit solution, if the binary is edge on. The dynamical mass function for an RC member also marginally permits the suggested WD mass. 

Similar radial-velocity measurements and orbital solutions are available for the 5 BSSs identified as having candidate ELM companions in NGC 7789 (\cite{Vaidya+2022}; see \cite{Nine+2020}). Again, 3 do not show radial-velocity variation. Another shows a very long orbital period of 4190 days, suggestive of AGB wind Roche-Lobe overflow which would not produce an ELM. Vaidya et al. suggest the latter system is a triple, with the ELM in a close inner orbit. 

Further comparison of derived masses for hot binary companions with dynamical mass constraints will be important.

\subsubsection{Masses}\label{sec:properties_OC_BSS_masses} 
Very often in the literature the masses of BSSs - and BLs and
YSSs - are determined from single-star evolutionary tracks, on the
presumption that, after a thermal relaxation time since their formation,
these stars settle into standard interior structures for their masses
and core composition. Innumerable cluster CMDs in the literature have such tracks superimposed
(e.g., Main Text Figure 1). Unfortunately, very few BSSs
have dynamically measured masses to test this
presumption.

The double-lined BSS binary WOCS 5078 ($P$ = 4.783 days) in NGC 188 provided a step in
this direction \citep{Mathieu+Geller2015}. Spectroscopic analysis of the
companion-star effective temperature yields a MS mass between 0.94
$\Msun$ and 1.08 $\Msun$ from a 
single-star isochrone.
The dynamical mass ratio
from the orbital solution then yields a BSS mass between 1.39 $\Msun$ and 1.59 $\Msun$.
Compared to the single-star evolutionary tracks of \citet{Marigo+2008},
the maximum inferred mass for the primary star is between 1.2
$\Msun$ and 1.3 $\Msun$ (and for the combined
light 1.3 $\Msun$). Thus in this analysis single-star evolution
models underestimate the BSS mass by $\approx 15\%$.

A particularly interesting system is the double-lined eclipsing BSS WOCS 2009 (S1082) in M67. \cite{Goranskij+1992}
showed WOCS 2009 to have partial eclipses with a
period of 1.07 days. Subsequent analyses found the combined light to be
from (at least) three stars, including a BSS in the
double-lined eclipsing binary and a more luminous BSS in a
single-lined binary with a 1189-day eccentric orbit \citep{vandenBerg+2001,Sandquist+2003}. Recent speckle observations find a companion at a projected separation of 340 AU, which may suggest the system is a quadruple in which each binary has a BSS (E. Horch, private communication).

Solution of the eclipsing binary shows the BSS to be on the ZAMS and the companion on the MS
\citep{Sandquist+2003}\footnote{The
  CMD locations found for the stars in the eclipsing binary
  are substantially different between \cite{vandenBerg+2001} and
  \cite{Sandquist+2003}, largely the result of Sandquist et al.~including spectroscopically derived effective temperatures in their
  analyses. This system merits additional careful study.}. 
The crucial
radial-velocity measurements for this system are difficult because of
both the luminous third star and rotational line broadening. Recently, D. Latham (private communication) has obtained a new SB1 orbital
solution which yields a primary mass of $1.28 \pm 0.12~\Msun$
and a secondary mass of $0.97 \pm 0.07~\Msun$ (based on the
Sandquist et al. inclination). The MIST single-star evolution models overestimate the dynamical BSS mass by $\approx 20\%$ (A. Quitral, private communication).

\cite{vernekarPhotometricVariabilityBlue2023} analyse ellipsoidal variability found in  K2 and TESS light curves for the BSS binary WOCS 1007 in M67 ($P=4.21$ days). They find the BSS mass to be $1.95 \pm 0.26~\Msun$ and the companion mass $0.22 \pm 0.05~\Msun$, with radii of $2.54 \pm 0.81~\Rsun$
and $0.078 \pm 0.027~\Rsun$, respectively. \cite{vernekarPhotometricVariabilityBlue2023} do not make a detailed comparison of the BSS mass with single-star evolution models, but note that their measurements are similar to others in the literature and from isochrone fitting (BASTI model; \cite{Sindhu+2018}). They argue that the secondary mass and radius indicate a helium WD and are consistent with results found from UOCS SED analyses \citep{Sindhu+2019}.

\cite{Brogaard+2018} determine the mass of the double-lined BSS binary
V106 ($P=1.4463$ days) in the old open cluster NGC 6791. Determining the inclination from the projected radius, they find a BSS mass of $1.62 \pm
0.21~\Msun$ and a companion mass $0.171
\pm 0.023~\Msun$, which they suggest is a
proto-ELM WD because of its currently large radius. They then use an
alternative stellar-model dependent approach to improve precision on the
mass measurement, and thus do not independently test single-star evolution models. They do use MESA mass-transfer modeling to reproduce the system, with a transformation age of only 40 Myr.

In summary, these assorted dynamical techniques yield masses at or above the cluster MSTO masses. Further, they suggest an agreement level of $\approx 20\%$ between various forms of dynamical mass determinations and masses from single-star evolution models, which themselves have systematic differences of $\approx 10\%$. 

All of these BSSs currently have close companions, and most likely have had prior interactions. That caution said, there is no evidence of current mass transfer, and their stars fit within their Roche radii as detached binaries.

\subsubsection{Rotation}\label{sec:properties_OC_BSS_rotation} 
\cite{Leiner+2018} examined the rotation periods among the NGC 188 BSSs (derived from \vsini) as a function of transformation age (from cooling ages of WD companions), combining them with three even younger BSSs and two older BSSs in the Galactic field. \cite{Nine+2023} added two BLs with WD companions to the sample. Main Text Figure 6 shows their finding of very rapid rotation (approaching 30\% of breakup) at transformation ages of 5-10 Myr, and subsequent spin-down over a Gyr. Leiner et al. concluded that BSSs form rapidly rotating, and then spin down at rates very similar to single solar-type MS stars. Such initial rapid rotation is consistent with expectations from all BSS formation mechanisms (see Section 4 in Main Text), while the subsequent spin-down opens opportunity for gyrochronological dating of BSSs. 

Similarly, BLs were discovered among M67 MS stars based on rotation rates much higher than expected at an age of 4 Gyr. The presence of such rapidly rotating MS stars also is consistent with theoretical expectations for BSS formation from lower-mass progenitor companions. 

\cite{Mathieu+Geller2009} also found that NGC 188 BSSs have higher projected rotation velocities than MS stars in the cluster. In addition, they found a decrease in projected rotation velocity with decreasing effective temperature among the BSSs in NGC 6819 and NGC 188, as also found for MS stars spanning the same temperature range. These several findings of  BSS rotation evolution suggest that the magnetic fields and winds of late-type BSSs are typical of late-type MS stars.

Finally, \cite{Subramaniam+2020} (see also \cite{Pandey+2021}) find a moderate correlation between lower projected rotation velocity and lower WD-companion effective temperature (higher transformation age) for M67 BSSs and BLs. This may be indicative of BSS spin-down with age, although the M67 BSSs have effective temperatures hotter than the Kraft break.


\subsubsection{Surface Abundances}\label{sec:properties_OC_BSS_abundances}

One of the first comprehensive abundance
studies of BSSs in an open cluster was done by \cite{Mathys1991} for 11 BSSs in M67,
with two BSSs particularly intensively analysed. The spectroscopic effective temperatures
and gravities are typical of similar A-type stars. Mathys found the
abundances of most elements to be solar to within a factor two. (See
also \cite{Shetrone+Sandquist2000}.) However, the heavy elements zinc,
strontium, yttrium, zirconium and barium are strongly overabundant.
Mathys does not find the abundance results to provide clear clues as to
the nature or origin of these BSS. No evidence for organized magnetic
fields was found. Curiously, Mathys suggested that the M67 BSSs are
rotating more slowly than other stars of similar spectral type. More
recently, \cite{BertelliMotta+2018}  have used APOGEE infrared spectra
to study the surface abundances of three BSS in M67. They find the
BSS abundances to be the same as the M67 MSTO, essentially solar.

AGB stars have thermal pulses that lead to the production of
\emph{s}-process elements that can be tracers of mass transfer. Excess barium was found in five BSS in NGC 6819 (1.6 Gyr) by \cite{Milliman+2015}. Surprisingly, these barium-excess stars are not in
long-period binaries as typical for barium dwarfs and giants \citep{Jorissen+2019}, and as expected from an AGB
mass-transfer origin for barium excesses. Also notable, these five BSS are localized within a
limited domain of the CMD.

Subsequently \cite{Nine+2024} completed a study of barium abundances in classical BSSs in
the open clusters NGC 7789, NGC 6819 and M67, spanning 2.4 Gyr. Fifteen of 35 BSSs in these three clusters show enhanced barium,
with a possible increased enrichment with younger age. The
Ba-enriched BSSs  lie in the same region of the HR diagram
irrespective of cluster age (Main Text Figure 5). Nine et al. note the increasing
separation of the Ba-enriched BSSs from the cluster MSTOs with
increasing cluster age, and suggest a link between AGB donor mass and
mass-transfer efficiency in the sense that less massive AGB donors tend to
undergo more conservative mass transfer. Some of the barium-enriched
BSSs are in long-period binaries, but again most are not detected as velocity variable. Nine et al. hypothesize that
these are binaries in periods longer than the detection limit of
$10^4$ days, indicative of AGB-driven wind mass transfer origins. The seemingly high frequency of mass-transfer formation from thermally pulsing AGB stars may reflect on mass transfer stability broadly, and calls for a population synthesis analysis.

\cite{Pal+2024} detected sodium and barium, ytrium and lanthanum (\emph{s}-process) excesses in the M67 BSS WOCS 9005, and concluded that this BSS formed as the result of wind mass transfer from an AGB star. Specifically, they find that 0.15 \Msun~was transferred onto a 0.45 \Msun ~MS progenitor companion. \cite{Nine+2024} did not detect a barium excess, which \cite{Pal+2024} attribute to different values employed for microturbulence. \cite{Pal+2024} argue for detection of a UV excess from broadband UV flux measurements, especially from UVIT, and derive a WD companion with a mass of 0.55 $\Msun$, a radius of 0.02 $\Rsun$, and a transformation age of $\approx$ 60 Myr. They note that \cite{Nine+2023} did not detect a UV excess using HST-derived narrow-band observations. The HST broadband flux measurements agree with the UVIT measurements, but \cite{Nine+2023} cautioned against flux contamination of these broadband UV observations from the BSS. The measurement and analysis of UV excesses in the presence of luminous F-type stars can be nuanced.


Lithium is a potentially interesting abundance diagnostic to distinguish between BSS formation mechanisms, although the predictions are not secure. \cite{Carney+2005} note that 5 of their 6 binary halo BSSs are depleted in lithium (although one lies in the lithium gap), which they argue is consistent with mass transfer. As yet, attempts to detect lithium in the M67 BSSs have yielded only upper limits (e.g., \cite{Shetrone+Sandquist2000}.  Shetrone \& Sandquist conclude that these limits are consistent with a formation process without mixing.

\subsubsection{Spatial Distributions}\label{sec:properties_OC_BSS_spatial}
\cite{Mathieu+Latham1986} showed that
the radial distribution of BSSs in M67 are centrally concentrated, and in
fact closely follow the distribution of
spectroscopic binary stars as well as a multi-mass King model for 2.0
\Msun~stars. Presuming that the stellar distributions are
in a relaxed equilibrium, they argue that the radial distributions are
dynamical evidence that the BSS are more massive than the cluster MSTO
mass.

The radial distribution of the BSS in NGC 188 is more complex in that it
shows a bimodal distribution much as found in many globular clusters
(Section \ref{sec:properties_GC_BSS_spatial}; \cite{Geller+2008}; discussed in more detail in \cite{Geller+2013}). This distribution has a centrally concentrated population
comprising two-thirds of the BSS, and a more extended distribution of
the remainder in the cluster halo. The studied NGC 188 radial
distribution extended to 13 core radii, while the studied M67 radial
distribution extended only to 5 core radii. Compared over similar
domains, they are not distinct at a statistically significant level.

\cite{Bhattacharya+2019} also find a bimodal distribution in the very
old (10 Gyr) cluster Berkeley 17, containing a population of 23 BSS
members. Thereafter, \cite{Vaidya+2020} used the Gaia DR2 release to
study the relative radial distributions of BSSs and RGBs in 7 rich open
clusters with ages from 1 -- 7 Gyr. They find that five of the clusters
(including NGC 188) show bimodal relative radial distributions. They also argue for a
positive correlation with the current central relaxation times.

\subsubsection{Summary}\label{sec:properties_OC_BSS_summary} 
Recent studies of BSSs in old open clusters have
established the following:

\begin{itemize}
\item
  A binary frequency of $\approx$ 80\% for orbital periods less than
  10\textsuperscript{4} days;
\item
  Primarily long orbital periods $\approx$ 1000 days, with important cases (including W UMa stars) at
  very short periods;
\item
  A dispersion of orbital eccentricities at long periods, from circular to e = 0.8;
\item
  Detected hot WD companion stars, some with spectroscopically measured masses;
\item
  Rapid rotation at formation, with subsequent spin-down with age;
\item
  Surface abundance variations from solar, most notably of the \emph{s}-process 
  element barium;
\item
  Bimodal radial spatial distributions.
\end{itemize}
  This is a remarkably complete characterization of the physical
  properties of the open cluster BSS population, which together point to a
  predominance of - but not necessarily solely - mass-transfer formation origins. These integrated results for one population permit highly detailed modeling of the formation and
  evolution of BSSs.

\subsection{Yellow Stragglers}\label{sec:properties_OC_YSS}

\subsubsection{Binary Properties}\label{sec:properties_OC_YSS_binary}
While less comprehensively surveyed than BSSs,
binaries are also frequently found among open cluster YSSs. Indeed all three YSSs
in the core of M67 (Main Text Figure 1) are single-lined spectroscopic binaries with notable
characteristics. WOCS 2002 is in a circular orbit with a period of 42.8
days. Based on the circularity of the orbit, \cite{Verbunt+Phinney1995}
predicted the system went through mass transfer and that it should have
a WD companion. The companion was found via FUV flux with the $\textit{Ultra-violet Imaging Telescope}$ \citep{Landsman+1997, Landsman+1998}. A FUV spectrum with HST/GHRS
yielded a mass of 0.22 \Msun~ and an age of 75 Myr for the WD. \cite{Landsman+1998} argue that the system has gone through RGB mass transfer, forming
a BSS that has evolved to its current YSS state.

Already known to be a binary with a period of 698 days and an orbital
eccentricity of 0.1, \cite{Leiner+2016} did an asteroseismic analysis
of the YSS WOCS 1015 with K2 photometric time-series data. Using asteroseismic scaling relations, they
find a mass of 2.9 ± 0.2 \Msun, to be compared with the
cluster MSTO mass of 1.3 \Msun. Due to the large mass, Leiner et al
suggested the system is an evolved BSS that formed from dynamical
encounters resulting in stellar collisions or a binary merger.

Finally, the photometric properties of WOCS 2008, with a period of 1513 days
and eccentricity of 0.30 \citep{Geller+2021a}, has presented a challenge for interpretation
for decades (see \cite{Mathieu+Latham1986}). Located 1.2 mag in V above the
MSTO yet a high-probability three-dimensional member in several studies, it has defied explanation as a combination of two or three cluster
members. \cite{BertelliMotta+2018} find it to have surface abundances consistent with M67. Mathieu \& Latham note the ``intriguing possibility remains
that S1072 is \ldots{} in an evolutionary phase leading to or from a
blue straggler''. \cite{Jadhav+2019} detect WOCS 2008 with UVIT, but do not find a secure UV excess.

After the launch of Kepler, \cite{Corsaro+2012} asteroseimically identified WOCS 1006 and WOCS 6002 as over-massive stars near the RC of NGC 6819 (Section \ref{sec:properties_OC_YSS_masses}). They conclude that these are YSSs. Both were also
suggested to be evolved BSSs by \cite{Rosvick+Vandenberg1998}
based on their CMD positions relative to the RC. WOCS 1006 and WOCS 6002 are both single-lined
binaries with periods of 3360 days and 1524 days and eccentricities of
0.24 and 0.72, respectively \citep{Milliman+2014}. 

While some YSSs are evident in CMDs, others lie at the boundaries of binary sequences, RCs and horizontal branches (HBs), making their photometric identification challenging. \cite{Linck+2024a} identify five YSSs in NGC 2506, and another 5 candidate YSSs. Four of the YSSs are identified based on CMD position; the fifth is based on a UV excess \citep{Panthi+2022}. Two of
these YSSs have long-period eccentric orbit solutions (with one having e=0.82), while another two show long-timescale radial-velocity variations without orbit
solutions yet. One of the orbit solutions marginally permits the companion WD mass suggested by \cite{Panthi+2022}). 


A different approach was taken by \cite{North2014} and \cite{VanderSwaelmen+2017}, who did statistical analyses of the mass functions of 124
RGB binaries in open clusters. They found that the mass function
distribution required 22\% of the giants to have WD companions, presumed
to be post-mass-transfer systems and thus YSSs.

\cite{Panthi+Vaidya2024} suggest sdB companions to both YSSs in NGC 6940, as well as a 26,000 K, 0.3 \Msun~companion to an RC star. Similarly, \cite{Rani+2021} suggest sdA companions to two spectroscopic-binary YSSs (and a MS star) in the younger cluster NGC 2818. Such companions are somewhat surprising, as the expected evolution time from BSS to YSS and RC can be long relative to the time that a hot subdwarf remains detectable.

\subsubsection{Masses}\label{sec:properties_OC_YSS_masses} 
In NGC 6819 \cite{Corsaro+2012} found asteroseismic masses for WOCS 1006 of 2.45
\Msun~(2.38 ± 0.24 \Msun, \cite{Handberg+2017}) and
for WOCS 6002 of 2.3 \Msun~(2.63 ± 0.1 \Msun,
Handberg et al). These masses are to be compared with masses of $\approx$ 1.65 \Msun~
(1.64 ± 0.2 \Msun, \cite{Handberg+2017}) for other RC
stars and the NGC 6819  MSTO mass of $\approx$ 1.4 M\textsubscript{o} \citep{Miglio+2012, Milliman+2015}.

Combining the orbit solutions and asteroseismology for these two binaries, \cite{Handberg+2017} find mass lower limits permitting WDs for the secondary stars (in the case of WOCS 6002, a CO WD). As yet no
direct detections of WD companions to these YSSs have been made.

\cite{Handberg+2017} also find in NGC 6819 that the luminous single RGB
(or AGB) WOCS 3003 is highly overmassive at 2.80 ± 0.25
\Msun, which they suggest is an evolved BSS of merger
origin. They find several other RGBs to be overmassive, albeit with
marginal significance. If NGC 6819 can be assumed representative of the
Galactic field population, they conclude that about 10$\%$ of
all giant stars have not evolved as single stars.

Similarly, \cite{Brogaard+2012} note the presence of several stars
above the RC in NGC 6791 which they conjecture to be evolved BSS,
in addition to one asteroseismically overmassive star within the RC \citep{Corsaro+2012}. Intriguingly, they also conjecture that the double-peaked WD
cooling sequence in NGC 6791 might be the consequence of evolved BSSs,
with the brighter peak perhaps being populated by more massive and bluer
WDs.

\subsubsection{Surface Abundances}\label{sec:properties_OC_YSS_abundances} 
The first open cluster barium giant was found in NGC 2420 (1 Gyr) by
\cite{McClure+1974}, and later was found to be velocity
variable \citep{McClure1983}. A second barium giant in the cluster was
confirmed by \cite{Smith+Suntzeff1987}. Detection of barium-selected YSSs in open
clusters was then quiet until \cite{KatimeSantrich+2013} found two
barium giants in the somewhat younger cluster NGC 5822 (0.7 Gyr) and \cite{VanderSwaelmen+2017} found \emph{s}-process overabundances in 4 of 12 open cluster giants.

\cite{BertelliMotta+2018} have used
APOGEE infrared spectra to study the surface abundances of two YSS in
M67. The two YSS show carbon depletions of $\approx$0.25 dex, similar to the
cluster RC. They argue that the depletion is more likely from
normal stellar evolution, but they cannot rule out an origin in mass transfer from an evolved star.

\subsection{Sub-dwarf B Stars}\label{sec:properties_OC_sdB}

sdB stars have been identified in three old, metal-rich open clusters.
The richest population of 5 (perhaps 6) sdB stars is in the open cluster
NGC 6791 \citep{Green+2001, Schindler+2015, Jadhav+2023}. One sdB candidate
has been identified in NGC 188 (Green et al. 2004; see also \cite{Rani+2021}) and another candidate
has been identified in Melotte 66 \citep{Zloczewski+2007, Rao+2022}. Among these are very short-period binaries, e.g. 0.399 days for NGC 6791/B4 \citep{Pablo+2011}, which notably is not tidally synchronized. 

Schindler et al. have done a comprehensive literature survey of
15 other populous old disk clusters somewhat younger and/or more metal
poor than NGC 6791 and NGC 188, and found no evidence of any other
sdBs. They suggest that very metal-rich open clusters produce a higher
fraction of sdB stars. \cite{Heber2016}, among others, has noted that both NGC 6791 and NGC 188 have other somewhat unique properties, such as large numbers of WDs and interacting binaries, respectively.

\subsection{Red Stragglers and Sub-Subgiants}\label{sec:properties_OC_RSS}

A complete literature census of SSGs and RSSs in both open and globular clusters
was taken by \cite{Geller+2017}, finding 43 SSGs and seven RSSs with secure cluster memberships. While some more examples have
been noted subsequently, the overall summary conclusions of Geller et
al. have not changed. Thus we concisely summarize here their findings in
open clusters without further reference to their paper. 

\subsubsection{Binary Properties}\label{sec:properties_OC_RSS_binary} 
Within the 4 open clusters with
comprehensive radial-velocity surveys, 8 of 11 SSGs and 1 RSS show
short-period radial-velocity variability. Six have orbit solutions, all
with periods of less than 20 days. Only the longest period system, one
of the prototype SSGs in M67, has an eccentric (e=0.206) orbit \citep{Mathieu+2003}. Two double-lined systems - one SSG and one RSS - indicate MS companions.
In several cases SSGs are identified as W UMa stars.

\subsubsection{Photometric Variability, Rotation and Stellar Activity}\label{sec:properties_OC_RSS_rotation}
Photometric variability is a ubiquitous property of both RSS and SSG
stars. Rapid rotation also is characteristic of SSGs, often detected via
photometric periods of $\le$ 15 days. In the shorter period
binaries such photometric periods often match the orbital periods,
indicating synchronous rotation. More than half of SSGs have detected
X-ray emission, with X-ray luminosities of order 10\textsuperscript{30}
-- 10\textsuperscript{31} erg s\textsuperscript{-1}, and often 
H$\alpha$
emission. Ca H and K emission has also been seen. Combined, these
properties are indicative of substantial stellar activity and spots,
likely linked to the rapid rotation of the convective layers.

\subsubsection{Spatial Distributions}\label{sec:properties_OC_RSS_spatial} 
SSGs as a whole are more frequent in
the cores of clusters. Combining both open and globular clusters, 93\%
are found to be within 3.3 core radii from their respective cluster
centers, and about 60\% within 1 core radius. Geller et al. conclude that these
results strengthen the argument for association with the clusters, but
do not yet speak to mass segregation.

\section{Globular Clusters}\label{sec:properties_GC}



\subsection{Catalogs and Populations}\label{sec:properties_GC_catalogs}

Catalogs of classical BSSs in globular clusters went through a revolution with the
launch of HST, both due to the higher angular resolution providing
access to the cluster cores and to the availability of UV
colors that are less influenced by confusion from giants (e.g., \cite{Paresce+1991}). The first reasonably complete catalogue of BSSs in
globular clusters was constructed and analyzed by \cite{Piotto+2004},
with the final catalog presented in \cite{Moretti+2008}. This
photometrically selected catalogue was derived from HST/WFPC2 B and V
CMDs for 74 clusters \citep{Piotto+2002}. BSSs were
selected in 56 of these clusters, yielding a total sample of nearly 3000
stars. Later, \cite{Leigh+2007} revisited this catalog to
create a catalog of BSSs in the cluster cores. More recently, \cite{Simunovic+Puzia2016} created a catalog of BSSs in 36 globular clusters based
on combining the \emph{ACS Globular Cluster Survey} in V and I using the
HST/ACS/WFC (\cite{Sarajedini+2007}) and the near-UV survey using
the HST/WFC3/UVIS \citep{Piotto+2015}.

\subsection{Blue Stragglers}\label{sec:properties_GC_BSS}

\subsubsection{Populations}\label{sec:properties_GC_BSS_populations}
\begin{figure}
\centerline{\includegraphics[width=0.7\textwidth]{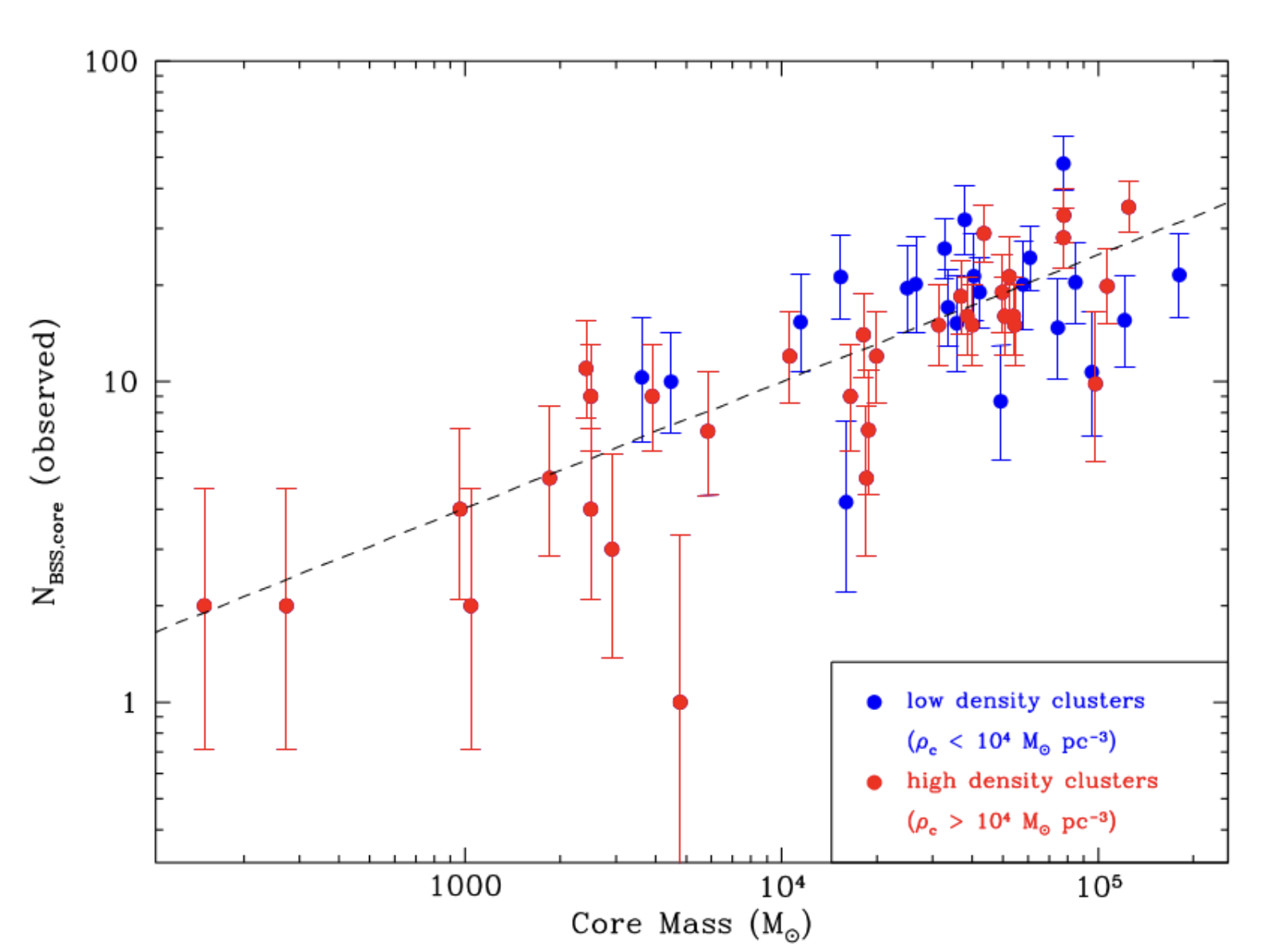}}
\caption{The number of blue stragglers found in a globular cluster core versus the total mass of the core. \cite{knigge2015}, adapted from \cite{Knigge+2009} (CC BY 4.0).}
\label{fig:KniggeLeighSills}
\end{figure}

There have been extensive efforts attempting to
connect frequencies of classical BSS with globular cluster properties. Early on,
\cite{Ferraro+2003} pointed out large variations in BSS specific
frequencies (relative to HB stars) from cluster to
cluster, without evident connection to cluster properties. Indeed, very
high specific frequencies were found in clusters at both extrema of
central density. \cite{Leigh+2007} and others found no
correlation between BSS frequency and the cluster collision rate.
Moreover, they found a weak anti-correlation between BSS frequency and
cluster luminosity (a proxy for total cluster mass). A similar anti-correlation between
the frequency of BSSs and cluster absolute luminosity 
was found by \cite{Piotto+2004} and \cite{Moretti+2008}. \cite{Knigge+2009}, however, found a positive correlation of the number
of BSSs with the mass of the globular cluster core (Supplemental Figure~\ref{fig:KniggeLeighSills}), from which they conclude that most blue stragglers, including in cluster cores, form from binary systems and not from single-star collisions. 

\cite{Milone+2012}
found a correlation in 24 clusters between the fraction of
BSSs and the cluster binary fraction. 
Interestingly, Milone et al.~find that two post-core collapse clusters do
not align with this correlation, perhaps indicative of a substantial
increase in dynamical formation of BSSs in these clusters. At the
other extreme, low-density globular clusters seem to show a very strong
correlation between the specific frequency of BSSs (relative to MS stars) and binary fraction \citep{Sollima+2008}.

\subsubsection{Color-Magnitude Diagrams}\label{sec:properties_GC_BSS_CMD} 
CMDs of globular clusters revealing
BSSs abound in the literature using an array of bandpasses, and we do
not attempt a review here. The CMDs of \cite{Piotto+2002} and
\cite{Sarajedini+2007} remain foundational references for researchers.

One of the most striking findings has been a double sequence
of BSSs in the CMD of the globular cluster M30 \citep{Ferraro+2009} (Main Text Figure 8). The numbers of
BSSs in both sequences are comparable. Notably, the BSSs in the red
sequence are more centrally concentrated, and have one of the highest
specific frequencies in any globular cluster. Ferraro et al. hypothesize
that the blue sequence represents BSSs created collisionally while the
red sequence formed (and may still be forming) from mass transfer,
based on theoretical models of both processes (\cite{Sills+2009}, \cite{Xin+2015}). (See models of \cite{PortegiesZwart2019}; but see also models of
\cite{Jiang+2017}, \cite{Jiang2022}.) They further hypothesize that both
populations were formed in a recent core collapse of the cluster.


Subsequent papers have suggested that a few other globular clusters have
similar BSS double sequences, with varying degrees of clarity - NGC 362
\citep{Dalessandro+2013}, {NGC 1261 (\cite{Simunovic+2014}; but see
\cite{Raso+2019}), M15 (\cite{Beccari+2019}; remarkably they identify two blue branches) and NGC 6256 \citep{Cadelano+2022}). Four of these globular clusters are post-core collapse, and again
the authors argue that the blue sequences are the result of
collisional BSS formation during that collapse, and indeed are
indicative of such collapse (recently discussed in \cite{Cadelano+2022}). The BSSs in identified red sequences tend to be more scattered in the CMD, which
authors attribute to mass transfer extending over long timescales.
In some clusters the red sequence is found to be more
centrally concentrated. Curiously, W UMa variables have been found
within the blue sequences of both M30 and NGC 362, perhaps consistent
with the models of \cite{Jiang+2017} and \cite{Jiang2022} which predict contamination of the
blue sequence with mass-transfer products.

Interestingly, FUV observations of M30 do not show the double sequence;
in fact the BSSs from the two sequences in the optical are mixed within a single linear distribution
in the FUV-UV CMD. \cite{Mansfield+2022} suggest that this may result from the sources on the red sequence experiencing active mass transfer
and emitting FUV radiation, shifting these sources blueward in the
ultraviolet CMD.

\subsubsection{Binary Properties}\label{sec:properties_GC_BSS_binary} 

\paragraph{Binary Frequency}\label{sec:properties_GC_BSS_frequency}

There are few studies of BSS binary frequencies in globular clusters. Here we briefly discuss globular cluster MS binary populations, which likely play an important role in globular cluster BSS formation rates. 

Historically, binary populations in
globular cluster main sequences have been identified and studied photometrically, both by searching for variability and by censuses of
photometric binaries to the red or above the MS. The latter has an extensive
history in the literature for individual clusters, but more recently has
been applied via HST data to large samples of globular clusters.

MS binary fractions in globular clusters inferred from photometric binaries are
typically lower (e.g., less than 20\%) than found in open clusters or in the field \citep{Milone+2012}, but occasionally reach values around $\approx$ 50$\%$ (e.g., \cite{Sollima+2007};
\cite{Milone+2012}). Photometric binary populations show mass
segregation, and anti-correlate with total luminosity} of the cluster (i.e., cluster mass), both likely to be dynamical effects. Importantly here, Sollima
et al. and Milone et al. find a correlation between
binary frequency and the number of BSSs in cluster cores.

Spectroscopic surveys for binary stars in globular clusters have been
few (e.g. \cite{Sommariva+2009}). 
In the core of NGC 3201, \cite{Giesers+2019} directly measure spectroscopically a core BSS binary frequency of $58\pm8$\%, substantially
higher than both their observed binary frequency of $20\pm0.2$\% for all core stars and their determined true cluster binary frequency of $6.7\pm0.7$\%. The BSS binary frequency is somewhat lower than has been found in open clusters, but
statistically consistent. Giesers et al.~also identify 5 SX Phoenicis pulsators among the BSSs with radial-velocity variation amplitudes
up to 18 kms\textsuperscript{-1}, whose binary status they cannot
empirically determine. They note however that SX Phoenicis stars are
thought to form from binary evolution.

\paragraph{Periods} 

Interestingly, 3 of eleven BSS binaries with highly constrained orbit
solutions in \cite{Giesers+2019} are quite hard, with periods of less than 1 day and small
minimum companion masses (0.11~\Msun~to 0.24 \Msun). One of these hard binaries is also a contact
system. The six longer period binaries (periods
greater than 100 days) have higher minimum secondary masses (0.35~\Msun~to 0.67 \Msun). The other two BSS binaries have orbital periods between
these two domains. If the companions are WDs, these results may be indicative of longer core growth times in longer-period BSSs.

Photometric variability surveys of globular clusters have revealed numerous short-period
contact binaries among BSSs. \cite{Rucinski2000} compiled and analyzed 20 W
UMa binaries among the BSSs of 14 globular clusters. The frequency among the BSS
stars was moderately well established at about 45$\pm$10 BSSs per
BSS contact binary. Thus, contact binaries are about three times more
common among globular cluster BSSs than among the disk dwarf population. Contact
binary systems with periods longer than 0.6 days are absent in the
sample, possibly because the more massive stars have left the contact
binary domain.

The Las Campanas Cluster AgeS Experiment (CASE; \cite{Kaluzny2005}) monitored
a large number of globular clusters for variability. Often the
photometric data extend well over a decade. Many
variables have been detected to be binary evolution products;
indeed, most of the discovered variable stars are BSSs, primarily SX Phoenicis
stars. In addition, the CASE team reports numerous BSS
eclipsing binaries, a treasure trove that has not yet been fully mined.

The CASE team has identified numerous mass-transfer products, including contact
binaries in many clusters, a semi-detached Algol in 47 Tuc \citep{Kaluzny+2007a}, a BSS undergoing rapid mass transfer in M55 \citep{Rozyczka+2013}, three eclipsing binaries in M55 with shallow eclipses indicative
of low mass ratios (one measured at q=0.16; see also \cite{Li+2017} and
\cite{Li2018}), and a semi-detached binary on the RGB of NGC 362
\citep{Rozyczka+2016}.

\paragraph{Companions and companion masses} 

Recently the GlobULeS survey \citep{Sahu+2022} has used UVIT and archival FUV data to identify FUV-bright BSSs in NGC 362, a suggested double-sequence cluster. \cite{Dattatrey+2023a} identified ELM companions to 12 BSSs, which they attribute to Case A or Case B mass transfer. They also note that for many the cooling ages are similar to the suggested time since core collapse. Subsequently, \cite{Dattatrey+2023} identified four FUV-bright stars at the MSTO, located in the outer part of the NGC 362 cluster. They suggest these to be BLs, with which they identify two ELM and two He WD companions. They derive remarkably small cooling ages, in one case less than 0.1 Myr.

\subsubsection{Masses}\label{sec:properties_GC_BSS_masses} 
The first direct mass measurements of globular cluster
BSS masses were done via pulsations for SX Phoenicis stars \citep{Nemec+1995, Gilliland+1998}, finding masses between 1.35 \Msun~and 1.6 \Msun, and via an atmospheric gravity
determination \citep {Shara+1997}. In the latter case, the mass of
1.7 ± 0.4 $\Msun$ for the second-brightest BSS in 47 Tuc is of order twice
the MSTO mass. \cite{DeMarco+2005} followed with atmospheric gravity determinations for 24 BSSs in four globular clusters, finding a mean BSS mass of 1.07 $\Msun$ and a
correlation with effective temperature and luminosity, albeit with large scatter. They note a mild correlation between their spectroscopic mass measurements and those from stellar evolution tracks, with a slightly higher mass estimate from the tracks. (DeMarco et al. find that most of these supposed BSSs lie somewhat beyond the TAMS. By the definition here, these are YSSs, albeit in a curious distribution of evolutionary state.) De Marco et al.~also find 4 of 13 HB stars to be overmassive, which they suggest may be high-luminosity BSSs.
Additional mass
determinations via pulsations by \cite{Fiorentino+2014} yielded an
average mass of 1.06 ± 0.09 $\Msun$ in NGC 6541, relative to a turnoff
mass of 0.75 \Msun. They find good agreement with single-star
evolutionary tracks. (See also \cite{Fiorentino+2015} for $\omega$ Cen.) \cite{Raso+2019} derive masses for 22 BSSs in 47 Tuc
based on spectral energy distributions. While the mass uncertainty for each BSS is large, they find a sequence of increasing mass with BSS luminosity.

A statistical dynamical estimate of BSS masses in globular clusters has been
obtained by \cite{Baldwin+2016} based on comparison of population velocity
distributions. They find an average mass ratio of \emph{M}$_{BSS}$/
\emph{M}$_{MSTO}$ = 1.50 ± 0.14 for 598 BSSs across 19 globular clusters.
Incorporating isochronal MSTO masses, this yields an average BSS mass of 1.22 ± 0.12 $\Msun$ across these globular clusters, in
good agreement with previous direct measures. A subsequent higher
proper-motion-precision study of NGC 362 by \cite{Libralato+2018}
found an average BSS mass of 1.96 $\Msun$ with a
large uncertainty. They suggest this very large mass may be the result
of an incompleteness-based bias in their BSS sample. However, \cite{Libralato+2019} again found an average mass of BSS of 1.82 ± 0.37
$\Msun$ in NGC 6352.  

\cite{Parada+2016a} also take a statistical dynamical approach to determining masses of
BSS populations within 47 Tuc, essentially using degree of mass
segregation as revealed in relative radial density profiles as a measure
of mass. They find the more luminous BSSs to have larger masses,
exceeding twice the MSTO mass. Lower luminosity BSSs have a radial
distribution similar to the MS photometric binaries. A compatible dynamical result was found for the BSSs in 47 Tuc by \cite{Cheng+2019} with comparison to X-ray data. A similar approach
was taken by \cite{Singh+Yadav2019} for NGC 6656, who find an average BSS
mass of 1.06 ± 0.09 $\Msun$.

Importantly, these statistical dynamical
techniques measure the total mass of BSS multiple systems rather than the masses of
the BSSs  themselves.


\subsubsection{Rotation}\label{sec:properties_GC_BSS_rotation}
Most measures of stellar rotation among globular
cluster BSSs are spectroscopic determinations of \vsini. The first such measurement was done for one of
the brightest BSS in 47 Tuc by \cite{Shara+1997}, who measured a
$\vsini$ of 155 ± 55 kms\textsuperscript{-1}. Subsequently \cite{DeMarco+2005} provided \vsini\ measurements for 5 BSSs and upper limits
for another 5 BSSs in 4 globular clusters. The measured values have average and median $\vsini$ 
values of 109 kms\textsuperscript{-1} and 100 kms\textsuperscript{-1}, respectively; the upper limits
range from similar values to as low as 25 kms\textsuperscript{-1}.

In contrast, using the VLT/FLAMES-GIRAFFE \cite{Ferraro+2006} found that the
majority of BSSs in 47 Tuc have $\vsini < 10$
km s$^{-1}$, much like found for the turnoff stars by\cite{Lucatello+Gratton2003}. Indeed, they find only one rapidly spinning
star, at approximately 100 kms\textsuperscript{-1}. \cite{Lovisi+2010a, Lovisi+2012} and \cite{Mucciarelli+2014} extended the
VLT study to 4 additional globular clusters, with observations of more
than 160 BSSs. They find three high-density clusters (including 47 Tuc) to have $\vsini$ distributions with predominantly small values, and two low-density
clusters, M4 and $\omega$ Cen, to have $\vsini$ distributions more
extended to higher values.

Simunovic \& Puzia (2014) obtained spectra for 116 BSSs in
three globular clusters. They find a similar $\vsini$ distribution
for $\omega$ Cen. However, their $\vsini$ distributions for NGC 3201 and
NGC 6218, both also low-density clusters, do not have extended
distributions to high $\vsini$. These authors find that, with the
exception of two BSSs in $\omega$ Cen, all fast-rotating BSSs are
significantly concentrated within $\approx$ 2 core radii, while BSSs with
$\vsini < 70$ km s$^{-1}$ populate the entire
spatial extent of their globular clusters. This relation also holds for
BSSs with \vsini\ below and above 50 kms\textsuperscript{-1} in
NGC 3201 and NGC 6218. Simunovic \& Puzia conclude that rapidly rotating
BSS preferentially form in the cores of globular clusters.

\cite{Giesers+2019} have five BSSs in common with Simunovic \& Puzia
(2014), with \vsini\  ranging from 34.6 kms\textsuperscript{-1}
to 135 kms\textsuperscript{-1}. They note that all of them are in binary
systems, albeit as yet with no clear connections between rotation and orbital
properties.

Recently \cite{Ferraro+2023} have completed a study of projected rotation velocities of 320 BSSs in 8 globular clusters. Half of the sample shows $\vsini > 20$
km s$^{-1}$, with some exceeding 200
km s$^{-1}$. They find the fraction of fast rotating BSSs (with $\vsini > 40$
km s$^{-1}$) to increase dramatically (one order of magnitude)  in loose clusters (Main Text Figure 7), and specifically with decreasing central density,  global (photometric) binary fraction, collision parameter and dynamical evolution (pre-core collapse).


\cite{Ferraro+2023} attribute the rapid rotation in low-density environments to spin-up at mass-transfer formation. They further suggest that the lack of rapid rotators in high-density environments indicates an as-yet-undefined efficient spin-down mechanism for (presumed to be collisional) BSSs, with a timescale of 1-2 Gyr. In this regard, it is interesting that the BSSs in the blue sequence of M30 (suggested to be collisional) are rotating somewhat faster than the BSSs in the red sequence (suggested to be mass transfer) \citep{Lovisi+2013a}.

\cite {Billi+2023} and \cite{Billi+2024} did detailed $\vsini$ studies in NGC 3201 and M55, respectively. They find the highest frequency of rapid rotators yet observed in M55, also the lowest density globular cluster yet studied. In both clusters they find that the BSS rotational velocities tend to decrease with lower luminosity and surface temperature. \cite{Kaluzny+2013} also found that the bluer BSSs in M4 are more rapidly rotating.

\subsubsection{Surface Abundances}\label{sec:properties_GC_BSS_abundances} 
Surface abundances in globular cluster BSSs have been suggested as a possible discriminant between collisional and mass-transfer formation mechanisms. Collisional BSSs are not expected to lead to chemical anomalies due to the lack of mixing \citep{Sills+2001a}, while mass-transfer material may contain tracers of the inner chemical structure of the donor star (e.g., \cite{Lovisi+2013b}).

\cite{Ferraro+2006} observed evidence
for C and O depletion in a subset of BSSs in 47 Tuc, which they suggest
is indicative of mass transfer of material from deep inside the donor
star where CNO burning was active. Notably, they find these stars to be
localized within the CMD. The qualitative morphology of this localization is similar
to that found in open cluster BSSs with barium enhancements, although
the luminosities of the BSSs are quite different in the two cases.
Interestingly, this domain of the CMD also contains the most rapidly
rotating BSSs in 47 Tuc (albeit not very rapid, \textgreater{} 20
kms\textsuperscript{-1}) as well as three W UMa stars. Ferraro et al.
suggest that together these characteristics indicate that this domain of the
CMD contains the most recently formed BSSs.

\cite{Lovisi+2013a} find a similar O depletion in 4 of 5 BSSs on the red sequence of M30 (based on upper limits), which they argue is compatible with the suggested mass-transfer origin of that sequence. However, they also note the ambiguities introduced by possible gravitational settling.

Observations of other globular clusters do not yield as rich a population of
depleted stars \citep{Lovisi+2010a,Lovisi+2013}. In total the frequency of observed CO depletion among
BSSs has been estimated to be $\approx$10\% \citep{Ferraro+2015}.

\begin{figure}
\centerline{\includegraphics[width=0.8\textwidth]{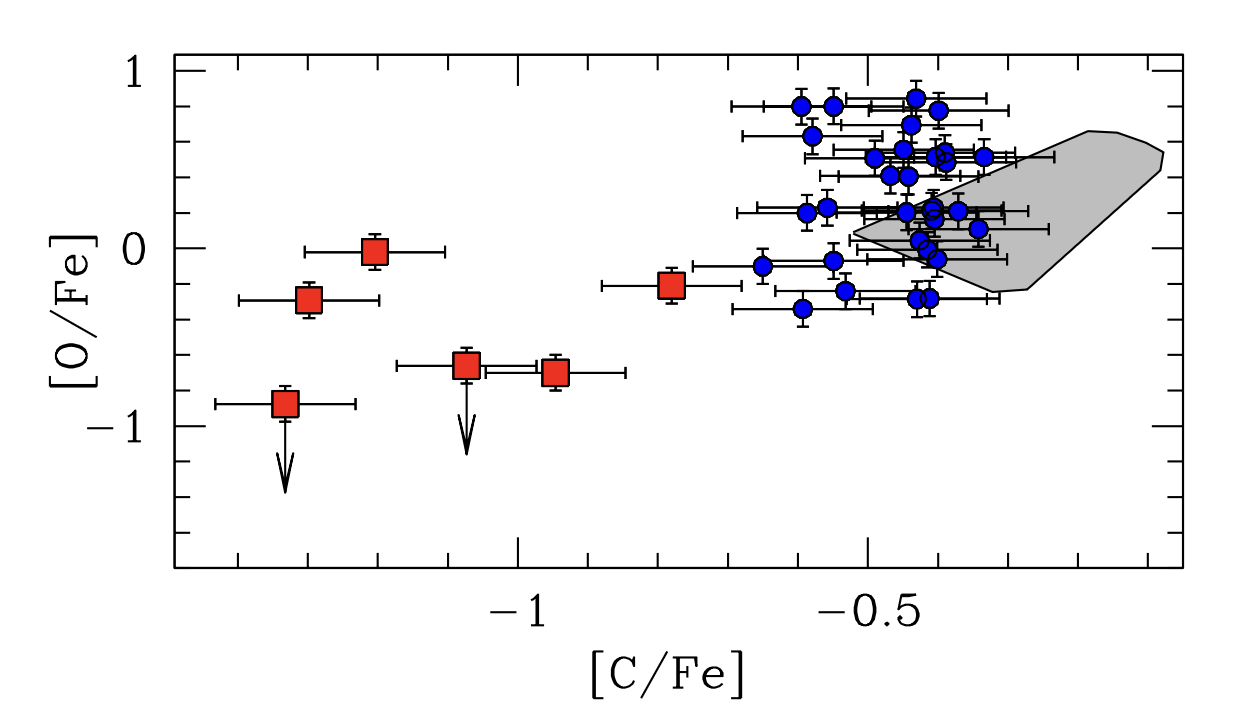}}
\caption{[O/Fe] ratio as a function of [C/Fe] for BSSs in the globular cluster 47
Tuc. Normal abundance BSSs are marked with blue circles, while CO-depleted BSSs are
marked with red squares. The gray regions
correspond to the location of the MSTO stars in 47 Tuc.  \citet{Ferraro+2015}, adapted from \cite{Ferraro+2006} (CC BY 4.0).} 
\label{fig:CO depletion}
\end{figure}


\subsubsection{Spatial Distributions}\label{sec:properties_GC_BSS_spatial}

Because BSSs are more massive than the other stars in their associated clusters, while at the same time being very minor contributors to the total cluster mass, they are invaluable tracers of the cluster potential. Similarly, they are very responsive to relaxation processes and in particular to dynamical friction. In short, in equilibrium they are expected to be centrally concentrated.

F. Ferraro and collaborators have used these properties to do detailed studies of the dynamical evolution of globular clusters. This work is beyond the scope of this review (see \citep{Ferraro+2012, Ferraro+2020}), but relevant here is their finding that the radial distributions of BSSs are often bimodal, with a central relaxed concentration and an extended unrelaxed halo. Ferraro and colleagues have argued that these two distributions reflect different BSS formation mechanisms, specifically with mass transfer dominant in the outer regions and dynamical processes (such as collisions) more prevalent in the cores, especially of post-core collapse clusters. Whether this hypothesis can be carried over to bimodal distributions seen in old open clusters (Section \ref{sec:properties_OC_BSS_spatial}) remains an open question \citep{Geller+2013}.

\subsection{Yellow Stragglers}\label{sec:properties_GC_YSS}

\cite{Renzini+FusiPecci1988} suggested searching for evolved BSSs between
the AGB and the HB (see also \cite{Sills+2009}), followed by a search
for such stars by \cite{FusiPecci+1992}. 

Candidate YSSs have
been identified in globular cluster CMDs. (Note that in globular clusters the domains of YSSs and extended HBs overlap.)  YSSs have been identified using
HST/WFPC2 in M3 (Ferraro et al. 1997), M13 \citep{Ferraro+1997a,Ferraro+1999} and M80 \citep{Ferraro+1997}. (Ferraro et al.\ call these
stars E-BSSs, presuming their origin to be evolved BSSs.) The YSSs are identified as
being slightly bluer than the RGBs and between 0.2 and
1 magnitudes brighter than the HBs in those clusters
\citep{Sills+2009}. The radial distributions of these candidates mimic
those of the BSSs, at the least indicative of similar masses. In the same vein, they are more centrally concentrated than RGB and HB stars, suggesting masses larger than evolved single stars.

\begin{figure}
\centerline{\includegraphics[width=1.0\textwidth]{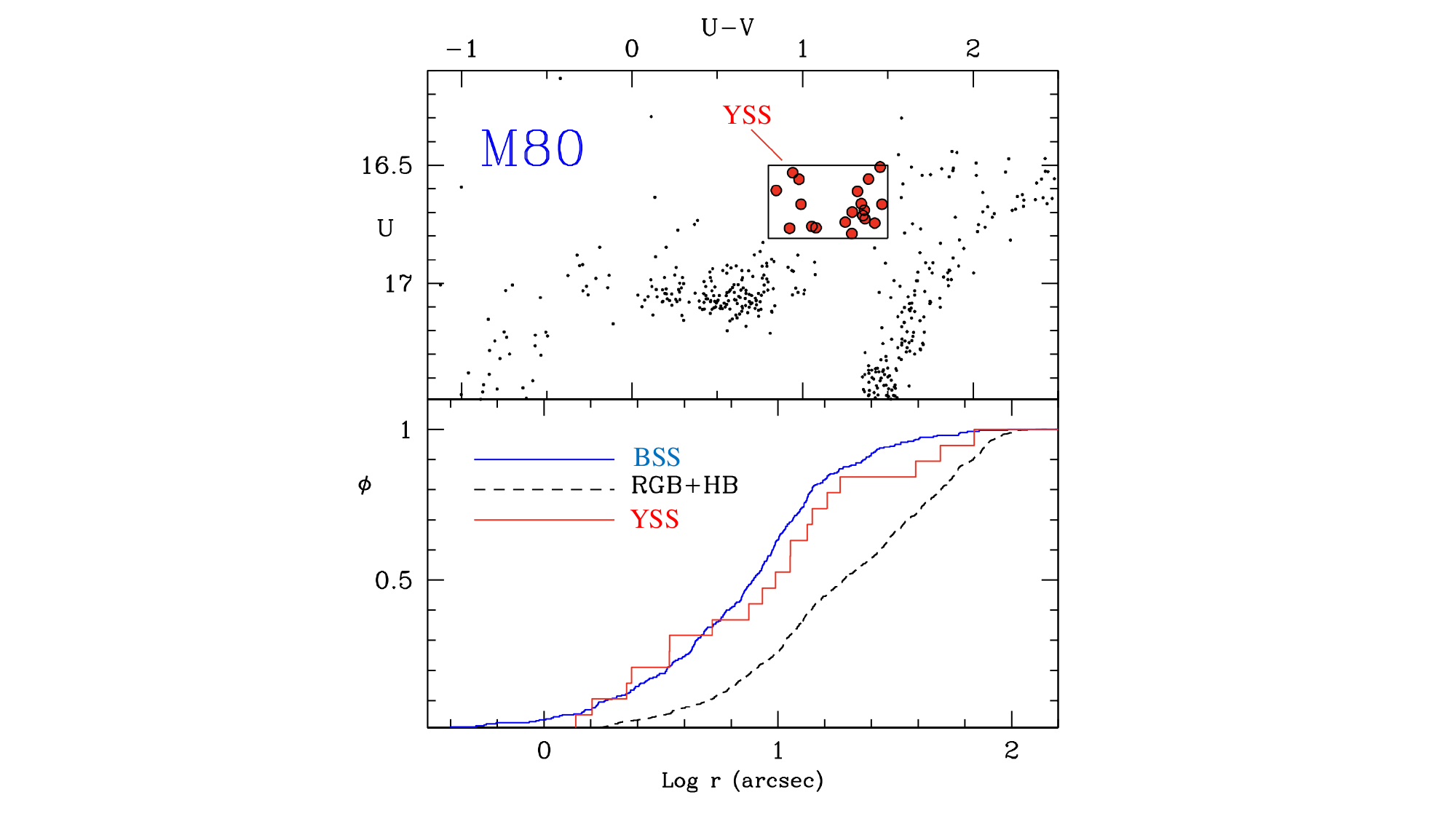}}
\caption{Upper panel: CMD of the globular cluster M80 showing the HB/AGB region. Red circles and the box mark the sample of YSSs. Lower panel: Cumulative radial distribution of BSSs (blue line), RGB+HB stars (black dashed line) and candidate YSSs (red line). candidate YSSs share the same radial distribution as BSSs, and are significantly more centrally concentrated than RGB and HB stars. Adapted from \cite{Ferraro+2015} (CC BY 4.0). }
\label{fig:GC YSSs}
\end{figure}

Similarly, \cite{Beccari+2006}
 find the stars on the AGB
of the globular cluster 47 Tuc to be centrally concentrated. This is
unexpected for evolved single stars, and they suggest the AGB also includes
evolved BSSs in that evolutionary phase. The derived ratio of the numbers of BSS and
YSSs are somewhat similar to the expected evolutionary ages in each
phase. In a subsequent detailed study of UV and visible images of 47 Tuc,
\cite{Parada+2016, Parada+2016a} argue for an evolved BSS population as the
origin of observed population excesses in the HB and the
AGB. They argue that the similar numbers of BSSs and evolved BSSs suggest
a BSS post-MS lifetime similar to single stars of similar mass and a BSS MS age
of 200-300 Myr, in contrast to prior estimates of much longer MS lifetimes.
\cite{Kravtsov+Caldern2021} use similar techniques to argue
for a population of over-massive red giants in NGC 3201, a third of
which they attribute to recently evolved BSSs.

\cite{Kaluzny+2015} identified three photometrically variable YSSs in M12, and a W UMa
that is a candidate YSS, with periods of less than 2 days. The CASE team also found a YSS with a W UMa-like light curve in M22
and a variable YSS X-ray source in NGC 3201 (\cite{Rozyczka+2017}, \cite{Rozyczka+2020}. \cite{Ulloa-Sols+2023} find a few variables
in NGC 3201 with sub-day periods, which they cannot associate with any
known type of variable stars.

\subsection{Sub-dwarf B Stars}\label{sec:properties_GC_sdB}

The observational literature on sdBs in globular clusters is deeply integrated with observations of extreme horizontal branch stars (and associated empirical phenomena such as blue tails and blue hooks). Often, but not always, the two are spoken of as the same.  Here we are agnostic regarding the nature of the overlap, and make no distinction.

One of the most striking findings related to sdBs in globular clusters is the marked lack of very close binaries, in contrast to field sdBs \citep[][but see the discovery of one close binary hot HB star in NGC 6752 \citep{MoniBidin+2015}]{MoniBidin+2009, MoniBidin+2011,Latour+2018}. \cite{MoniBidin+2011} suggest that younger globular clusters, such as NGC 2808 and NGC 5986, have higher close-binary frequencies among hot HB stars than older clusters such as NGC 6752. They also point to a discovered higher frequency of intermediate-period (10 - 200 days) binaries as of possible importance in understanding these differences.

\cite{DAntona+2010}  propose that the properties of hot subdwarf stars may depend on helium content, particularly relevant given the mutiple stellar populations within globular clusters thought to be associated with various helium abundances. The multiple populations in NGC 2808 and $\omega$ Cen have been connected to He-abundance differences among their sdBs. In $\omega$ Cen \cite{Latour+2018} find a much higher frequency of He-sdOB stars ($\approx$ 52\%) than in the field ($\approx$ 5\%), which they attribute to the $\omega$ Cen multiple populations and environment. Marino et al. (2014) found an enrichment of helium in the blue HB stars of the massive cluster NGC 2808, which they note is consistent with a second generation of stars.

Recently, pulsating hot subdwarfs have been found found in the massive clusters NGC 2808 and $\omega$ Cen. The rather homogenous set of pulsators in $\omega$ Cen are very hot, He-poor sdO stars, and beyond the scope of this review \citep{Randall+2016}. (Note that $\omega$ Cen is likely not a Galactic globular cluster but the remnant nucleus of a dwarf galaxy \citep{Bekki+Freeman2003}.) On the other hand, \cite{Brown+2013} stress that the inhomogeneity of the six sdB pulsators in NGC 2808 - including both H-rich and He-rich atmospheres - is strikingly different from both $\omega$ Cen and the field. (Although they caution that the $\omega$ Cen pulsators were discovered in the cluster halo in the optical versus in the cluster core and in the FUV for NGC 2808.)
Their one curious commonality is very similar FUV luminosities.

Finally, \cite{Kaluzny+2007} found the eclipsing binary V209 in $\omega$ Cen to be a
candidate post-CE product. In the CMD
the primary star lies between the tip of the BSS region and
the extended HB, while the secondary star is located
close to the red border of the area occupied by hot subdwarfs. However,
Kaluzny et al. find its radius too large and its effective temperature
too low for V209 to be an sdB star. They suggest that the primary star
is ``reborn'' by mass transfer from the secondary onto a white dwarf,
while the current secondary has lost most of its envelope.

\subsection{Red Stragglers and Sub-Subgiants}\label{sec:properties_GC_RSS}

The first 6 SSGs in a globular cluster were identified in the optical CMD of 47 Tuc by \cite{Albrow+2001}; four more were added by \cite{Edmonds+2003} who also provide UV and X-ray analyses. Albrow et al.\ note that the 47 Tuc SSG properties are very similar to those in the open cluster M67. Interestingly, one of their SSGs is a 1.1-day period eclipsing binary with an optically identified WD and a UV excess, which they cite as a long-period CV candidate. (They also note that the long-period field CV BV Centauri falls within their SSG domain in the 47 Tuc CMD.)

Geller et al. (2017) find 26
secure SSGs and 6 secure RSSs in globular clusters. More recently, \cite{Gottgens+2019} identify 12 SSGs and 4 RSSs in 26 globular clusters, selected by H$\alpha$ emission. Again, many of
these show other signatures of stellar activity, as well as
radial-velocity variation. For four of the
SSGs identified in NGC 3201, \cite{Giesers+2019} found orbital
solutions. One is an eclipsing binary with a period of 10.0037 d and a
circular orbit. Assuming a primary mass of 0.82 $\Msun$, the companion mass
is 0.53 ± 0.04 $\Msun$. Based on Balmer line filling, Giesers et al. suggest the system may be a proto-BSS currently undergoing mass transfer.
Two shorter period SSGs show X-ray emission and Balmer-line activity
indicators typical of SSGs. The one long-period SSG (17.2 d) has an
eccentric orbit, making it rather similar to one of the prototype SSGs,
WOCS 13008, in M67.

\cite{Kaluzny+2010} found two
variable RSSs in M55. Both of the latter are periodic variables, and one
is an X-ray source and a candidate magnetically active binary star.

\section{Galactic Field}\label{sec:properties_field}

The identification of BSSs, YSSs, etc.\ in the Galactic field is often
hampered by the lack of a reference population of the same age and
metallicity, such that identifying alternative evolution products from
their locations in a CMD is not possible. A notable
exception to this has been BSSs in the Galactic halo, where stars that have gained
appreciable mass may stand out in a CMD, especially in their effective
temperatures \citep{Carney+1994, Preston+Sneden2000}. 

However, mass gainers in
the field may be identified in other ways, in some cases by direct mass measurements using asteroseismology but most frequently by their surface
abundances. 
The latter is the case with several different classes of stars, such as as barium stars, CH stars, carbon- and \emph{s}-process enhanced metal-poor (CEMP-\emph{s}) stars, dwarf carbon (dC) stars, and extrinsic S stars.
Despite the wide variety in terminology, high-resolution spectroscopy has revealed that these stars all share the property that their surfaces are enriched in carbon and \emph{s}-process elements (with the possible exception of the dC stars, for which detailed abundance measurements have generally not been possible).
They can all be understood within a single framework of mass gain from a companion AGB star that underwent thermal pulses and produced these elements by internal nucleosynthesis. 

\begin{marginnote}[]
\entry{Extrinsic S star}{Evolved giant of spectral type S, enriched in \emph{s}-process elements but lacking the radioactive element technetium}
\end{marginnote}

It is primarily the initial mass and metallicity, as well as the current evolution state, of the mass gainer that determines its spectroscopic appearance. 
For example, a metal-poor star will more easily become C-enriched than a metal-rich star. 
If the surface C/O ratio increases above unity, strong bands of CH (and possibly other C-rich molecules) appear in the spectrum. In more metal-rich stars, strong Ba absorption lines often become the most notable feature. Thus the distinction between Ba, CH and CEMP-\emph{s} stars is primarily caused by decreasing metallicity. Even though mass gain during the MS is by far most likely, if the gainer mass is sufficiently high it can evolve off the MS and become a similarly enriched RGB or RC star. 
In the case of extrinsic S stars, the absence of technetium (a radioactive element produced by the \emph{s}-process that decays within several million years) reveals that their \emph{s}-process enhancements originate from prior mass gain.

\subsection{Populations and Catalogs}\label{sec:properties_field_catalogs}

\begin{figure}
    \centering
    \includegraphics[width=0.9\textwidth]{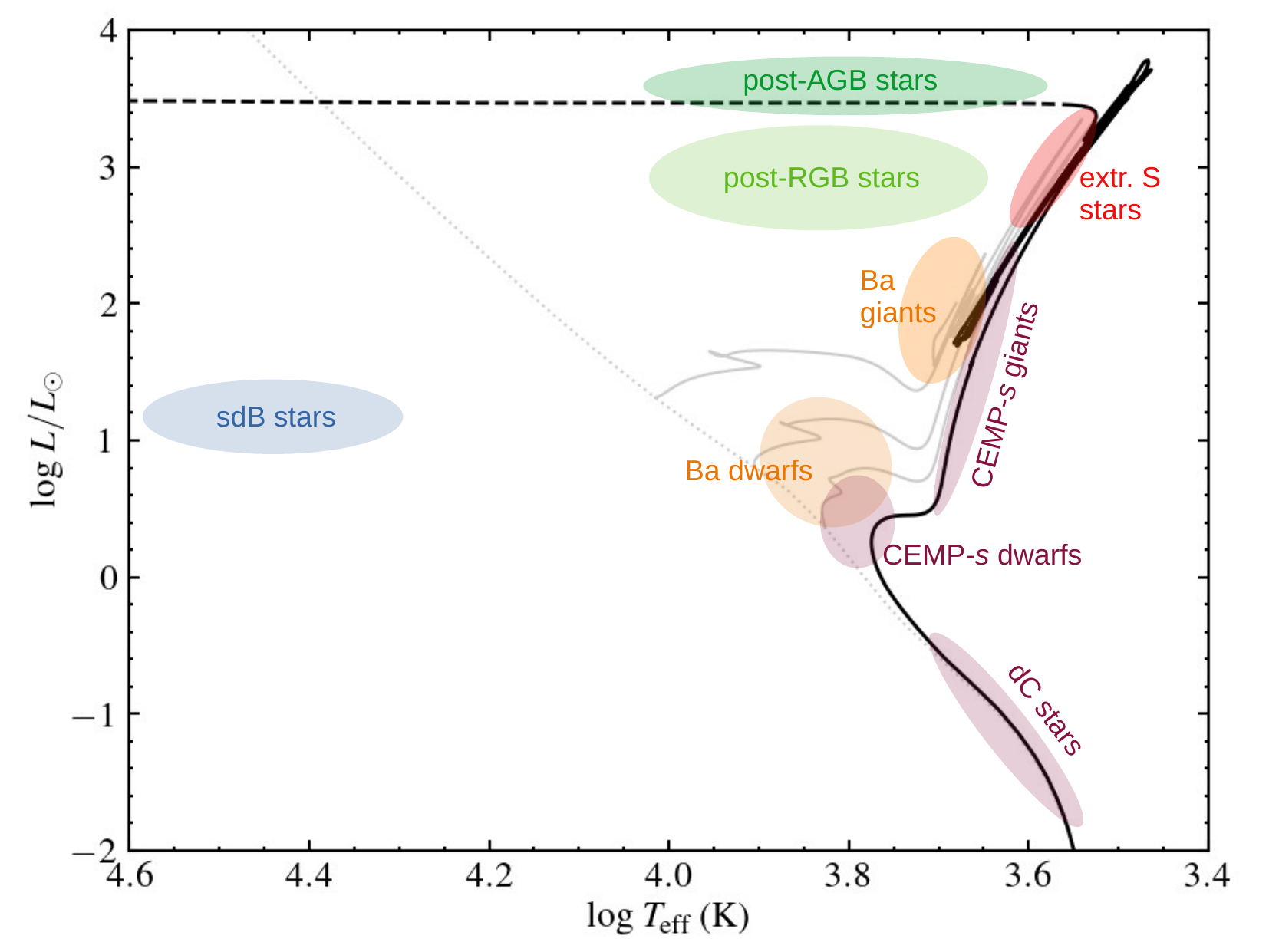}
    \caption{Schematic H-R diagram of the Galactic field products discussed in this review. Halo BSSs roughly coincide in position with the Ba and CEMP-\emph{s} dwarfs. The thick solid line is a 10~Gyr MIST \citep{Choi+2016} isochrone at [Fe/H] $=-0.25$, with the dashed part indicating the post-AGB phase. Also shown in light grey are MIST evolution tracks for 1.2, 1.5 and 2.0~\Msun\ (solid lines) and the ZAMS (dotted) at the same metallicity.}
    \label{fig:field_HRD}
\end{figure}

Supplemental Figure~\ref{fig:field_HRD} gives a schematic overview of the location in the H-R diagram of the 
stars discussed in this section.

Classical BSSs have been identified in the halo (and thick disk) as blue metal-poor stars using two-color photometry. Often called "field BSSs", we use the term "halo BSSs" in this review.) 
The specific frequency of halo BSSs (relative to halo HB stars) is significantly higher than that in globular clusters, by up to an order of magnitude.
Initially , this was interpreted as evidence that most halo BSSs originated from accretion of a younger population in satellite dwarf galaxies rather than from binary mass transfer \citep{Preston+1994}. 
However, the work of \citet{Preston+Sneden2000} made it clear that the binary frequency among halo BSSs is much higher than among normal halo stars, and that the specific frequency of halo BSSs is similar to that of the least luminous and lowest-concentration globular clusters. \cite{Carney+2001} find a binary frequency of $60 \pm 24\%$ for $P < 1600$~days, and derive a frequency of $69 \pm 12\%$ based on orbital solutions (and possible solutions) from \cite{Preston+Sneden2000}.
This suggests that halo BSSs are predominantly mass gainers, like classical BSSs in open and globular clusters.
A compilation of 8000 halo BSSs was made from Sloan Digital Sky Survey (SDSS) spectra by \citet{Santucci+2015}.

\citet{Sneden+2003} analysed the abundances of six halo BSSs with [Fe/H] $<-2$ from the \citeauthor{Preston+Sneden2000} sample, finding that the binaries among this sample are strongly enhanced in carbon and \emph{s}-process elements (Sr and Ba). 
Some of the halo BSSs are therefore also CEMP-\emph{s} stars.
\cite{Carney+2005} note that 5 of their 6 binary halo BSSs are depleted in lithium (although one lies in the lithium gap), which they argue is consistent with mass transfer. 

Carbon-enhanced metal-poor (CEMP) stars have been discovered in great numbers in low-resolution sectroscopic surveys for very metal-poor halo stars, starting with the HK survey and Hamburg-ESO survey \citep{Beers+1992,Christlieb+2001} and subsequently SDSS-SEGUE, LAMOST and PRISTINE \citep{Lee+2013,Li+2018,Arentsen+2021,Fang+2025}. 
The heavy-element abundance patterns of CEMP stars show a large variety, leading to a further division into subclasses (see e.g.\ \citealp{Beers+Christlieb2005} for a review). 
Of these, only the CEMP-\emph{s} stars which show enhancements of \emph{s}-process elements are likely mass gainers and comprise YSSs (and some BSSs) for this review. 

CH stars and CEMP-\emph{s} stars are traditionally treated as separate classes, with CH stars occupying a more limited (and somewhat higher) metallicity range, but they are hard to distinguish on objective criteria \citep[e.g.][]{Jorissen+2016}. 
Most known CH and CEMP-\emph{s} stars are YSSs, but equivalents on the MS have been found as well. 
Some 200 CH BSSs and YSSs in the field have been catalogued by \citet{Bartkevicius1996}. 
The proportion of CEMP stars among metal-poor stars is a matter of uncertainty and debate \citep{Arentsen+2022}, but it is certainly high: estimates range from $\sim$8\% to $\sim$30\% \citep[e.g.][]{Lucatello+2006,Lee+2013,Placco+2014,Li+2022} at [Fe/H] $\lsim -2$, and increase at lower metallicities.
The literature does not provide such fractions for the CEMP-\emph{s} stars as a separate subclass, but the more limited sample of metal-poor stars studied at high spectral resolution \citep[e.g. the SAGA database;][]{Suda+2008} shows that CEMP-\emph{s} stars consititute the majority, roughly 80\%, of CEMP stars.

Carbon dwarfs (dC stars) were first identified with the discovery by \citet{Dahn+1977} that the M-type dwarf G77-61 exhibits carbon-star chemistry, and they have since been discovered in large numbers in spectroscopic surveys.
They are predominantly a high-proper-motion halo population \citep{Farihi+2018}. 
dC stars partly overlap with CH and CEMP BSSs, although they are cooler and of lower mass. 
They can be considered as BLs within the lower MS, and as such fall within our definition of BSSs.
\citet{Green2013} presented a catalog of about 1200 high-latitude carbon stars from SDSS, about 60\% of which (729) are dC stars, making them the most common type of carbon-rich star in the Galaxy.
The lack of high-resolution spectroscopy, and therefore of measured heavy-element abundances, for all but one dC star makes their identification as mass gainers somewhat tentative, although their binary properties (Section~\ref{sec:properties_field_BSS_binary}) do support this interpretation.

Barium stars populate the Galactic disk, and are the metal-rich counterparts of the CH and CEMP-\emph{s} stars. 
A catalogue of 389 barium stars was compiled by \citet{Lu1991}, which has served as a basis for further studies. 
This sample comprises primarily Ba giant YSSs, with only a few Ba dwarf BSSs. 
Ba-rich YSSs have been estimated by \citet{MacConnell+1972} to constitute about 1\% of the population of normal G-K giant stars. 
\citet{North+Duquennoy1991} estimate the fraction of Ba-rich BSSs among their parent population (F-type MS stars) to be about 1\% as well.

Combining spectroscopic and asteroseismic measurements, a sub-population of $\alpha$-element-enhanced giant stars in the thick disk has been uncovered with masses much larger than those of typical old, $\alpha$-rich stars \citep{Chiappini+2015,Martig+2015}. Termed 'young, alpha-rich stars' these were initially interpreted as having formation ages younger than their parent population. However, radial-velocity monitoring has shown a higher than typical binary frequency, while their C/N ratios are more in line with an old thick-disk population, supporting an interpretation as mass-transfer products, that is, YSSs \citep{Jofre+2016,Jofre+2023}.

\subsection{Blue Stragglers}\label{sec:properties_field_BSS}

\subsubsection{Hertzsprung-Russell Diagrams}\label{sec:properties_field_BSS_CMD}

\begin{figure}
\centerline{\includegraphics[width=4.8875in,height=1.92361in]{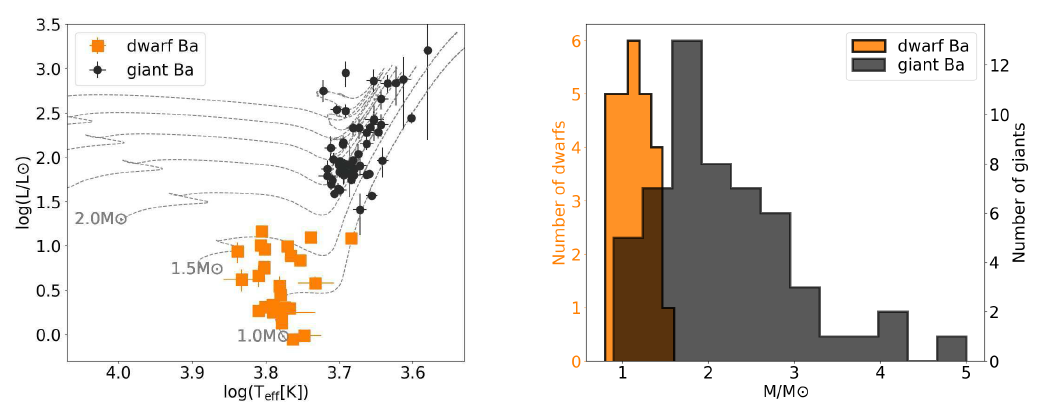}}

\caption{Location in the HRD (left) and derived mass distribution (right) of Ba BSSs and Ba YSSs (denoted in the figures as 'dwarf Ba' and 'giant Ba', respectively).
The sample of Ba BSSs includes the subgiant CH stars. \cite{Escorza+2019a}, adapted from \cite{Escorza+2017,Escorza+2019}; copyright ESO with permission from A\&A.}
\label{fig:Ba_stars_HRD}
\end{figure}

Using Gaia parallaxes, \citet{Escorza+2017} constructed a Hertzsprung-Russell diagram (HRD) of barium and CH stars (Supplemental Figure~\ref{fig:Ba_stars_HRD}). 
Dwarf barium stars and so-called `subgiant' CH stars are almost indistinguishable in terms of their HRD locations (both are distributed similarly among the orange squares shown) and their metallicities (somewhat subsolar), another indication that these groups of BSSs are manifestations of the same phenomenon. 
The detections of these BSSs are limited to effective temperatures $\lsim$ 7000 K due to the difficulty of obtaining accurate abundances of early-F and A stars \citep{Escorza+2019}.\footnote{Earlier-type (and higher-mass) counterparts may be hiding among the binaries detected among pulsating ($\delta$ Scuti) stars of type A-F using the pulsation-timing technique \citep{Murphy+2018}, some 20\% of which appear to have WD companions and similar orbital properties as the Ba BSSs.} 

CEMP-\emph{s} stars classified as dwarfs based on their surface gravities are located in, or somewhat to the blue of, the MSTO region of a 12\,Gyr halo isochrone \citep{Masseron+2010}. 

\subsubsection{Masses}\label{sec:properties_field_BSS_masses}

Masses of the various types of BSSs in the field have not been dynamically measured. 
\citet{Carney+2005} used isochrone fitting to estimate the masses of a subset of halo BSSs from the \citet{Preston+Sneden2000} sample, obtaining values between 0.83\,\Msun\ (just above the halo MSTO) and 1.28\,\Msun. 
As noted above, some of these are MS CEMP-\emph{s} stars (Supplemental Figure~\ref{fig:field_HRD}). 
Their counterparts among lower-mass dwarfs are the dC stars, with estimated masses (based on infrared photometry) between 0.2~\Msun\ and 0.8~\Msun\ \citep{Roulston+2021}.
\citet{Escorza+2017} derived the masses of barium stars by fitting stellar evolution tracks to their locations in the HRD, assuming an average [Fe/H] = $-0.25$. 
\citet{Escorza+2019} refined these results specifically for Ba and CH dwarfs, making use of measured metallicities. The masses of these BSSs range from 0.8~\Msun~to 1.5~\Msun\ (Supplemental Figure~\ref{fig:Ba_stars_HRD}).




\subsubsection{Binary and Multiplicity Properties}\label{sec:properties_field_BSS_binary}

\paragraph{Binary Frequency}

About 2/3 of the halo BSSs are spectroscopic binaries with $P < 3000$~d \citep{Preston+Sneden2000}. Presuming the presence of longer-period binaries, the true binary fraction is higher.

Among the 60 Ba and CH BSSs analysed by \citet{Escorza+2019}, about 2/3 are confirmed spectroscopic binaries (with orbits determined for 27 systems). 
Again, this is a lower limit to the true binary fraction, due to
insufficient data and/or phase coverage for many stars. 

\citet{Whitehouse+2018} reported 21 radial-velocity variables among 28 monitored dC stars, which they find consistent with a 100\% binary frequency. 
\citet{Roulston+2019} reached a similar conclusion based on a larger radial-velocity survey of 240 dC stars. 

\paragraph{Periods and Eccentricities}

\citet{Carney+2001,Carney+2005} determined periods and eccentricities for 20 halo BSSs. 
These are shown in panel c of Main Text Figure~4, where they are compared with other binary evolution products. 
The period-eccentricity distribution of the halo BSSs is similar to that of BSSs in NGC 188 (panel a of the same figure).

Two of these halo BSSs are classified as CEMP-\emph{s} stars, and therefore also appear in Main Text Figure~4e.
Only a handful of other CEMP-\emph{s} BSSs have measured orbits. 
A clearer picture of carbon-rich BSSs emerges with the inclusion of the orbital properties of the dC stars, also shown in Main Text Figure~4e as red symbols. The first-discovered dC star, G77-61, turned out to be a spectroscopic binary with a 245-day circular orbit \citep{Dearborn+1986}.
For a long time this was the only confirmed binary dC star, until the discovery of a dC star in a circular 3.14-day orbit inside the Necklace planetary nebula \citep{Miszalski+2013}. 
Since then populations of both short- and long-period dC binaries have been found. 
\citet{Harris+2018} discovered three astrometric dC binaries in long-period (400-4000 d) and mildly eccentric ($0.1 < e < 0.3$) orbits. 
The radial-velocity surveys by \citet{Whitehouse+2018} and \citet{Roulston+2019}, although too sparse to determine orbital solutions, have radial-velocity amplitudes that indicate most dC stars have typical periods of around one year. 
Prompted by the discovery of stellar activity and photometric variability among a subset of dC stars (Section~\ref{sec:properties_field_BSS_rotation}), an additional population of short-period spectroscopic dC binaries has recently been uncovered with circular orbits and periods between 0.2 and 4.4 days \citep{Margon+2018,Roulston+2021,Whitehouse+2021}.


Orbital solutions for 27 Ba BSSs were determined by \citet{Escorza+2019}, with an additional handful of orbits added by \citet{North+2020}. Their periods range from 100 days to 20,000 days and eccentricities again span a wide range from nearly circular up to $e=0.8$, as in NGC 188 (Main Text Figure~4g).
Remarkably, the population of short-period circular binaries found among C-rich BSSs is not present among Ba-selected BSSs.

\paragraph{Companions and companion masses}

The clearest evidence for companions to BSSs has been found among dC stars, about a dozen of which have a composite DA-type WD spectrum in the optical \citep[see e.g.][]{Green2013}. One of these dC-DA systems is an SB1 with a period of 0.3~days, in which the WD spectrum indicates a temperature of 31,000~K and a mass of 0.57~\Msun\ \citep{Roulston+2021}.

For other field BSS samples, WD companions have been detected in the UV.
\citet{Ekanayake+Wilhelm2018} cross-correlated SDSS-selected halo BSSs with GALEX data, finding several UV-excess stars. 
Using SED fitting to one FUV flux, they estimate effective temperatures and masses of the companions, finding them to be mostly hot (20,000--40,000 K), young WDs with a mass distribution peaking at 0.4-0.45~\Msun. 
\citet{Panthi+2023} observed 27 halo BSSs in the FUV with UVIT, finding evidence for hot (likely WD) companions for 12 BSSs, 8 of which are known spectroscopic binaries. 
SED fitting yielded estimated effective temperatures of 10,000--40,000~K and inferred masses in a wide range, including 6 ELM WDs.
\citet{Gray+2011} analysed GALEX observations of a small sample of Ba BSSs and concluded that, as an ensemble, they show excess UV flux compared to their parent population, indicative of WD companions.

The presence of WD companions can be inferred from distributions of observed mass functions, which indicate companions with typical WD masses \citep[e.g.][]{Escorza+2019}. 
\citet{Escorza+DeRosa2023} combined Hipparcos and Gaia astrometric observations with radial-velocity data to measure orbital inclinations of about 60 Ba stars, among which were 17 Ba BSSs, allowing a measurement of the companion masses (making use of the inferred primary star masses).
This reveals masses mostly between 0.5 and 0.8~\Msun, consistent with a WD companion population and with evidence for a few more-massive WDs.

\paragraph{High-order multiplicity}

\citet{Escorza+2019} found four likely triples in their sample of 40 spectroscopic-binary Ba-dwarf BSSs (a frequency of 10\%). 
One system comprises an inner 73-day and an outer 10,500-day (29-yr) orbit, both eccentric (blue triangles in Main Text Figure~4g). 
One of the two companions is likely the former mass donor and currently a WD, but it is not clear which.
Another triple is an SB2 composed of the Ba BSS and a lower-mass MS star ($q = 0.73$) in a 437-day eccentric orbit (blue square in Main Text Figure~4g), with a likely WD (revealed by a UV excess) in a wider orbit. 
Two further Ba BSSs are also SB2 binaries but without determined orbits. 
Although there is no direct evidence of third stars, tertiary WDs are expected in the mass-transfer formation scenario. Finally, the SB2 CEMP-\emph{s} star CS22964-161 \citep{Thompson+2008} consists of two polluted turnoff stars with a mass ratio of 0.68 in an eccentric 252-day orbit, in which the mass gained must have come from a third star. (See Main Text Section 4.2.4 for a possible formation scenario.)


\subsubsection{Surface Abundances}\label{sec:properties_field_BSS_abundances}


Compared to the intrinsically more luminous YSSs, high spectral-resolution abundance studies of BSSs in the Galactic Field are relatively less common. As noted in Section~\ref{sec:properties_field_catalogs}, \citep{Sneden+2003} found strongly elevated C- and \emph{s}-process abundances in the three metal-poor binary halo BSSs they studied. In one of these, they also measured a very high lead (Pb) overabundance. Similar surface abundance patterns of enhanced C, \emph{s}-process elements and especially Pb were found in the CEMP-s BSSs HE~0024--2523 \citep{Lucatello+2003} and CS~22964--161 \citep{Thompson+2008}. In all of these case, the abundance patterns are well matched by low-mass AGB nucleosynthesis models at the measured metallicities.

dC stars are typically too faint to allow high-resolution spectroscopy, and the lack of reliable atmosphere models for C-rich dwarfs makes detailed abundance studies difficult. The one exception is the prototype G77-61, which was analysed by \citet{Plez+Cohen2005} and is extremely metal-poor ([Fe/H] = $-4$), very rich in C and N, but lacking strong \emph{s}-process enhancements. This surprising result may be a single case akin to CEMP-\emph{no} stars -- C-enhanced metal-poor stars without heavy-element overabundances -- which at very low metallicity ([Fe/H] $<-3$) are the most common type of C-rich star.

Surface abundance analyses of Ba-rich BSSs (including CH subgiants) have been presented by, among others, \citet{North+1994}, \citet{Allen+Barbuy2006} and \citet{Roriz+2024}, finding enhanced abundances of carbon and various \emph{s}-process elements, including Pb. The latter authors also present an overview of abundance measurements for 71 Ba-rich BSSs from the literature. Their findings confirm that the observational distinction between 'Ba dwarfs' and 'CH subgiants' is related to the C/O ratio being smaller or greater than 1, respectively, but that these stars are very similar otherwise.

\subsubsection{Rotation and Stellar Activity}\label{sec:properties_field_BSS_rotation}

\citet{Preston+Sneden2000} and \citet{Carney+2005} measured projected rotation velocities of halo BSSs, with both studies finding significantly higher \vsini\ values for confirmed binaries than for non-velocity-variable stars, in some cases above 100 km/s. (They note that the non-velocity-variable stars may not be true BSSs but interlopers from an intermediate-age population.)

Also potentially interesting in terms of post-formation rapid rotation is the system UCAC2 46706450, consisting of a very hot WD (105,000 K) and a rapidly rotating (P $\approx$ 2 days; $v_\mathrm{rot} = 150$ km/s) K subgiant, but the orbit is unknown \citep{Werner+2020}.

On the other hand, among the so-called 'F str $\lambda$4077' stars, many of which later turned out to be Ba BSSs, \citet{North+Duquennoy1991} found no evidence for rotation more rapid than normal F stars.

Signs of stellar activity have been found for a small subset of dC stars, in the form of H$\alpha$ and/or X-ray emission \citep{Green2013,Green+2019}. A similar fraction, about 3\%, of dC stars show short-period photometric variability \citep{Roulston+2021}.
Some of these variable dC stars have been confirmed as spectroscopic binaries with orbital periods very close to the photometric periods \citep{Whitehouse+2021,Roulston+2021}. 
This finding suggests these short-period binaries have been (almost) tidally synchronised.

\subsection{Yellow Stragglers}\label{sec:properties_field_YSS}


\subsubsection{Hertzsprung-Russell Diagrams}\label{sec:properties_field_YSS_CMD}

Most of the Ba YSSs occupy the RC region of the HRD (Supplemental Figure~\ref{fig:Ba_stars_HRD}), consistent with this being a relatively long-lived phase compared to the RGB. 
The extrinsic (Tc-poor) S-star YSSs are located at higher luminosities along the giant branch than the Ba giants (Supplemental Figure~\ref{fig:field_HRD}), but are less luminous than the Tc-rich S-stars which are genuine AGB stars \citep{Shetye+2018}.

\subsubsection{Masses}\label{sec:properties_field_YSS_masses}

Masses of C- and \emph{s}-process-enriched YSSs have not been measured dynamically; mass determinations are only available from stellar evolution models. 
The locations of CEMP-\emph{s} YSSs along metal-poor halo isochrones suggest that they have masses somewhat in excess of the halo MSTO mass of 0.8-0.9~\Msun. However, \citet{Karinkuzhi+2021} showed that the location in the HRD of RGB evolution models is affected by a strongly increased C abundance, making accurate mass determinations by this method difficult.

The mass distribution of Ba YSSs derived from their HRD location ranges from 1~\Msun\ to almost 5~\Msun, peaking around 2~\Msun\ \citep[][see Supplemental Figure~\ref{fig:Ba_stars_HRD}]{Escorza+2017}. 
Compared to non-Ba-enriched giants, as well as to Ba BSSs, Ba YSSs show a deficit of low-mass stars ($<2~\Msun$). 
\citet{Escorza+2020} used binary evolution modelling to show that low-mass Ba BSSs with orbital periods up to about 750 days fill their Roche lobes during their ascent of the RGB, preventing them from reaching the RC region as Ba YSSs. This may at least partly explain the deficit of low-mass Ba YSSs. 
Extrinsic S-type YSSs have a similar mass distribution to the Ba YSSs, but peaking at lower masses \citep{Shetye+2018}.


\subsubsection{Binary and Multiplicity Properties}\label{sec:properties_field_YSS_binary}

From long-term radial-velocity monitoring of $\sim$110 Ba YSSs and extrinsic-S YSSs, \citet{Jorissen+1998,Jorissen+2019} found that all are spectroscopic binaries. 
Orbital parameters have now been determined for the vast majority of this sample. 
The binary frequency of the CH YSSs is similarly high \citep{McClure+Woodsworth1990}.

The situation for the CEMP-\emph{s} YSSs is somewhat less clear. \citet{Lucatello+2005} found a large fraction of spectroscopic binaries among the CEMP-\emph{s} YSSs known at the time, consistent with a 100\% true binary fraction. 
\citet{Starkenburg+2014} show that the distribution of radial-velocity variations is consistent with all of them being binaries with periods $\lsim$ 10,000 days. 
However, from long-term precise monitoring of 22 CEMP-\emph{s} YSSs \citet{Hansen+2016} found variable radial velocities for only 18 of the sample, the other four having constant radial velocity over a timescale of several years. 
While much longer periods or very small inclination angles cannot be excluded, this is unlikely to be the case for all four, so single CEMP-\emph{s} YSSs appear to exist as well. 
There thus may  be multiple formation scenarios at play for these C-rich YSSs stars.

\paragraph{Periods and eccentricities}

\begin{figure}

\centerline{\includegraphics[width=\textwidth]{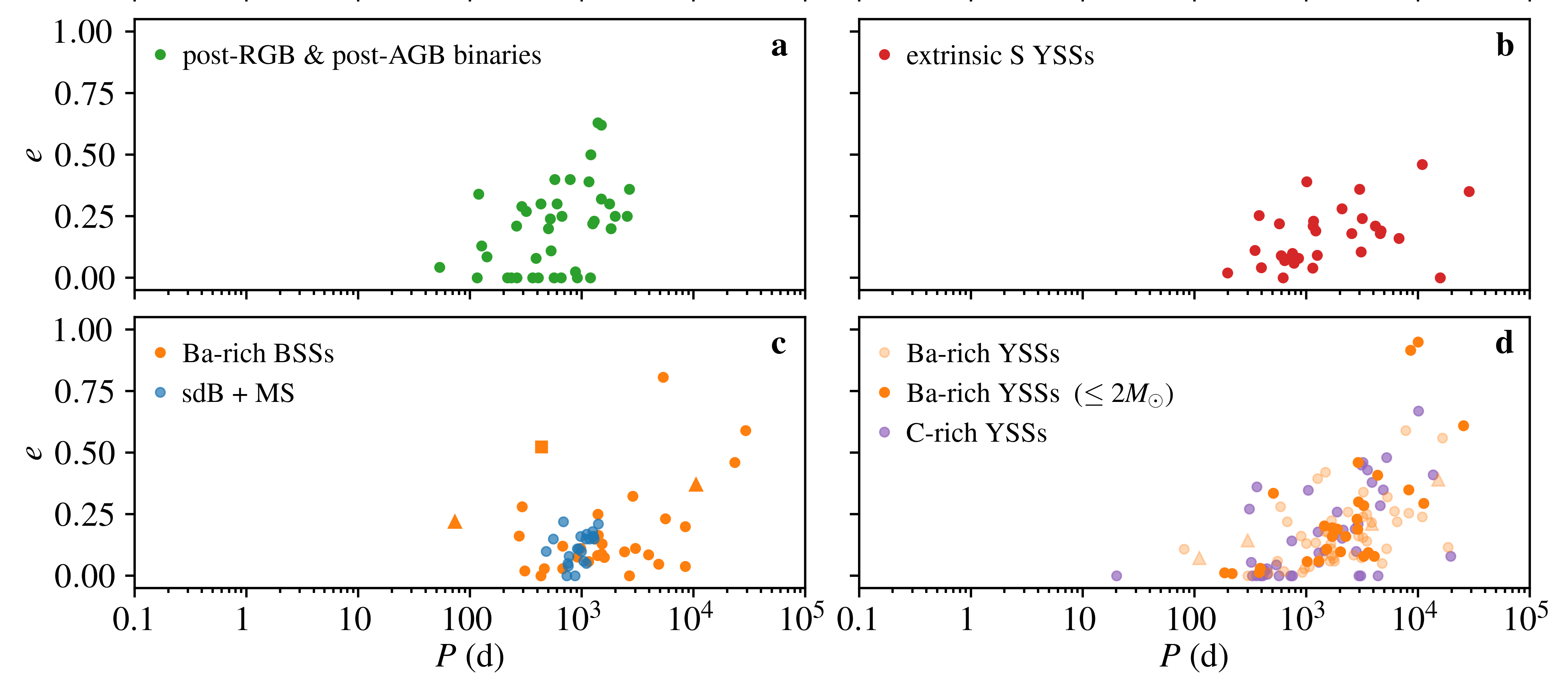} \\[1ex]}
\caption{Periods and eccentricities for different samples of
post-mass-transfer binaries: post-AGB and post-RGB binaries (a) and extrinsic S stars (b). These are compared in panel (c) with barium BSSs and sdB + MS binaries, and in panel (d) with Ba-rich and C-rich YSSs. The color-coding corresponds to that used for the same samples in Supplemental Figure~\ref{fig:field_HRD};
\emph{square symbols} are double-lined spectroscopic binaries; \emph{triangles} indicate the inner and outer orbits of triple systems.
Data taken from: (a) \citet{Oomen+2018,KLuska+2022}; (b) \citet{Jorissen+1998,Jorissen+2019}; (c, d) \citet{Jorissen+1998,Jorissen+2016,Jorissen+2019,Hansen+2016,VanderSwaelmen+2017,Escorza+DeRosa2023}.
}
\label{fig:e_logP_field}
\end{figure}

The distributions of the orbital periods and eccentricities of CH and CEMP-\emph{s} YSSs \citep{Jorissen+2016,Hansen+2016} and of Ba and extrinsic-S YSSs \citep{Jorissen+1998,Jorissen+2019,Escorza+DeRosa2023} 
are very similar (Supplemental Figure~\ref{fig:e_logP_field}b and d).
Periods range from about 100 to 40,000 days, with circular or mildly eccentric orbits dominating below $\sim$1000 days, and somewhat larger eccentricities and a near absence of circular orbits at longer periods. 


While these similarities are the most striking, there are is one possible difference of interest. The Ba-rich YSSs with masses less than 2~\Msun\ appear to have fewer binaries with eccentric orbits at $P < 1000$ days, for example compared to the Ba-rich BSSs. If confirmed with larger samples, this may be explained by the fact that most Ba YSSs are RC stars: in binaries with periods up to 1000 days the orbits of low-mass Ba RC stars are expected to be circularized as a result of tidal interactions during the preceding RGB phase when they reach very large radii \citep{Escorza+2020}.


In the same vein, the extrinsic-S YSSs (Supplemental Figure~\ref{fig:e_logP_field}b) show a larger fraction of eccentric orbits at $P < 1000$ days than the Ba-rich YSSs. 
Although these stars are cooler and more luminous than Ba YSSs, most are likely to be RGB stars in a less evolved (pre-RC) stage. They would then have suffered less from tidal circularization and may have preserved their MS orbital properties.

Finally, the two Ba-rich YSSs with eccentricities of $\approx$0.95 are notable. Achieving such high eccentricity through a mass-transfer process would be remarkable. Perhaps these are indicative of orbital evolution in a higher-order multiple.

Although not shown in Supplemental Figure~\ref{fig:e_logP_field}, field S-type symbiotic binaries (consisting of an RGB star transferring mass to a WD, and not to be confused with S stars) also have similar orbital properties \citep{Mikoajewska2003,Mikoajewska2012}.
They likely correspond to a more advanced stage of YSS evolution, in which the giant is now close to filling its Roche lobe, but their exact evolutionary relation to the polluted stars is not yet clear.

\paragraph{Companions and companion masses}

Direct evidence for WD companions in the form of excess UV flux has been found for several Ba YSSs, with inferred effective temperatures ranging between 10,000 and 35,000~K \citep[e.g.][]{Bohm-Vitense1980,Fekel+1993,Bohm-Vitense+2000}.

Indirect statistical evidence for WD companions has been derived from the distribution of spectroscopic mass functions, assuming primary masses inferred from HRD locations. 
This indicates companion mass distributions peaking around 0.6~\Msun\ for CH and CEMP-\emph{s} YSSs \citep{Jorissen+2016} and between 0.6 and 0.9~\Msun\ for Ba YSSs \citep{Jorissen+2019}. 
Using measured orbital inclinations, \citet{Escorza+DeRosa2023} found the companion masses of Ba YSSs to lie mostly in the range 0.5--0.8~\Msun, with a tail extending to higher masses.

\paragraph{High-order multiplicity}

\citet{Escorza+DeRosa2023} found one Ba YSS in a triple system among their sample of 60 Ba stars. 
It has a 111-day inner and $\approx\,$15,000-day (41-year) outer orbit. The outer companion is suspected of being a WD while the inner companion is a low-mass star.

\subsubsection{Surface Abundances}\label{sec:properties_field_YSS_abundances}

Surface abundances of a large sample of 182 barium YSSs were determined in a homogeneous way by \citet{deCastro+2016}. 
They confirm and strengthen earlier results by finding a wide range of overall \emph{s}-process enhancements (between [s/Fe] = 0.25 and 2.0) and ratios of heavy (Ba, Ce, La, Nd) to light (Sr, Zr, Y) \emph{s}-process elements, with these quantities being strongly correlated with each other and weakly correlated with metallicity.
These abundance measurements, in combination with AGB nucleosynthesis models, have been used to derive constraints on the efficiency of mass transfer from the former AGB donor star (see Main Text Section 3.5).
\citet{Jorissen+2019} showed that the average level of \emph{s}-process enhancement of Ba YSSs anti-correlates with orbital period, and that so-called strong Ba stars (those with the highest heavy-element abundances, [s/Fe] $\gsim 1$) only occur in orbits with $P < 10,000$~days.

C-enriched YSSs show a larger and somewhat more puzzling variety in heavy-element abundance patterns, which have been the subject of many studies. 
\citet{Masseron+2010} compiled abundance measurements from the literature of some 100 C-enriched YSSs and BSSs and attempted a comprehensive interpretation of their diverse abundance patterns. CEMP-\emph{s} stars show a strong correlation between the overabundances of C and \emph{s}-process elements, consistent with nucleosynthesis models of low-mass, metal-poor AGB stars. However, a roughly equally numerous subgroup -- named CEMP-\emph{rs} stars -- shows overabundances of elements (such as europium, Eu) typically associated with the \emph{r}-process, alongside enhancements of \emph{s}-process elements. They have very similar binary properties as the CEMP-\emph{s} stars (and are subsumed in this category in Main Text Figure~4 and Supplemental Figure~\ref{fig:e_logP_field}). The possible origin of their abundances has led to much speculation, but \citet{Hampel+2016,Hampel+2019} demonstrated that their abundance pattern can most naturally result from a neutron-capture process (named the \emph{i}-process) operating at neutron densities intermediate between those of the \emph{s}- and \emph{r}-processes. From a homogeneous analysis of 24 such CEMP-\emph{s} and CEMP-\emph{rs} stars, \citet{Karinkuzhi+2021} showed that their abundances form a continuum rather than being clearly separated, and that recent nucleosynthesis models of low-mass, metal-poor AGB stars undergoing proton ingestion during thermal pulses can well match the \emph{i}-process abundance pattern of the latter group.


\subsection{Subdwarf B Stars and Post-AGB Stars}\label{sec:properties_field_sdB}

As noted in Main Text Section 3.4.2, sdB stars appear to fall into three groups: single sdB stars, short-period binary sdB stars, and binary sdB stars in wide orbits. 
Here we focus on the latter category, in which the companions are MS stars of spectral type F-K.
\citet{Vos+2017} find mass ratios ($M_\mathrm{sdB}/M_\mathrm{MS}$) in the range 0.3--0.7, indicating MS companion masses between 0.7~\Msun\ and 1.3~\Msun\ (assuming a canonical sdB mass of 0.47~\Msun). The MS stars have $\vsini$ values between 8~km s\textsuperscript{-1} and 80~km s$^{-1}$ \citep{Vos+2018}, suggestive of spin-up by mass transfer. If so, then these companions may classify as BSSs.
Orbital elements have been determined for 23 wide sdB systems \citep[][and references therein]{Vos+2017,Vos+2019} and are shown as blue symbols in Supplemental Figure~\ref{fig:e_logP_field}c. 
The periods range between 500 and 1400 days and the eccentricities are mild but mostly non-zero, similar to those of the field BSSs and YSSs.
The limited period range of wide sdB binaries may be partly understood from the fact that a relatively small range of initial separations is required to ignite helium in a RGB star at the same time that it is losing its envelope near the tip of the RGB \citep{Han+2002}.

Post-AGB stars are in a short-lived phase after the end of the AGB, contracting towards the WD stage (Supplemental Figure~\ref{fig:field_HRD}). 
Many of them are in binary systems surrounded by circumbinary disks \citep[see][for reviews]{VanWinckel2003,VanWinckel2018}. 
Equivalents with luminosities below the RGB tip have also been found and named post-RGB stars, since they are too faint to have originated from an AGB progenitor \citep{Kamath+2014}.
Binary post-AGB and post-RGB stars may hold important clues to the prior interactions taking place during mass transfer on the AGB and upper RGB, since they have so recently emerged from it.
The distribution of their spectroscopic mass functions suggests that they have MS companions, with masses distributed around a mean of 1.1~\Msun\ \citep{Oomen+2018}.
Their periods and eccentricities \citep{Oomen+2018} are very similar to those of BSSs and YSSs (Supplemental Figure~\ref{fig:e_logP_field}a), although with periods only up to about 3000 days. The latter is not simply a selection effect because the timescale of RV monitoring extends well beyond 10 years \citep{VanWinckel2018}. 

The orbital properties of binary post-AGB stars are suggestive of an evolutionary connection with the C- and Ba-rich BSSs, but whether they are the actual progenitor systems of these AGB-polluted BSSs is not clear. There are indications (mostly indirect, from the circumstellar material) that in many post-AGB stars no strong enrichment in carbon- and \emph{s}-process elements had taken place during the preceding AGB phase \citep{VanWinckel2018}. 
This suggests the AGB donors were relatively unevolved when mass transfer started, which implies they would have had relatively small radii and may be consistent with the lack of post-AGB orbital periods longer than 3000 d.
Perhaps the mass transfer process responsible for C- and Ba-rich BSSs in wider orbits is ineffective at producing the circumbinary disks that characterize the binary post-AGB stars.

\subsection{Red Stragglers and Sub-Subgiants}\label{sec:properties_field_RSS}

\citet{Leiner+2022} recently used a catalog of active giant binaries (RS CVn's) in the field to identify and study a field SSG population. They used Gaia EDR3 to place the stars on the HR diagram, and identify as SSGs those stars that fall below or to the red of a 14-Gyr isochrone. They find a large population of 448 SSGs, roughly a quarter of the original sample. Most of the population lies to the red of the base of the giant branch. Almost all (95$\%$) of the SSG candidates have rotation periods of 2-20 days, with a higher yield at smaller rotation period. They conclude that SSGs are a typical outcome of giant evolution in close binaries.

Building upon a sample of 7,286 RGB stars from APOGEE DR17 with measurable rotation, \cite{Dixon+2025} also define a field sample of SSGs. They find these stars to be rapidly rotating, which they attribute to tidal synchronization with binary companions, and to show high magnetic activity indices. They argue that these SSGS are close to a maximum of RS CVn activity cycles.

RSSs have not been heavily studied in the field. Recently, \cite{Stassun+2023}  serendipitously discovered that the eclipsing binary 2M0056–08 (period 33.9 days) comprises two RSSs of identical mass (1.419 $\Msun$). Both stars are rotating synchronously with each other with a period of 30.2 days, slightly less than the orbital period, but a rapid rotation for giant stars. The latter presumably is the origin of substantial magnetic activity in the system. An FUV excess may be either chromospheric or indicative of a tertiary WD. The HRD position of the giants may be the result of mass transfer or of extensive photospheric spots. \cite{Stassun+2023} are able to model the slightly different HRD locations of the two stars with reduced convective efficiencies. The origin of the identical masses, whether through mass transfer from a tertiary \citep{PortegiesZwart+Leigh2019} or formation, remains uncertain.

\bibliography{references}

\begin{thebibliography}{}
\expandafter\ifx\csname natexlab\endcsname\relax\def\natexlab#1{#1}\fi

\bibitem[{Ahumada \& Lapasset(1995)}]{Ahumada+Lapasset1995}
Ahumada J, Lapasset E. 1995.
\textit{Astronomy and Astrophysics Supplement Series} 109:375--382

\bibitem[{Ahumada \& Lapasset(2007)}]{Ahumada+Lapasset2007}
Ahumada JA, Lapasset E. 2007.
\textit{Astronomy \& Astrophysics} 463(2):789--797

\bibitem[{Albrow et~al.(2001)Albrow, Gilliland, Brown, Edmonds, Guhathakurta \& Sarajedini}]{Albrow+2001}
Albrow MD, Gilliland RL, Brown TM, Edmonds PD, Guhathakurta P, Sarajedini A. 2001.
\textit{The Astrophysical Journal} 559:1060--1081

\bibitem[{Allen \& Barbuy(2006)}]{Allen+Barbuy2006}
Allen DM, Barbuy B. 2006.
\textit{Astronomy and Astrophysics} 454(3):895--915

\bibitem[{Arentsen et~al.(2022)Arentsen, Placco, Lee, Aguado, Martin et~al.}]{Arentsen+2022}
Arentsen A, Placco VM, Lee YS, Aguado DS, Martin NF, et~al. 2022.
\textit{Monthly Notices of the Royal Astronomical Society} 515:4082--4098

\bibitem[{Arentsen et~al.(2021)Arentsen, Starkenburg, Aguado, Martin, Placco et~al.}]{Arentsen+2021}
Arentsen A, Starkenburg E, Aguado DS, Martin NF, Placco VM, et~al. 2021.
\textit{Monthly Notices of the Royal Astronomical Society} 505:1239--1253

\bibitem[{Baldwin et~al.(2016)Baldwin, Watkins, {van der Marel}, Bianchini, Bellini \& Anderson}]{Baldwin+2016}
Baldwin AT, Watkins LL, {van der Marel} RP, Bianchini P, Bellini A, Anderson J. 2016.
\textit{The Astrophysical Journal} 827:12

\bibitem[{Baliunas \& Guinan(1985)}]{Baliunas+Guinan1985}
Baliunas SL, Guinan EF. 1985.
\textit{The Astrophysical Journal} 294:207--215

\bibitem[{Bartkevicius(1996)}]{Bartkevicius1996}
Bartkevicius A. 1996.
\textit{Baltic Astronomy} 5:217--229

\bibitem[{Beccari et~al.(2019)Beccari, Ferraro, Dalessandro, Lanzoni, Raso et~al.}]{Beccari+2019}
Beccari G, Ferraro FR, Dalessandro E, Lanzoni B, Raso S, et~al. 2019.
\textit{The Astrophysical Journal} 876:87

\bibitem[{Beccari et~al.(2006)Beccari, Ferraro, Lanzoni \& Bellazzini}]{Beccari+2006}
Beccari G, Ferraro FR, Lanzoni B, Bellazzini M. 2006.
\textit{The Astrophysical Journal} 652:L121--L124

\bibitem[{Beers \& Christlieb(2005)}]{Beers+Christlieb2005}
Beers TC, Christlieb N. 2005.
\textit{Annual Review of Astronomy and Astrophysics} 43:531--580

\bibitem[{Beers et~al.(1992)Beers, Preston \& Shectman}]{Beers+1992}
Beers TC, Preston GW, Shectman SA. 1992.
\textit{The Astronomical Journal} 103:1987

\bibitem[{Bekki \& Freeman(2003)}]{Bekki+Freeman2003}
Bekki K, Freeman KC. 2003.
\textit{Monthly Notices of the Royal Astronomical Society} 346:L11--L15

\bibitem[{Bertelli~Motta et~al.(2018)Bertelli~Motta, Pasquali, Caffau \& Grebel}]{BertelliMotta+2018}
Bertelli~Motta C, Pasquali A, Caffau E, Grebel EK. 2018.
\textit{Monthly Notices of the Royal Astronomical Society} 480:4314--4326

\bibitem[{Bhattacharya et~al.(2019)Bhattacharya, Vaidya, Chen \& Beccari}]{Bhattacharya+2019}
Bhattacharya S, Vaidya K, Chen WP, Beccari G. 2019.
\textit{Astronomy and Astrophysics} 624:A26

\bibitem[{Billi et~al.(2024)Billi, Ferraro, Mucciarelli, Lanzoni, Cadelano \& Monaco}]{Billi+2024}
Billi A, Ferraro FR, Mucciarelli A, Lanzoni B, Cadelano M, Monaco L. 2024.
\textit{Astronomy and Astrophysics} 690:A156

\bibitem[{Billi et~al.(2023)Billi, Ferraro, Mucciarelli, Lanzoni, Cadelano et~al.}]{Billi+2023}
Billi A, Ferraro FR, Mucciarelli A, Lanzoni B, Cadelano M, et~al. 2023.
\textit{The Astrophysical Journal} 956:124

\bibitem[{{B{\"o}hm-Vitense}(1980)}]{Bohm-Vitense1980}
{B{\"o}hm-Vitense} E. 1980.
\textit{The Astrophysical Journal} 239:L79--L83

\bibitem[{{B{\"o}hm-Vitense} et~al.(2000){B{\"o}hm-Vitense}, Carpenter, Robinson, Ake \& Brown}]{Bohm-Vitense+2000}
{B{\"o}hm-Vitense} E, Carpenter K, Robinson R, Ake T, Brown J. 2000.
\textit{The Astrophysical Journal} 533:969--983

\bibitem[{Brogaard et~al.(2018)Brogaard, Christiansen, Grundahl, Miglio, Izzard et~al.}]{Brogaard+2018}
Brogaard K, Christiansen SM, Grundahl F, Miglio A, Izzard RG, et~al. 2018.
\textit{Monthly Notices of the Royal Astronomical Society} 481:5062--5072

\bibitem[{Brogaard et~al.(2012)Brogaard, VandenBerg, Bruntt, Grundahl, Frandsen et~al.}]{Brogaard+2012}
Brogaard K, VandenBerg DA, Bruntt H, Grundahl F, Frandsen S, et~al. 2012.
\textit{Astronomy and Astrophysics} 543:A106

\bibitem[{Brown et~al.(2013)Brown, Landsman, Randall, Sweigart \& Lanz}]{Brown+2013}
Brown TM, Landsman WB, Randall SK, Sweigart AV, Lanz T. 2013.
\textit{The Astrophysical Journal} 777:L22

\bibitem[{Cadelano et~al.(2022)Cadelano, Ferraro, Dalessandro, Lanzoni, Pallanca \& Saracino}]{Cadelano+2022}
Cadelano M, Ferraro FR, Dalessandro E, Lanzoni B, Pallanca C, Saracino S. 2022.
\textit{The Astrophysical Journal} 941:69

\bibitem[{Carney et~al.(2005)Carney, Latham \& Laird}]{Carney+2005}
Carney BW, Latham DW, Laird JB. 2005.
\textit{The Astronomical Journal} 129:466--479

\bibitem[{Carney et~al.(1994)Carney, Latham, Laird \& Aguilar}]{Carney+1994}
Carney BW, Latham DW, Laird JB, Aguilar LA. 1994.
\textit{The Astronomical Journal} 107:2240

\bibitem[{Carney et~al.(2001)Carney, Latham, Laird, Grant \& Morse}]{Carney+2001}
Carney BW, Latham DW, Laird JB, Grant CE, Morse JA. 2001.
\textit{The Astronomical Journal} 122:3419--3435

\bibitem[{Cheng et~al.(2019)Cheng, Li, Li, Xu \& Fang}]{Cheng+2019}
Cheng Z, Li Z, Li X, Xu X, Fang T. 2019.
\textit{The Astrophysical Journal} 876:59

\bibitem[{Chiappini et~al.(2015)Chiappini, Anders, Rodrigues, Miglio, Montalb{\'a}n et~al.}]{Chiappini+2015}
Chiappini C, Anders F, Rodrigues TS, Miglio A, Montalb{\'a}n J, et~al. 2015.
\textit{Astronomy and Astrophysics} 576:L12

\bibitem[{Choi et~al.(2016)Choi, Dotter, Conroy, Cantiello, Paxton \& Johnson}]{Choi+2016}
Choi J, Dotter A, Conroy C, Cantiello M, Paxton B, Johnson BD. 2016.
\textit{The Astrophysical Journal} 823(2):102

\bibitem[{Christlieb et~al.(2001)Christlieb, Green, Wisotzki \& Reimers}]{Christlieb+2001}
Christlieb N, Green PJ, Wisotzki L, Reimers D. 2001.
\textit{Astronomy and Astrophysics} 375:366--374

\bibitem[{Cordoni et~al.(2023)Cordoni, Milone, Marino, Vesperini, Dondoglio et~al.}]{Cordoni+2023}
Cordoni G, Milone AP, Marino AF, Vesperini E, Dondoglio E, et~al. 2023.
\textit{Astronomy \& Astrophysics} 672:A29

\bibitem[{Corsaro et~al.(2012)Corsaro, Stello, Huber, Bedding, Bonanno et~al.}]{Corsaro+2012}
Corsaro E, Stello D, Huber D, Bedding TR, Bonanno A, et~al. 2012.
\textit{The Astrophysical Journal} 757:190

\bibitem[{Dahn et~al.(1977)Dahn, Liebert, Kron, Spinrad \& Hintzen}]{Dahn+1977}
Dahn CC, Liebert J, Kron RG, Spinrad H, Hintzen PM. 1977.
\textit{The Astrophysical Journal} 216:757--766

\bibitem[{Dalessandro et~al.(2013)Dalessandro, Ferraro, Massari, Lanzoni, Miocchi et~al.}]{Dalessandro+2013}
Dalessandro E, Ferraro FR, Massari D, Lanzoni B, Miocchi P, et~al. 2013.
\textit{The Astrophysical Journal} 778:135

\bibitem[{D'Antona et~al.(2010)D'Antona, Caloi \& Ventura}]{DAntona+2010}
D'Antona F, Caloi V, Ventura P. 2010.
\textit{Monthly Notices of the Royal Astronomical Society} :no--no

\bibitem[{Dattatrey et~al.(2023{\natexlab{a}})Dattatrey, Yadav, Kumawat, Rani, Singh et~al.}]{Dattatrey+2023}
Dattatrey AK, Yadav RKS, Kumawat G, Rani S, Singh G, et~al. 2023{\natexlab{a}}.
\textit{Monthly Notices of the Royal Astronomical Society} 523:L58--L63

\bibitem[{Dattatrey et~al.(2023{\natexlab{b}})Dattatrey, Yadav, Rani, Subramaniam, Singh et~al.}]{Dattatrey+2023a}
Dattatrey AK, Yadav RKS, Rani S, Subramaniam A, Singh G, et~al. 2023{\natexlab{b}}.
\textit{The Astrophysical Journal} 943:130

\bibitem[{{de Castro} et~al.(2016){de Castro}, Pereira, Roig, Jilinski, Drake et~al.}]{deCastro+2016}
{de Castro} DB, Pereira CB, Roig F, Jilinski E, Drake NA, et~al. 2016.
\textit{Monthly Notices of the Royal Astronomical Society} 459:4299--4324

\bibitem[{De~Marchi et~al.(2007)De~Marchi, Poretti, Montalto, Piotto, Desidera et~al.}]{DeMarchi+2007}
De~Marchi F, Poretti E, Montalto M, Piotto G, Desidera S, et~al. 2007.
\textit{Astronomy \& Astrophysics} 471(2):515--526

\bibitem[{De~Marco et~al.(2005)De~Marco, Shara, Zurek, Ouellette, Lanz et~al.}]{DeMarco+2005}
De~Marco O, Shara MM, Zurek D, Ouellette JA, Lanz T, et~al. 2005.
\textit{The Astrophysical Journal} 632:894--919

\bibitem[{Dearborn et~al.(1986)Dearborn, Liebert, Aaronson, Dahn, Harrington et~al.}]{Dearborn+1986}
Dearborn DSP, Liebert J, Aaronson M, Dahn CC, Harrington R, et~al. 1986.
\textit{The Astrophysical Journal} 300:314

\bibitem[{Dixon et~al.(2025)Dixon, Stassun, Mathieu, Tayar \& Cao}]{Dixon+2025}
Dixon D, Stassun KG, Mathieu RD, Tayar J, Cao L. 2025.
\textit{The Astronomical Journal} 169(6):309

\bibitem[{Edmonds et~al.(2003)Edmonds, Gilliland, Heinke \& Grindlay}]{Edmonds+2003}
Edmonds PD, Gilliland RL, Heinke CO, Grindlay JE. 2003.
\textit{The Astrophysical Journal} 596:1177--1196

\bibitem[{Ekanayake \& Wilhelm(2018)}]{Ekanayake+Wilhelm2018}
Ekanayake G, Wilhelm R. 2018.
\textit{Monthly Notices of the Royal Astronomical Society} 479:2623--2629

\bibitem[{Escorza et~al.(2017)Escorza, Boffin, Jorissen, Van~Eck, Siess et~al.}]{Escorza+2017}
Escorza A, Boffin HMJ, Jorissen A, Van~Eck S, Siess L, et~al. 2017.
\textit{Astronomy and Astrophysics} 608:A100

\bibitem[{Escorza \& De~Rosa(2023)}]{Escorza+DeRosa2023}
Escorza A, De~Rosa RJ. 2023.
\textit{Astronomy and Astrophysics} 671:A97

\bibitem[{Escorza et~al.(2019{\natexlab{a}})Escorza, Jorissen \& Boffin}]{Escorza+2019a}
Escorza A, Jorissen A, Boffin HMJ. 2019{\natexlab{a}}.
\textit{Memorie della Societa Astronomica Italiana} 90:407

\bibitem[{Escorza et~al.(2019{\natexlab{b}})Escorza, Karinkuzhi, Jorissen, Siess, Van~Winckel et~al.}]{Escorza+2019}
Escorza A, Karinkuzhi D, Jorissen A, Siess L, Van~Winckel H, et~al. 2019{\natexlab{b}}.
\textit{Astronomy and Astrophysics} 626:A128

\bibitem[{Escorza et~al.(2020)Escorza, Siess, Van~Winckel \& Jorissen}]{Escorza+2020}
Escorza A, Siess L, Van~Winckel H, Jorissen A. 2020.
\textit{Astronomy and Astrophysics} 639:A24

\bibitem[{Fang et~al.(2025)Fang, Li \& Li}]{Fang+2025}
Fang Z, Li X, Li H. 2025.
\textit{The Astrophysical Journal Supplement Series} 277:30

\bibitem[{Farihi et~al.(2018)Farihi, Arendt, Machado \& Whitehouse}]{Farihi+2018}
Farihi J, Arendt AR, Machado HS, Whitehouse LJ. 2018.
\textit{Monthly Notices of the Royal Astronomical Society} 477:3801--3806

\bibitem[{Fekel et~al.(1993)Fekel, Henry, Busby \& Eitter}]{Fekel+1993}
Fekel FC, Henry GW, Busby MR, Eitter JJ. 1993.
\textit{The Astronomical Journal} 106:2370

\bibitem[{Ferraro et~al.(2009)Ferraro, Beccari, Dalessandro, Lanzoni, Sills et~al.}]{Ferraro+2009}
Ferraro FR, Beccari G, Dalessandro E, Lanzoni B, Sills A, et~al. 2009.
\textit{Nature} 462:1028--1031

\bibitem[{Ferraro et~al.(2020)Ferraro, Lanzoni \& Dalessandro}]{Ferraro+2020}
Ferraro FR, Lanzoni B, Dalessandro E. 2020.
\textit{Rendiconti Lincei. Scienze Fisiche e Naturali} 31:19--31

\bibitem[{Ferraro et~al.(2012)Ferraro, Lanzoni, Dalessandro, Beccari, Pasquato et~al.}]{Ferraro+2012}
Ferraro FR, Lanzoni B, Dalessandro E, Beccari G, Pasquato M, et~al. 2012.
\textit{Nature} 492:393--395

\bibitem[{Ferraro et~al.(2015)Ferraro, Lanzoni, Dalessandro, Mucciarelli \& Lovisi}]{Ferraro+2015}
Ferraro FR, Lanzoni B, Dalessandro E, Mucciarelli A, Lovisi L. 2015.
\textit{Blue {{Straggler Stars}} in {{Globular Clusters}}: {{A Powerful Tool}} to {{Probe}} the {{Internal Dynamical Evolution}} of {{Stellar Systems}}}. In \textit{Ecology of {{Blue Straggler Stars}}}, eds. H~Boffin, G~Carraro, G~Beccari, vol. 413 of \textit{Astrophysics and {{Space Science Library}}}. Berlin, Heidelberg: Springer, ~99

\bibitem[{Ferraro et~al.(2023)Ferraro, Mucciarelli, Lanzoni, Pallanca, Cadelano et~al.}]{Ferraro+2023}
Ferraro FR, Mucciarelli A, Lanzoni B, Pallanca C, Cadelano M, et~al. 2023.
\textit{Nature Communications} 14(1):2584

\bibitem[{Ferraro et~al.(1997{\natexlab{a}})Ferraro, Paltrinieri, Fusi~Pecci, Cacciari, Dorman \& Rood}]{Ferraro+1997a}
Ferraro FR, Paltrinieri B, Fusi~Pecci F, Cacciari C, Dorman B, Rood RT. 1997{\natexlab{a}}.
\textit{The Astrophysical Journal} 484:L145--L148

\bibitem[{Ferraro et~al.(1997{\natexlab{b}})Ferraro, Paltrinieri, Fusi~Pecci, Cacciari, Dorman et~al.}]{Ferraro+1997}
Ferraro FR, Paltrinieri B, Fusi~Pecci F, Cacciari C, Dorman B, et~al. 1997{\natexlab{b}}.
\textit{Astronomy and Astrophysics} 324:915--928

\bibitem[{Ferraro et~al.(1999)Ferraro, Paltrinieri, Rood \& Dorman}]{Ferraro+1999}
Ferraro FR, Paltrinieri B, Rood RT, Dorman B. 1999.
\textit{The Astrophysical Journal} 522:983--990

\bibitem[{Ferraro et~al.(2006)Ferraro, Sabbi, Gratton, Piotto, Lanzoni et~al.}]{Ferraro+2006}
Ferraro FR, Sabbi E, Gratton R, Piotto G, Lanzoni B, et~al. 2006.
\textit{The Astrophysical Journal} 647:L53--L56

\bibitem[{Ferraro et~al.(2003)Ferraro, Sills, Rood, Paltrinieri \& Buonanno}]{Ferraro+2003}
Ferraro FR, Sills A, Rood RT, Paltrinieri B, Buonanno R. 2003.
\textit{The Astrophysical Journal} 588:464--477

\bibitem[{Fiorentino et~al.(2014)Fiorentino, Lanzoni, Dalessandro, Ferraro, Bono \& Marconi}]{Fiorentino+2014}
Fiorentino G, Lanzoni B, Dalessandro E, Ferraro FR, Bono G, Marconi M. 2014.
\textit{The Astrophysical Journal} 783:34

\bibitem[{Fiorentino et~al.(2015)Fiorentino, Marconi, Bono, Dalessandro, Ferraro et~al.}]{Fiorentino+2015}
Fiorentino G, Marconi M, Bono G, Dalessandro E, Ferraro FR, et~al. 2015.
\textit{The Astrophysical Journal} 810:15

\bibitem[{Fusi~Pecci et~al.(1992)Fusi~Pecci, Ferraro, Corsi, Cacciari \& Buonanno}]{FusiPecci+1992}
Fusi~Pecci F, Ferraro FR, Corsi CE, Cacciari C, Buonanno R. 1992.
\textit{The Astronomical Journal} 104:1831

\bibitem[{Geller et~al.(2013)Geller, Hurley \& Mathieu}]{Geller+2013}
Geller AM, Hurley JR, Mathieu RD. 2013.
\textit{The Astronomical Journal} 145:8

\bibitem[{Geller et~al.(2015)Geller, Latham \& Mathieu}]{Geller+2015}
Geller AM, Latham DW, Mathieu RD. 2015.
\textit{The Astronomical Journal} 150(3):97

\bibitem[{Geller et~al.(2017)Geller, Leiner, Bellini, Gleisinger, Haggard et~al.}]{Geller+2017}
Geller AM, Leiner EM, Bellini A, Gleisinger R, Haggard D, et~al. 2017.
\textit{The Astrophysical Journal} 840(2):66

\bibitem[{Geller \& Mathieu(2011)}]{Geller+Mathieu2011}
Geller AM, Mathieu RD. 2011.
\textit{Nature} 478:356--359

\bibitem[{Geller \& Mathieu(2012)}]{Geller+Mathieu2012}
Geller AM, Mathieu RD. 2012.
\textit{The Astronomical Journal} 144(2):54

\bibitem[{Geller et~al.(2008)Geller, Mathieu, Harris \& McClure}]{Geller+2008}
Geller AM, Mathieu RD, Harris HC, McClure RD. 2008.
\textit{The Astronomical Journal} 135(6):2264--2278

\bibitem[{Geller et~al.(2009)Geller, Mathieu, Harris \& McClure}]{Geller+2009}
Geller AM, Mathieu RD, Harris HC, McClure RD. 2009.
\textit{The Astronomical Journal} 137(4):3743--3760

\bibitem[{Geller et~al.(2021)Geller, Mathieu, Latham, Pollack, Torres \& Leiner}]{Geller+2021a}
Geller AM, Mathieu RD, Latham DW, Pollack M, Torres G, Leiner EM. 2021.
\textit{The Astronomical Journal} 161:190

\bibitem[{Giesers et~al.(2019)Giesers, Kamann, Dreizler, Husser, Askar et~al.}]{Giesers+2019}
Giesers B, Kamann S, Dreizler S, Husser TO, Askar A, et~al. 2019.
\textit{Astronomy and Astrophysics} 632:A3

\bibitem[{Gilliland et~al.(1998)Gilliland, Bono, Edmonds, Caputo, Cassisi et~al.}]{Gilliland+1998}
Gilliland RL, Bono G, Edmonds PD, Caputo F, Cassisi S, et~al. 1998.
\textit{The Astrophysical Journal} 507:818--845

\bibitem[{Goranskij et~al.(1992)Goranskij, Kusakin, Mironov, Moshkaljov \& Pastukhova}]{Goranskij+1992}
Goranskij VP, Kusakin AV, Mironov AV, Moshkaljov VG, Pastukhova EN. 1992.
\textit{Astronomical and Astrophysical Transactions} 2:201--208

\bibitem[{Gosnell et~al.(2019)Gosnell, Leiner, Mathieu, Geller, Knigge et~al.}]{Gosnell+2019}
Gosnell NM, Leiner EM, Mathieu RD, Geller AM, Knigge C, et~al. 2019.
\textit{The Astrophysical Journal} 885:45

\bibitem[{Gosnell et~al.(2014)Gosnell, Mathieu, Geller, Sills, Leigh \& Knigge}]{Gosnell+2014}
Gosnell NM, Mathieu RD, Geller AM, Sills A, Leigh N, Knigge C. 2014.
\textit{The Astrophysical Journal} 783:L8

\bibitem[{Gosnell et~al.(2015)Gosnell, Mathieu, Geller, Sills, Leigh \& Knigge}]{Gosnell+2015}
Gosnell NM, Mathieu RD, Geller AM, Sills A, Leigh N, Knigge C. 2015.
\textit{The Astrophysical Journal} 814(2):163

\bibitem[{G{\"o}ttgens et~al.(2019)G{\"o}ttgens, Husser, Kamann, Dreizler, Giesers et~al.}]{Gottgens+2019}
G{\"o}ttgens F, Husser TO, Kamann S, Dreizler S, Giesers B, et~al. 2019.
\textit{Astronomy and Astrophysics} 631:A118

\bibitem[{Gray et~al.(2011)Gray, McGahee, Griffin \& Corbally}]{Gray+2011}
Gray RO, McGahee CE, Griffin REM, Corbally CJ. 2011.
\textit{The Astronomical Journal} 141:160

\bibitem[{Green et~al.(2001)Green, Liebert \& Saffer}]{Green+2001}
Green EM, Liebert J, Saffer RA. 2001.
\textit{On {{The Origin Of Subdwarf B Stars}} and {{Related Metal-Rich Binaries}}}. In \textit{12th {{European Workshop}} on {{White Dwarfs}}}, vol. 226 of \textit{{{ASP Conference Proceedings}}}. San Francisco: Astronomical Society of the Pacific,  192

\bibitem[{Green(2013)}]{Green2013}
Green P. 2013.
\textit{The Astrophysical Journal} 765:12

\bibitem[{Green et~al.(2019)Green, Montez, Mazzoni, Filippazzo, Anderson et~al.}]{Green+2019}
Green PJ, Montez R, Mazzoni F, Filippazzo J, Anderson SF, et~al. 2019.
\textit{The Astrophysical Journal} 881:49

\bibitem[{Hampel et~al.(2019)Hampel, Karakas, Stancliffe, Meyer \& Lugaro}]{Hampel+2019}
Hampel M, Karakas AI, Stancliffe RJ, Meyer BS, Lugaro M. 2019.
\textit{The Astrophysical Journal} 887(1):11

\bibitem[{Hampel et~al.(2016)Hampel, Stancliffe, Lugaro \& Meyer}]{Hampel+2016}
Hampel M, Stancliffe RJ, Lugaro M, Meyer BS. 2016.
\textit{The Astrophysical Journal} 831(2):171

\bibitem[{Han et~al.(2002)Han, Podsiadlowski, Maxted, Marsh \& Ivanova}]{Han+2002}
Han Z, Podsiadlowski {\relax Ph}, Maxted PFL, Marsh TR, Ivanova N. 2002.
\textit{Monthly Notices of the Royal Astronomical Society} 336:449--466

\bibitem[{Handberg et~al.(2017)Handberg, Brogaard, Miglio, Bossini, Elsworth et~al.}]{Handberg+2017}
Handberg R, Brogaard K, Miglio A, Bossini D, Elsworth Y, et~al. 2017.
\textit{Monthly Notices of the Royal Astronomical Society} 472:979--997

\bibitem[{Hansen et~al.(2016)Hansen, Andersen, Nordstr{\"o}m, Beers, Placco et~al.}]{Hansen+2016}
Hansen TT, Andersen J, Nordstr{\"o}m B, Beers TC, Placco VM, et~al. 2016.
\textit{Astronomy and Astrophysics} 586:A160

\bibitem[{Harris et~al.(2018)Harris, Dahn, Subasavage, Munn, Canzian et~al.}]{Harris+2018}
Harris HC, Dahn CC, Subasavage JP, Munn JA, Canzian BJ, et~al. 2018.
\textit{The Astronomical Journal} 155:252

\bibitem[{Heber(2016)}]{Heber2016}
Heber U. 2016.
\textit{Publications of the Astronomical Society of the Pacific} 128:082001

\bibitem[{Jadhav et~al.(2019)Jadhav, Sindhu \& Subramaniam}]{Jadhav+2019}
Jadhav VV, Sindhu N, Subramaniam A. 2019.
\textit{The Astrophysical Journal} 886:13

\bibitem[{Jadhav \& Subramaniam(2021)}]{Jadhav+Subramaniam2021}
Jadhav VV, Subramaniam A. 2021.
\textit{Monthly Notices of the Royal Astronomical Society} 507(2):1699--1709

\bibitem[{Jadhav et~al.(2023)Jadhav, Subramaniam \& Sagar}]{Jadhav+2023}
Jadhav VV, Subramaniam A, Sagar R. 2023.
\textit{Astronomy and Astrophysics} 676:A47

\bibitem[{Jiang(2022)}]{Jiang2022}
Jiang D. 2022.
\textit{The Astrophysical Journal} 940:97

\bibitem[{Jiang et~al.(2017)Jiang, Chen, Li \& Han}]{Jiang+2017}
Jiang D, Chen X, Li L, Han Z. 2017.
\textit{The Astrophysical Journal} 849:100

\bibitem[{Jofr{\'e} et~al.(2023)Jofr{\'e}, Jorissen, {Aguilera-G{\'o}mez}, Van~Eck, Tayar et~al.}]{Jofre+2023}
Jofr{\'e} P, Jorissen A, {Aguilera-G{\'o}mez} C, Van~Eck S, Tayar J, et~al. 2023.
\textit{Astronomy and Astrophysics} 671:A21

\bibitem[{Jofr{\'e} et~al.(2016)Jofr{\'e}, Jorissen, Van~Eck, Izzard, Masseron et~al.}]{Jofre+2016}
Jofr{\'e} P, Jorissen A, Van~Eck S, Izzard RG, Masseron T, et~al. 2016.
\textit{Astronomy and Astrophysics} 595:A60

\bibitem[{Jorissen et~al.(2019)Jorissen, Boffin, Karinkuzhi, Van~Eck, Escorza et~al.}]{Jorissen+2019}
Jorissen A, Boffin HMJ, Karinkuzhi D, Van~Eck S, Escorza A, et~al. 2019.
\textit{Astronomy and Astrophysics} 626:A127

\bibitem[{Jorissen et~al.(1998)Jorissen, Van~Eck, Mayor \& Udry}]{Jorissen+1998}
Jorissen A, Van~Eck S, Mayor M, Udry S. 1998.
\textit{Astronomy and Astrophysics} 332:877--903

\bibitem[{Jorissen et~al.(2016)Jorissen, Van~Eck, Van~Winckel, Merle, Boffin et~al.}]{Jorissen+2016}
Jorissen A, Van~Eck S, Van~Winckel H, Merle T, Boffin HMJ, et~al. 2016.
\textit{Astronomy and Astrophysics} 586:A158

\bibitem[{Kaluzny(2005)}]{Kaluzny2005}
Kaluzny J. 2005.
\textit{Eclipsing Binaries in Globular Clusters as Age and Distance Indicators}. In \textit{{{AIP Conference Proceedings}}}, vol. 752. Torun (Poland): AIP

\bibitem[{Kaluzny et~al.(2007{\natexlab{a}})Kaluzny, Rucinski, Thompson, Pych \& Krzeminski}]{Kaluzny+2007}
Kaluzny J, Rucinski SM, Thompson IB, Pych W, Krzeminski W. 2007{\natexlab{a}}.
\textit{The Astronomical Journal} 133:2457--2463

\bibitem[{Kaluzny \& Shara(1987)}]{Kaluzny+Shara1987}
Kaluzny J, Shara MM. 1987.
\textit{The Astrophysical Journal} 314:585

\bibitem[{Kaluzny et~al.(2010)Kaluzny, Thompson, Krzeminski \& Zloczewski}]{Kaluzny+2010}
Kaluzny J, Thompson IB, Krzeminski W, Zloczewski K. 2010.
\textit{Acta Astronomica} 60:245--260

\bibitem[{Kaluzny et~al.(2015)Kaluzny, Thompson, Narloch, Pych \& Rozyczka}]{Kaluzny+2015}
Kaluzny J, Thompson IB, Narloch W, Pych W, Rozyczka M. 2015.
\textit{Acta Astronomica} 65:267--282

\bibitem[{Kaluzny et~al.(2013)Kaluzny, Thompson, Rozyczka \& Krzeminski}]{Kaluzny+2013}
Kaluzny J, Thompson IB, Rozyczka M, Krzeminski W. 2013.
\textit{Acta Astronomica} 63:181--201

\bibitem[{Kaluzny et~al.(2007{\natexlab{b}})Kaluzny, Thompson, Rucinski, Pych, Stachowski et~al.}]{Kaluzny+2007a}
Kaluzny J, Thompson IB, Rucinski SM, Pych W, Stachowski G, et~al. 2007{\natexlab{b}}.
\textit{The Astronomical Journal} 134:541--546

\bibitem[{Kamath et~al.(2014)Kamath, Wood \& Van~Winckel}]{Kamath+2014}
Kamath D, Wood PR, Van~Winckel H. 2014.
\textit{Monthly Notices of the Royal Astronomical Society} 439:2211--2270

\bibitem[{Karinkuzhi et~al.(2021)Karinkuzhi, Van~Eck, Goriely, Siess, Jorissen et~al.}]{Karinkuzhi+2021}
Karinkuzhi D, Van~Eck S, Goriely S, Siess L, Jorissen A, et~al. 2021.
\textit{Astronomy and Astrophysics} 645:A61

\bibitem[{Katime~Santrich et~al.(2013)Katime~Santrich, Pereira \& {de Castro}}]{KatimeSantrich+2013}
Katime~Santrich OJ, Pereira CB, {de Castro} DB. 2013.
\textit{The Astronomical Journal} 146:39

\bibitem[{Kluska et~al.(2022)Kluska, Van~Winckel, Copp{\'e}e, Oomen, Dsilva et~al.}]{KLuska+2022}
Kluska J, Van~Winckel H, Copp{\'e}e Q, Oomen GM, Dsilva K, et~al. 2022.
\textit{Astronomy and Astrophysics} 658:A36

\bibitem[{Knigge(2015)}]{knigge2015}
Knigge C. 2015.
\textit{Blue {{Stragglers}} in {{Globular Clusters}}: {{Observations}}, {{Statistics}} and {{Physics}}}. In \textit{Ecology of {{Blue Straggler Stars}}}, eds. H~Boffin, G~Carraro, G~Beccari, vol. 413 of \textit{Astrophysics and {{Space Science Library}}}. Berlin, Heidelberg: Springer,  295

\bibitem[{Knigge et~al.(2009)Knigge, Leigh \& Sills}]{Knigge+2009}
Knigge C, Leigh N, Sills A. 2009.
\textit{Nature} 457:288--290

\bibitem[{Kravtsov \& Calder{\'o}n(2021)}]{Kravtsov+Caldern2021}
Kravtsov V, Calder{\'o}n FA. 2021.
\textit{The Astronomical Journal} 161:7

\bibitem[{Landsman et~al.(1997)Landsman, Aparicio, Bergeron, Di~Stefano \& Stecher}]{Landsman+1997}
Landsman W, Aparicio J, Bergeron P, Di~Stefano R, Stecher TP. 1997.
\textit{The Astrophysical Journal} 481:L93--L96

\bibitem[{Landsman et~al.(1998)Landsman, Bohlin, Neff, O'Connell, Roberts et~al.}]{Landsman+1998}
Landsman W, Bohlin RC, Neff SG, O'Connell RW, Roberts MS, et~al. 1998.
\textit{The Astronomical Journal} 116:789--800

\bibitem[{Latham(2007)}]{Latham2007}
Latham DW. 2007.
\textit{Highlights of Astronomy} 14:444--445

\bibitem[{Latham \& Milone(1996)}]{Latham+Milone1996}
Latham DW, Milone AAE. 1996.
\textit{Spectroscopic {{Binaries Among}} the {{M67 Blue Stragglers}}}. In \textit{The Origins, Evolution, and Destinies of Binary Stars in Clusters}, vol.~90 of \textit{{{ASP Conference Series}}}. San Francisco: Astronomical Society of the Pacific,  385

\bibitem[{Latour et~al.(2018)Latour, Randall, Calamida, Geier \& Moehler}]{Latour+2018}
Latour M, Randall SK, Calamida A, Geier S, Moehler S. 2018.
\textit{Astronomy and Astrophysics} 618:A15

\bibitem[{Lee et~al.(2013)Lee, Beers, Masseron, Plez, Rockosi et~al.}]{Lee+2013}
Lee YS, Beers TC, Masseron T, Plez B, Rockosi CM, et~al. 2013.
\textit{The Astronomical Journal} 146:132

\bibitem[{Leigh et~al.(2007)Leigh, Sills \& Knigge}]{Leigh+2007}
Leigh N, Sills A, Knigge C. 2007.
\textit{The Astrophysical Journal} 661:210--221

\bibitem[{Leiner et~al.(2018)Leiner, Mathieu, Gosnell \& Sills}]{Leiner+2018}
Leiner E, Mathieu RD, Gosnell NM, Sills A. 2018.
\textit{The Astrophysical Journal} 869:L29

\bibitem[{Leiner et~al.(2016)Leiner, Mathieu, Stello, Vanderburg \& Sandquist}]{Leiner+2016}
Leiner E, Mathieu RD, Stello D, Vanderburg A, Sandquist E. 2016.
\textit{The Astrophysical Journal} 832:L13

\bibitem[{Leiner et~al.(2019)Leiner, Mathieu, Vanderburg, Gosnell \& Smith}]{Leiner+2019}
Leiner E, Mathieu RD, Vanderburg A, Gosnell NM, Smith JC. 2019.
\textit{The Astrophysical Journal} 881(1):47

\bibitem[{Leiner \& Geller(2021)}]{Leiner+Geller2021}
Leiner EM, Geller A. 2021.
\textit{The Astrophysical Journal} 908(2):229

\bibitem[{Leiner et~al.(2022)Leiner, Geller, {Gully-Santiago}, Gosnell \& Tofflemire}]{Leiner+2022}
Leiner EM, Geller AM, {Gully-Santiago} MA, Gosnell NM, Tofflemire BM. 2022.
\textit{The Astrophysical Journal} 927(2):222

\bibitem[{Leiner et~al.(2025)Leiner, Gosnell, Geller, Sun, Mathieu \& Sills}]{Leiner+2025}
Leiner EM, Gosnell NM, Geller AM, Sun M, Mathieu RD, Sills A. 2025.
\textit{The Astrophysical Journal} 979(1):L1

\bibitem[{Leonard(1989)}]{Leonard1989}
Leonard PJT. 1989.
\textit{The Astronomical Journal} 98:217

\bibitem[{Li et~al.(2022)Li, Aoki, Matsuno, Xing, Suda et~al.}]{Li+2022}
Li H, Aoki W, Matsuno T, Xing Q, Suda T, et~al. 2022.
\textit{The Astrophysical Journal} 931:147

\bibitem[{Li(2018)}]{Li2018}
Li K. 2018.
\textit{New Astronomy} 59:60--64

\bibitem[{Li et~al.(2017)Li, Hu, Chen \& Guo}]{Li+2017}
Li K, Hu S, Chen X, Guo D. 2017.
\textit{Publications of the Astronomical Society of Japan} 69:79

\bibitem[{Li et~al.(2018)Li, Luo, Du, Zuo, Wang et~al.}]{Li+2018}
Li YB, Luo AL, Du CD, Zuo F, Wang MX, et~al. 2018.
\textit{The Astrophysical Journal Supplement Series} 234:31

\bibitem[{Libralato et~al.(2019)Libralato, Bellini, Piotto, Nardiello, {van der Marel} et~al.}]{Libralato+2019}
Libralato M, Bellini A, Piotto G, Nardiello D, {van der Marel} RP, et~al. 2019.
\textit{The Astrophysical Journal} 873:109

\bibitem[{Libralato et~al.(2018)Libralato, Bellini, {van der Marel}, Anderson, Watkins et~al.}]{Libralato+2018}
Libralato M, Bellini A, {van der Marel} RP, Anderson J, Watkins LL, et~al. 2018.
\textit{The Astrophysical Journal} 861:99

\bibitem[{Linck et~al.(2024)Linck, Mathieu \& Latham}]{Linck+2024a}
Linck E, Mathieu RD, Latham DW. 2024.
\textit{The Astronomical Journal} 168:205

\bibitem[{Lovisi et~al.(2013{\natexlab{a}})Lovisi, Mucciarelli, Dalessandro, Ferraro \& Lanzoni}]{Lovisi+2013}
Lovisi L, Mucciarelli A, Dalessandro E, Ferraro FR, Lanzoni B. 2013{\natexlab{a}}.
\textit{The Astrophysical Journal} 778:64

\bibitem[{Lovisi et~al.(2010)Lovisi, Mucciarelli, Ferraro, Lucatello, Lanzoni et~al.}]{Lovisi+2010a}
Lovisi L, Mucciarelli A, Ferraro FR, Lucatello S, Lanzoni B, et~al. 2010.
\textit{The Astrophysical Journal} 719:L121--L125

\bibitem[{Lovisi et~al.(2013{\natexlab{b}})Lovisi, Mucciarelli, Lanzoni, Ferraro \& Dalessandro}]{Lovisi+2013b}
Lovisi L, Mucciarelli A, Lanzoni B, Ferraro FR, Dalessandro E. 2013{\natexlab{b}}.
\textit{Memorie della Societa Astronomica Italiana} 84:232

\bibitem[{Lovisi et~al.(2013{\natexlab{c}})Lovisi, Mucciarelli, Lanzoni, Ferraro, Dalessandro \& Monaco}]{Lovisi+2013a}
Lovisi L, Mucciarelli A, Lanzoni B, Ferraro FR, Dalessandro E, Monaco L. 2013{\natexlab{c}}.
\textit{The Astrophysical Journal} 772:148

\bibitem[{Lovisi et~al.(2012)Lovisi, Mucciarelli, Lanzoni, Ferraro, Gratton et~al.}]{Lovisi+2012}
Lovisi L, Mucciarelli A, Lanzoni B, Ferraro FR, Gratton R, et~al. 2012.
\textit{The Astrophysical Journal} 754:91

\bibitem[{Lu(1991)}]{Lu1991}
Lu PK. 1991.
\textit{The Astronomical Journal} 101:2229

\bibitem[{Lucatello et~al.(2006)Lucatello, Beers, Christlieb, Barklem, Rossi et~al.}]{Lucatello+2006}
Lucatello S, Beers TC, Christlieb N, Barklem PS, Rossi S, et~al. 2006.
\textit{The Astrophysical Journal} 652:L37--L40

\bibitem[{Lucatello et~al.(2003)Lucatello, Gratton, Cohen, Beers, Christlieb et~al.}]{Lucatello+2003}
Lucatello S, Gratton R, Cohen JG, Beers TC, Christlieb N, et~al. 2003.
\textit{The Astronomical Journal} 125:875--893

\bibitem[{Lucatello \& Gratton(2003)}]{Lucatello+Gratton2003}
Lucatello S, Gratton RG. 2003.
\textit{Astronomy and Astrophysics} 406:691--702

\bibitem[{Lucatello et~al.(2005)Lucatello, Tsangarides, Beers, Carretta, Gratton \& Ryan}]{Lucatello+2005}
Lucatello S, Tsangarides S, Beers TC, Carretta E, Gratton RG, Ryan SG. 2005.
\textit{The Astrophysical Journal} 625(2):825--832

\bibitem[{MacConnell et~al.(1972)MacConnell, Frye \& Upgren}]{MacConnell+1972}
MacConnell DJ, Frye RL, Upgren AR. 1972.
\textit{The Astronomical Journal} 77:384--391

\bibitem[{Mansfield et~al.(2022)Mansfield, Dieball, Kroupa, Knigge, Zurek et~al.}]{Mansfield+2022}
Mansfield S, Dieball A, Kroupa P, Knigge C, Zurek DR, et~al. 2022.
\textit{Monthly Notices of the Royal Astronomical Society} 513:3022--3034

\bibitem[{Margon et~al.(2018)Margon, Kupfer, Burdge, Prince, Kulkarni \& Shupe}]{Margon+2018}
Margon B, Kupfer T, Burdge K, Prince TA, Kulkarni SR, Shupe DL. 2018.
\textit{The Astrophysical Journal} 856:L2

\bibitem[{Marigo et~al.(2008)Marigo, Girardi, Bressan, Groenewegen, Silva \& Granato}]{Marigo+2008}
Marigo P, Girardi L, Bressan A, Groenewegen MAT, Silva L, Granato GL. 2008.
\textit{Astronomy and Astrophysics} 482:883--905

\bibitem[{Martig et~al.(2015)Martig, Rix, Silva~Aguirre, Hekker, Mosser et~al.}]{Martig+2015}
Martig M, Rix HW, Silva~Aguirre V, Hekker S, Mosser B, et~al. 2015.
\textit{Monthly Notices of the Royal Astronomical Society} 451:2230--2243

\bibitem[{Masseron et~al.(2010)Masseron, Johnson, Plez, {van Eck}, Primas et~al.}]{Masseron+2010}
Masseron T, Johnson JA, Plez B, {van Eck} S, Primas F, et~al. 2010.
\textit{Astronomy and Astrophysics} 509:A93

\bibitem[{Mathieu(2000)}]{Mathieu2000}
Mathieu RD. 2000.
\textit{Stellar Clusters and Associations: Convection, Rotation, and Dynamos. Proceedings from ASP Conference} 198:517

\bibitem[{Mathieu \& Geller(2009)}]{Mathieu+Geller2009}
Mathieu RD, Geller AM. 2009.
\textit{Nature} 462:1032--1035

\bibitem[{Mathieu \& Geller(2015)}]{Mathieu+Geller2015}
Mathieu RD, Geller AM. 2015.
\textit{The {{Blue Stragglers}} of the {{Old Open Cluster NGC}} 188}. In \textit{Ecology of {{Blue Straggler Stars}}}, eds. {Boffin, H.}, {Carraro, G.}, {Beccari, G.}, vol. 413 of \textit{Astrophysics and {{Space Science Library}}}. Berlin, Heidelberg: Springer,  29--63

\bibitem[{Mathieu \& Latham(1986)}]{Mathieu+Latham1986}
Mathieu RD, Latham DW. 1986.
\textit{The Astronomical Journal} 92:1364--1371

\bibitem[{Mathieu et~al.(2003)Mathieu, Van Den~Berg, Torres, Latham, Verbunt \& Stassun}]{Mathieu+2003}
Mathieu RD, Van Den~Berg M, Torres G, Latham D, Verbunt F, Stassun K. 2003.
\textit{The Astronomical Journal} 125(1):246--259

\bibitem[{Mathys(1991)}]{Mathys1991}
Mathys G. 1991.
\textit{Astronomy and Astrophysics} 245:467

\bibitem[{McClure(1983)}]{McClure1983}
McClure RD. 1983.
\textit{The Astrophysical Journal} 268:264

\bibitem[{McClure et~al.(1974)McClure, Forrester \& Gibson}]{McClure+1974}
McClure RD, Forrester WT, Gibson J. 1974.
\textit{The Astrophysical Journal} 189:409

\bibitem[{McClure \& Woodsworth(1990)}]{McClure+Woodsworth1990}
McClure RD, Woodsworth AW. 1990.
\textit{The Astrophysical Journal} 352:709

\bibitem[{McCrea(1964)}]{McCrea1964}
McCrea WH. 1964.
\textit{Monthly Notices of the Royal Astronomical Society} 128(2):147--155

\bibitem[{Meibom \& Mathieu(2005)}]{Meibom+Mathieu2005}
Meibom S, Mathieu RD. 2005.
\textit{The Astrophysical Journal} 620:970--983

\bibitem[{Miglio et~al.(2012)Miglio, Brogaard, Stello, Chaplin, D'Antona et~al.}]{Miglio+2012}
Miglio A, Brogaard K, Stello D, Chaplin WJ, D'Antona F, et~al. 2012.
\textit{Monthly Notices of the Royal Astronomical Society} 419:2077--2088

\bibitem[{Miko{\l}ajewska(2003)}]{Mikoajewska2003}
Miko{\l}ajewska J. 2003.
\textit{Orbital and {{Stellar Parameters}} of {{Symbiotic Stars}}}. In \textit{Symbiotic {{Stars Probing Stellar Evolution}}}, vol. 303 of \textit{{{ASP Conference Series}}}. San Francisco: Astronomical Society of the Pacific, ~9

\bibitem[{Miko{\l}ajewska(2012)}]{Mikoajewska2012}
Miko{\l}ajewska J. 2012.
\textit{Baltic Astronomy} 21:5--12

\bibitem[{Milliman et~al.(2016)Milliman, Leiner, Mathieu, Tofflemire \& Platais}]{Milliman+2016}
Milliman KE, Leiner E, Mathieu RD, Tofflemire BM, Platais I. 2016.
\textit{The Astronomical Journal} 151(6):152

\bibitem[{Milliman et~al.(2014)Milliman, Mathieu, Geller, Gosnell, Meibom \& Platais}]{Milliman+2014}
Milliman KE, Mathieu RD, Geller AM, Gosnell NM, Meibom S, Platais I. 2014.
\textit{The Astronomical Journal} 148:38

\bibitem[{Milliman et~al.(2015)Milliman, Mathieu \& Schuler}]{Milliman+2015}
Milliman KE, Mathieu RD, Schuler SC. 2015.
\textit{The Astronomical Journal} 150:84

\bibitem[{Milone \& Latham(1992)}]{Milone+Latham1992}
Milone AAE, Latham DW. 1992.
\textit{The {{Blue Straggler F}}:190 - a {{Case}} for {{Mass Transfer}}}. In \textit{Evolutionary {{Processes}} in {{Interacting Binary Stars}}}, eds. Y~Kondo, R~Sistero, R~Polidan, vol. 151 of \textit{{{IAU Symposium}}}. Dordrecht: Kluwer,  475

\bibitem[{Milone \& Latham(1994)}]{Milone+Latham1994}
Milone AAE, Latham DW. 1994.
\textit{The Astronomical Journal} 108:1828

\bibitem[{Milone et~al.(1992)Milone, Latham, Mathieu, Morse \& Davis}]{Milone+1992}
Milone AAE, Latham DW, Mathieu RD, Morse JA, Davis RJ. 1992.
\textit{Can {{Evolution}} in {{Close Binaries Account}} for the {{Blue Stragglers}} in {{M67}}}. In \textit{Evolutionary {{Processes}} in {{Interacting Binary Stars}}}, vol. 151 of \textit{{{IAU Symposium}}}. Dordrecht: Kluwer,  473

\bibitem[{Milone et~al.(2012)Milone, Piotto, Bedin, Aparicio, Anderson et~al.}]{Milone+2012}
Milone AP, Piotto G, Bedin LR, Aparicio A, Anderson J, et~al. 2012.
\textit{Astronomy and Astrophysics} 540:A16

\bibitem[{Miszalski et~al.(2013)Miszalski, Boffin \& Corradi}]{Miszalski+2013}
Miszalski B, Boffin HMJ, Corradi RLM. 2013.
\textit{Monthly Notices of the Royal Astronomical Society} 428:L39--L43

\bibitem[{Moni~Bidin et~al.(2009)Moni~Bidin, Moehler, Piotto, Momany \& {Recio-Blanco}}]{MoniBidin+2009}
Moni~Bidin C, Moehler S, Piotto G, Momany Y, {Recio-Blanco} A. 2009.
\textit{Astronomy and Astrophysics} 498(3):737--751

\bibitem[{Moni~Bidin et~al.(2015)Moni~Bidin, Momany, Montalto, Catelan, Villanova et~al.}]{MoniBidin+2015}
Moni~Bidin C, Momany Y, Montalto M, Catelan M, Villanova S, et~al. 2015.
\textit{The Astrophysical Journal} 812:L31

\bibitem[{Moni~Bidin et~al.(2011)Moni~Bidin, Villanova, Piotto \& Momany}]{MoniBidin+2011}
Moni~Bidin C, Villanova S, Piotto G, Momany Y. 2011.
\textit{Astronomy and Astrophysics} 528:A127

\bibitem[{Moretti et~al.(2008)Moretti, {de Angeli} \& Piotto}]{Moretti+2008}
Moretti A, {de Angeli} F, Piotto G. 2008.
\textit{Astronomy and Astrophysics} 483:183--197

\bibitem[{Mucciarelli et~al.(2014)Mucciarelli, Lovisi, Ferraro, Dalessandro, Lanzoni \& Monaco}]{Mucciarelli+2014}
Mucciarelli A, Lovisi L, Ferraro FR, Dalessandro E, Lanzoni B, Monaco L. 2014.
\textit{The Astrophysical Journal} 797:43

\bibitem[{Murphy et~al.(2018)Murphy, Moe, Kurtz, Bedding, Shibahashi \& Boffin}]{Murphy+2018}
Murphy SJ, Moe M, Kurtz DW, Bedding TR, Shibahashi H, Boffin HMJ. 2018.
\textit{Monthly Notices of the Royal Astronomical Society} 474:4322--4346

\bibitem[{Nemec et~al.(1995)Nemec, Mateo, Burke \& Olszewski}]{Nemec+1995}
Nemec JM, Mateo M, Burke M, Olszewski EW. 1995.
\textit{The Astronomical Journal} 110:1186

\bibitem[{Nine et~al.(2023)Nine, Mathieu, Gosnell \& Leiner}]{Nine+2023}
Nine AC, Mathieu RD, Gosnell NM, Leiner EM. 2023.
\textit{The Astrophysical Journal} 944(2):145

\bibitem[{Nine et~al.(2024)Nine, Mathieu, Schuler \& Milliman}]{Nine+2024}
Nine AC, Mathieu RD, Schuler SC, Milliman KE. 2024.
\textit{The Astrophysical Journal} 970(2):187

\bibitem[{Nine et~al.(2020)Nine, Milliman, Mathieu, Geller, Leiner et~al.}]{Nine+2020}
Nine AC, Milliman KE, Mathieu RD, Geller AM, Leiner EM, et~al. 2020.
\textit{The Astronomical Journal} 160(4):169

\bibitem[{North(2014)}]{North2014}
North P. 2014.
\textit{Multiplicity of {{A-type}} and Related Stars}. In \textit{Putting {{A Stars}} into {{Context}}: {{Evolution}}, {{Environment}}, and {{Related Stars}}}, eds. G~Mathys, E~Griffin, O~Kochukhov, P~Monier, G~Wahlgren. Moscow: Publishing house "Pero"

\bibitem[{North et~al.(1994)North, Berthet \& Lanz}]{North+1994}
North P, Berthet S, Lanz T. 1994.
\textit{Astronomy and Astrophysics} 281:775--796

\bibitem[{North \& Duquennoy(1991)}]{North+Duquennoy1991}
North P, Duquennoy A. 1991.
\textit{Astronomy and Astrophysics} 244:335

\bibitem[{North et~al.(2020)North, Jorissen, Escorza, Miszalski \& Mikolajewska}]{North+2020}
North PL, Jorissen A, Escorza A, Miszalski B, Mikolajewska J. 2020.
\textit{The Observatory} 140:11--20

\bibitem[{Oomen et~al.(2018)Oomen, Van~Winckel, Pols, Nelemans, Escorza et~al.}]{Oomen+2018}
Oomen GM, Van~Winckel H, Pols O, Nelemans G, Escorza A, et~al. 2018.
\textit{Astronomy and Astrophysics} 620:A85

\bibitem[{Pablo et~al.(2011)Pablo, Kawaler \& Green}]{Pablo+2011}
Pablo H, Kawaler SD, Green EM. 2011.
\textit{The Astrophysical Journal} 740:L47

\bibitem[{Pal et~al.(2024)Pal, Subramaniam, Reddy \& Jadhav}]{Pal+2024}
Pal H, Subramaniam A, Reddy ABS, Jadhav VV. 2024.
\textit{The Astrophysical Journal} 970:L39

\bibitem[{Pandey et~al.(2021)Pandey, Subramaniam \& Jadhav}]{Pandey+2021}
Pandey S, Subramaniam A, Jadhav VV. 2021.
\textit{Monthly Notices of the Royal Astronomical Society} 507(2):2373--2382

\bibitem[{Panthi et~al.(2023)Panthi, Subramaniam, Vaidya, Jadhav, Rani et~al.}]{Panthi+2023}
Panthi A, Subramaniam A, Vaidya K, Jadhav V, Rani S, et~al. 2023.
\textit{Monthly Notices of the Royal Astronomical Society} 525:1311--1328

\bibitem[{Panthi \& Vaidya(2024)}]{Panthi+Vaidya2024}
Panthi A, Vaidya K. 2024.
\textit{Monthly Notices of the Royal Astronomical Society} 527:10335--10347

\bibitem[{Panthi et~al.(2022)Panthi, Vaidya, Jadhav, Rao, Subramaniam et~al.}]{Panthi+2022}
Panthi A, Vaidya K, Jadhav V, Rao KK, Subramaniam A, et~al. 2022.
\textit{Monthly Notices of the Royal Astronomical Society} 516(4):5318--5330

\bibitem[{Panthi et~al.(2024)Panthi, Vaidya, Vernekar, Subramaniam, Jadhav \& Agarwal}]{Panthi+2024}
Panthi A, Vaidya K, Vernekar N, Subramaniam A, Jadhav V, Agarwal M. 2024.
\textit{Monthly Notices of the Royal Astronomical Society} 527:8325--8336

\bibitem[{Parada et~al.(2016{\natexlab{a}})Parada, Richer, Heyl, Kalirai \& Goldsbury}]{Parada+2016}
Parada J, Richer H, Heyl J, Kalirai J, Goldsbury R. 2016{\natexlab{a}}.
\textit{The Astrophysical Journal} 826:88

\bibitem[{Parada et~al.(2016{\natexlab{b}})Parada, Richer, Heyl, Kalirai \& Goldsbury}]{Parada+2016a}
Parada J, Richer H, Heyl J, Kalirai J, Goldsbury R. 2016{\natexlab{b}}.
\textit{The Astrophysical Journal} 830:139

\bibitem[{Paresce et~al.(1991)Paresce, Shara, Meylan, Baxter, Greenfield et~al.}]{Paresce+1991}
Paresce F, Shara M, Meylan G, Baxter D, Greenfield P, et~al. 1991.
\textit{Nature} 352:297--301

\bibitem[{Paxton et~al.(2011)Paxton, Bildsten, Dotter, Herwig, Lesaffre \& Timmes}]{paxton+2011}
Paxton B, Bildsten L, Dotter A, Herwig F, Lesaffre P, Timmes F. 2011.
\textit{The Astrophysical Journal Supplement Series} 192:3

\bibitem[{Piotto et~al.(2004)Piotto, De~Angeli, King, Djorgovski, Bono et~al.}]{Piotto+2004}
Piotto G, De~Angeli F, King IR, Djorgovski SG, Bono G, et~al. 2004.
\textit{The Astrophysical Journal} 604:L109--L112

\bibitem[{Piotto et~al.(2002)Piotto, King, Djorgovski, Sosin, Zoccali et~al.}]{Piotto+2002}
Piotto G, King IR, Djorgovski SG, Sosin C, Zoccali M, et~al. 2002.
\textit{Astronomy and Astrophysics} 391:945--965

\bibitem[{Piotto et~al.(2015)Piotto, Milone, Bedin, Anderson, King et~al.}]{Piotto+2015}
Piotto G, Milone AP, Bedin LR, Anderson J, King IR, et~al. 2015.
\textit{The Astronomical Journal} 149:91

\bibitem[{Placco et~al.(2014)Placco, Frebel, Beers \& Stancliffe}]{Placco+2014}
Placco VM, Frebel A, Beers TC, Stancliffe RJ. 2014.
\textit{The Astrophysical Journal} 797:21

\bibitem[{Plez \& Cohen(2005)}]{Plez+Cohen2005}
Plez B, Cohen JG. 2005.
\textit{Astronomy and Astrophysics} 434:1117--1124

\bibitem[{Portegies~Zwart(2019)}]{PortegiesZwart2019}
Portegies~Zwart S. 2019.
\textit{Astronomy and Astrophysics} 621:L10

\bibitem[{Portegies~Zwart \& Leigh(2019)}]{PortegiesZwart+Leigh2019}
Portegies~Zwart S, Leigh NWC. 2019.
\textit{The Astrophysical Journal} 876(2):L33

\bibitem[{Preston et~al.(1994)Preston, Beers \& Shectman}]{Preston+1994}
Preston GW, Beers TC, Shectman SA. 1994.
\textit{The Astronomical Journal} 108:538

\bibitem[{Preston \& Sneden(2000)}]{Preston+Sneden2000}
Preston GW, Sneden C. 2000.
\textit{The Astronomical Journal} 120:1014--1055

\bibitem[{Raghavan et~al.(2010)Raghavan, McAlister, Henry, Latham, Marcy et~al.}]{Raghavan+2010}
Raghavan D, McAlister HA, Henry TJ, Latham DW, Marcy GW, et~al. 2010.
\textit{The Astrophysical Journal Supplement Series} 190:1--42

\bibitem[{Rain et~al.(2021{\natexlab{a}})Rain, Ahumada \& Carraro}]{Rain+2021}
Rain MJ, Ahumada JA, Carraro G. 2021{\natexlab{a}}.
\textit{Astronomy and Astrophysics} 650:A67

\bibitem[{Rain et~al.(2021{\natexlab{b}})Rain, Carraro, Ahumada, Villanova, Boffin \& Monaco}]{Rain+2021a}
Rain MJ, Carraro G, Ahumada JA, Villanova S, Boffin H, Monaco L. 2021{\natexlab{b}}.
\textit{The Astronomical Journal} 161(1):37

\bibitem[{Rain et~al.(2020)Rain, Carraro, Ahumada, Villanova, Boffin et~al.}]{Rain+2020}
Rain MJ, Carraro G, Ahumada JA, Villanova S, Boffin H, et~al. 2020.
\textit{The Astronomical Journal} 159:59

\bibitem[{Randall et~al.(2016)Randall, Calamida, Fontaine, Monelli, Bono et~al.}]{Randall+2016}
Randall SK, Calamida A, Fontaine G, Monelli M, Bono G, et~al. 2016.
\textit{Astronomy and Astrophysics} 589:A1

\bibitem[{Rani et~al.(2021)Rani, Subramaniam, Pandey, Sahu, Mondal \& Pandey}]{Rani+2021}
Rani S, Subramaniam A, Pandey S, Sahu S, Mondal C, Pandey G. 2021.
\textit{Journal of Astrophysics and Astronomy} 42:47

\bibitem[{Rao et~al.(2022)Rao, Vaidya, Agarwal, Panthi, Jadhav \& Subramaniam}]{Rao+2022}
Rao KK, Vaidya K, Agarwal M, Panthi A, Jadhav V, Subramaniam A. 2022.
\textit{Monthly Notices of the Royal Astronomical Society} 516:2444--2454

\bibitem[{Rappaport et~al.(1995)Rappaport, Podsiadlowski, Joss, Di~Stefano \& Han}]{Rappaport+1995}
Rappaport S, Podsiadlowski {\relax Ph}, Joss PC, Di~Stefano R, Han Z. 1995.
\textit{Monthly Notices of the Royal Astronomical Society} 273:731--741

\bibitem[{Raso et~al.(2019)Raso, Pallanca, Ferraro, Lanzoni, Mucciarelli et~al.}]{Raso+2019}
Raso S, Pallanca C, Ferraro FR, Lanzoni B, Mucciarelli A, et~al. 2019.
\textit{The Astrophysical Journal} 879:56

\bibitem[{Renzini \& Fusi~Pecci(1988)}]{Renzini+FusiPecci1988}
Renzini A, Fusi~Pecci F. 1988.
\textit{Annual Review of Astronomy and Astrophysics} 26(1):199--244

\bibitem[{Roriz et~al.(2024)Roriz, Holanda, {da Concei{\c c}{\~a}o}, Junqueira, Drake et~al.}]{Roriz+2024}
Roriz MP, Holanda N, {da Concei{\c c}{\~a}o} LV, Junqueira S, Drake NA, et~al. 2024.
\textit{The Astronomical Journal} 167(4):184

\bibitem[{Rosvick \& Vandenberg(1998)}]{Rosvick+Vandenberg1998}
Rosvick JM, Vandenberg DA. 1998.
\textit{The Astronomical Journal} 115:1516--1523

\bibitem[{Roulston et~al.(2019)Roulston, Green, Ruan, MacLeod, Anderson et~al.}]{Roulston+2019}
Roulston BR, Green PJ, Ruan JJ, MacLeod CL, Anderson SF, et~al. 2019.
\textit{The Astrophysical Journal} 877:44

\bibitem[{Roulston et~al.(2021)Roulston, Green, Toonen \& Hermes}]{Roulston+2021}
Roulston BR, Green PJ, Toonen S, Hermes JJ. 2021.
\textit{The Astrophysical Journal} 922:33

\bibitem[{Rozyczka et~al.(2013)Rozyczka, Kaluzny, Thompson, Rucinski, Pych \& Krzeminski}]{Rozyczka+2013}
Rozyczka M, Kaluzny J, Thompson IB, Rucinski SM, Pych W, Krzeminski W. 2013.
\textit{Acta Astronomica} 63:67--78

\bibitem[{Rozyczka et~al.(2020)Rozyczka, Pych, Thompson \& Mazur}]{Rozyczka+2020}
Rozyczka M, Pych W, Thompson IB, Mazur B. 2020.
\textit{Acta Astronomica} 70:291--299

\bibitem[{Rozyczka et~al.(2016)Rozyczka, Thompson, Narloch, Pych \& {Schwarzenberg-Czerny}}]{Rozyczka+2016}
Rozyczka M, Thompson IB, Narloch W, Pych W, {Schwarzenberg-Czerny} A. 2016.
\textit{Acta Astronomica} 66:307--332

\bibitem[{Rozyczka et~al.(2017)Rozyczka, Thompson, Pych, Narloch, Poleski \& {Schwarzenberg-Czerny}}]{Rozyczka+2017}
Rozyczka M, Thompson IB, Pych W, Narloch W, Poleski R, {Schwarzenberg-Czerny} A. 2017.
\textit{Acta Astronomica} 67:203--224

\bibitem[{Rucinski(2000)}]{Rucinski2000}
Rucinski SM. 2000.
\textit{The Astronomical Journal} 120:319--332

\bibitem[{Sahu et~al.(2022)Sahu, Subramaniam, Singh, Yadav, Valcarce et~al.}]{Sahu+2022}
Sahu S, Subramaniam A, Singh G, Yadav R, Valcarce AR, et~al. 2022.
\textit{Monthly Notices of the Royal Astronomical Society} 514:1122--1139

\bibitem[{Sandquist et~al.(2003)Sandquist, Latham, Shetrone \& Milone}]{Sandquist+2003}
Sandquist EL, Latham DW, Shetrone MD, Milone AAE. 2003.
\textit{The Astronomical Journal} 125:810--824

\bibitem[{Santucci et~al.(2015)Santucci, Placco, Rossi, Beers, Reggiani et~al.}]{Santucci+2015}
Santucci RM, Placco VM, Rossi S, Beers TC, Reggiani HM, et~al. 2015.
\textit{The Astrophysical Journal} 801:116

\bibitem[{Sarajedini et~al.(2007)Sarajedini, Bedin, Chaboyer, Dotter, Siegel et~al.}]{Sarajedini+2007}
Sarajedini A, Bedin LR, Chaboyer B, Dotter A, Siegel M, et~al. 2007.
\textit{The Astronomical Journal} 133:1658--1672

\bibitem[{Schindler et~al.(2015)Schindler, Green \& Arnett}]{Schindler+2015}
Schindler JT, Green EM, Arnett WD. 2015.
\textit{The Astrophysical Journal} 806:178

\bibitem[{Shara et~al.(1997)Shara, Saffer \& Livio}]{Shara+1997}
Shara MM, Saffer RA, Livio M. 1997.
\textit{The Astrophysical Journal} 489:L59--L62

\bibitem[{Shetrone \& Sandquist(2000)}]{Shetrone+Sandquist2000}
Shetrone MD, Sandquist EL. 2000.
\textit{The Astronomical Journal} 120:1913--1924

\bibitem[{Shetye et~al.(2018)Shetye, Van~Eck, Jorissen, Van~Winckel, Siess et~al.}]{Shetye+2018}
Shetye S, Van~Eck S, Jorissen A, Van~Winckel H, Siess L, et~al. 2018.
\textit{Astronomy and Astrophysics} 620:A148

\bibitem[{Sills et~al.(2001)Sills, Faber, Lombardi, Rasio \& Warren}]{Sills+2001a}
Sills A, Faber JA, Lombardi Jr. JC, Rasio FA, Warren AR. 2001.
\textit{The Astrophysical Journal} 548:323--334

\bibitem[{Sills et~al.(2009)Sills, Karakas \& Lattanzio}]{Sills+2009}
Sills A, Karakas A, Lattanzio J. 2009.
\textit{The Astrophysical Journal} 692:1411--1420

\bibitem[{Simunovic \& Puzia(2016)}]{Simunovic+Puzia2016}
Simunovic M, Puzia TH. 2016.
\textit{Monthly Notices of the Royal Astronomical Society} 462:3401--3418

\bibitem[{Simunovic et~al.(2014)Simunovic, Puzia \& Sills}]{Simunovic+2014}
Simunovic M, Puzia TH, Sills A. 2014.
\textit{The Astrophysical Journal} 795:L10

\bibitem[{Sindhu et~al.(2019)Sindhu, Subramaniam, Jadhav, Chatterjee, Geller et~al.}]{Sindhu+2019}
Sindhu N, Subramaniam A, Jadhav VV, Chatterjee S, Geller AM, et~al. 2019.
\textit{The Astrophysical Journal} 882:43

\bibitem[{Sindhu et~al.(2018)Sindhu, Subramaniam \& Radha}]{Sindhu+2018}
Sindhu N, Subramaniam A, Radha CA. 2018.
\textit{Monthly Notices of the Royal Astronomical Society} 481(1):226--243

\bibitem[{Singh \& Yadav(2019)}]{Singh+Yadav2019}
Singh G, Yadav RKS. 2019.
\textit{Monthly Notices of the Royal Astronomical Society} 482:4874--4882

\bibitem[{Smith \& Suntzeff(1987)}]{Smith+Suntzeff1987}
Smith VV, Suntzeff NB. 1987.
\textit{The Astronomical Journal} 93:359

\bibitem[{Sneden et~al.(2003)Sneden, Preston \& Cowan}]{Sneden+2003}
Sneden C, Preston GW, Cowan JJ. 2003.
\textit{The Astrophysical Journal} 592:504--515

\bibitem[{Sollima et~al.(2007)Sollima, Beccari, Ferraro, Fusi~Pecci \& Sarajedini}]{Sollima+2007}
Sollima A, Beccari G, Ferraro FR, Fusi~Pecci F, Sarajedini A. 2007.
\textit{Monthly Notices of the Royal Astronomical Society} 380:781--791

\bibitem[{Sollima et~al.(2008)Sollima, Lanzoni, Beccari, Ferraro \& Fusi~Pecci}]{Sollima+2008}
Sollima A, Lanzoni B, Beccari G, Ferraro FR, Fusi~Pecci F. 2008.
\textit{Astronomy and Astrophysics} 481:701--704

\bibitem[{Sommariva et~al.(2009)Sommariva, Piotto, Rejkuba, Bedin, Heggie et~al.}]{Sommariva+2009}
Sommariva V, Piotto G, Rejkuba M, Bedin LR, Heggie DC, et~al. 2009.
\textit{Astronomy and Astrophysics} 493:947--958

\bibitem[{Starkenburg et~al.(2014)Starkenburg, Shetrone, McConnachie \& Venn}]{Starkenburg+2014}
Starkenburg E, Shetrone MD, McConnachie AW, Venn KA. 2014.
\textit{Monthly Notices of the Royal Astronomical Society} 441:1217--1229

\bibitem[{Stassun et~al.(2023)Stassun, Torres, Kounkel, Tofflemire, Leiner et~al.}]{Stassun+2023}
Stassun KG, Torres G, Kounkel M, Tofflemire BM, Leiner E, et~al. 2023.
\textit{The Astrophysical Journal} 950:99

\bibitem[{Subramaniam et~al.(2020)Subramaniam, Pandey, Jadhav \& Sahu}]{Subramaniam+2020}
Subramaniam A, Pandey S, Jadhav VV, Sahu S. 2020.
\textit{Journal of Astrophysics and Astronomy} 41:45

\bibitem[{Suda et~al.(2008)Suda, Katsuta, Yamada, Suwa, Ishizuka et~al.}]{Suda+2008}
Suda T, Katsuta Y, Yamada S, Suwa T, Ishizuka C, et~al. 2008.
\textit{Publications of the Astronomical Society of Japan} 60:1159

\bibitem[{Sun \& Mathieu(2023)}]{Sun+Mathieu2023}
Sun M, Mathieu RD. 2023.
\textit{The Astrophysical Journal} 944:89

\bibitem[{Sun et~al.(2021)Sun, Mathieu, Leiner \& Townsend}]{Sun+2021}
Sun M, Mathieu RD, Leiner EM, Townsend RHD. 2021.
\textit{The Astrophysical Journal} 908:7

\bibitem[{Thompson et~al.(2008)Thompson, Ivans, Bisterzo, Sneden, Gallino et~al.}]{Thompson+2008}
Thompson IB, Ivans II, Bisterzo S, Sneden C, Gallino R, et~al. 2008.
\textit{The Astrophysical Journal} 677:556--571

\bibitem[{Tofflemire et~al.(2014)Tofflemire, Gosnell, Mathieu \& Platais}]{tofflemire+2014}
Tofflemire BM, Gosnell NM, Mathieu RD, Platais I. 2014.
\textit{The Astronomical Journal} 148(4):61

\bibitem[{{Ulloa-Sol{\'i}s} et~al.(2023){Ulloa-Sol{\'i}s}, Cort{\'e}s, Villanova, Albornoz, Ahumada \& Parisi}]{Ulloa-Sols+2023}
{Ulloa-Sol{\'i}s} D, Cort{\'e}s CC, Villanova S, Albornoz {\'A}L, Ahumada JA, Parisi C. 2023.
\textit{The Astronomical Journal} 166:261

\bibitem[{Vaidya et~al.(2022)Vaidya, Panthi, Agarwal, Pandey, Rao et~al.}]{Vaidya+2022}
Vaidya K, Panthi A, Agarwal M, Pandey S, Rao KK, et~al. 2022.
\textit{Monthly Notices of the Royal Astronomical Society} 511:2274--2284

\bibitem[{Vaidya et~al.(2020)Vaidya, Rao, Agarwal \& Bhattacharya}]{Vaidya+2020}
Vaidya K, Rao KK, Agarwal M, Bhattacharya S. 2020.
\textit{Monthly Notices of the Royal Astronomical Society} 496(2):2402--2421

\bibitem[{{van den Berg} et~al.(2001){van den Berg}, Orosz, Verbunt \& Stassun}]{vandenBerg+2001}
{van den Berg} M, Orosz J, Verbunt F, Stassun K. 2001.
\textit{Astronomy and Astrophysics} 375:375--386

\bibitem[{{Van der Swaelmen} et~al.(2017){Van der Swaelmen}, Boffin, Jorissen \& Van~Eck}]{VanderSwaelmen+2017}
{Van der Swaelmen} M, Boffin HMJ, Jorissen A, Van~Eck S. 2017.
\textit{Astronomy and Astrophysics} 597:A68

\bibitem[{Van~Winckel(2003)}]{VanWinckel2003}
Van~Winckel H. 2003.
\textit{Annual Review of Astronomy and Astrophysics} 41:391--427

\bibitem[{Van~Winckel(2019)}]{VanWinckel2018}
Van~Winckel H. 2019.
\textit{Binary Post-{{AGB}} Stars as Tracers of Stellar Evolution}. In \textit{The {{Impact}} of {{Binaries}} on {{Stellar Evolution}}}, eds. G~Beccary, H~Boffin. Cambridge: Cambridge University Press,  92--105

\bibitem[{Verbunt \& Phinney(1995)}]{Verbunt+Phinney1995}
Verbunt F, Phinney ES. 1995.
\textit{Astronomy and Astrophysics} 296:709

\bibitem[{Vernekar et~al.(2023)Vernekar, Subramaniam, Jadhav \& Bowman}]{vernekarPhotometricVariabilityBlue2023}
Vernekar N, Subramaniam A, Jadhav VV, Bowman DM. 2023.
\textit{Monthly Notices of the Royal Astronomical Society} 524(1):1360--1373

\bibitem[{Vos et~al.(2018)Vos, N{\'e}meth, Vu{\v c}kovi{\'c}, {\O}stensen \& Parsons}]{Vos+2018}
Vos J, N{\'e}meth P, Vu{\v c}kovi{\'c} M, {\O}stensen R, Parsons S. 2018.
\textit{Monthly Notices of the Royal Astronomical Society} 473:693--709

\bibitem[{Vos et~al.(2017)Vos, {\O}stensen, Vu{\v c}kovi{\'c} \& Van~Winckel}]{Vos+2017}
Vos J, {\O}stensen RH, Vu{\v c}kovi{\'c} M, Van~Winckel H. 2017.
\textit{Astronomy and Astrophysics} 605:A109

\bibitem[{Vos et~al.(2019)Vos, Vu{\v c}kovi{\'c}, Chen, Han, Boudreaux et~al.}]{Vos+2019}
Vos J, Vu{\v c}kovi{\'c} M, Chen X, Han Z, Boudreaux T, et~al. 2019.
\textit{Monthly Notices of the Royal Astronomical Society} 482:4592--4605

\bibitem[{Werner et~al.(2020)Werner, Reindl, L{\"o}bling, Pelisoli, Schaffenroth et~al.}]{Werner+2020}
Werner K, Reindl N, L{\"o}bling L, Pelisoli I, Schaffenroth V, et~al. 2020.
\textit{Astronomy and Astrophysics} 642:A228

\bibitem[{Whitehouse et~al.(2018)Whitehouse, Farihi, Green, Wilson \& Subasavage}]{Whitehouse+2018}
Whitehouse LJ, Farihi J, Green PJ, Wilson TG, Subasavage JP. 2018.
\textit{Monthly Notices of the Royal Astronomical Society} 479:3873--3878

\bibitem[{Whitehouse et~al.(2021)Whitehouse, Farihi, Howarth, Mancino, Walters et~al.}]{Whitehouse+2021}
Whitehouse LJ, Farihi J, Howarth ID, Mancino S, Walters N, et~al. 2021.
\textit{Monthly Notices of the Royal Astronomical Society} 506:4877--4892

\bibitem[{Xin et~al.(2015)Xin, Ferraro, Lu, Deng, Lanzoni et~al.}]{Xin+2015}
Xin Y, Ferraro FR, Lu P, Deng L, Lanzoni B, et~al. 2015.
\textit{The Astrophysical Journal} 801:67

\bibitem[{Zhang et~al.(2004)Zhang, Deng, Zhou \& Xin}]{Zhang+2004}
Zhang XB, Deng L, Zhou X, Xin Y. 2004.
\textit{Monthly Notices of the Royal Astronomical Society} 355:1369--1377

\bibitem[{Zloczewski et~al.(2007)Zloczewski, Kaluzny, Krzeminski, Olech \& Thompson}]{Zloczewski+2007}
Zloczewski K, Kaluzny J, Krzeminski W, Olech A, Thompson IB. 2007.
\textit{Monthly Notices of the Royal Astronomical Society} 380:1191--1197

\end{thebibliography}

\end{document}